\documentclass{ar-1col-S2O}
\usepackage[T1]{fontenc}
\usepackage{url}
\usepackage{natbib}
\usepackage{amsmath,amssymb}
\usepackage{xcolor}
\usepackage{wasysym}

\jname{Annu. Rev. Astron. Astrophys.}
\jvol{63 / page}
\jyear{2025}
\doi{10.1146/((please add article doi))}

\newcommand{\mission}[1]{\textit{#1}}
\newcommand{\MEarth}{M_\oplus}
\newcommand{\MJupiter}{M_\mathrm{J}}
\newcommand{\la}{\lesssim}
\newcommand{\ga}{\gtrsim}
\newcommand{\rev}[1]{\textcolor{red}{#1}}

\begin{document}

% Page header
\markboth{M.~Ikoma \& H.~Kobayashi}{}

% Title
\title{Formation of Giant Planets}

%Authors, affiliations address.
%\author{Masahiro Ikoma$^{1,2,3}$, Hiroshi Kobayashi$^4$
%\affil{$^1$Division of Science, National Astronomical Observatory of Japan (NAOJ), Tokyo 181-8588, Japan; email: masahiro.ikoma@nao.ac.jp}
%\affil{$^2$Department of Earth and Planetary Science, The University of Tokyo, Tokyo 113-0033, Japan}
%\affil{$^3$The Graduate University for Advanced Studies (SOKENDAI), Kanagawa 240-0193, Japan}
%\affil{$^4$Department of Physics, Nagoya University, Aichi 464-8602, Japan}
%}

\author{
  Masahiro Ikoma\footnote{Division of Science, National Astronomical Observatory of Japan (NAOJ)} \& Hiroshi Kobayashi\footnote{Department of Physics, Nagoya University}
}
%\affil{{$^1$Division of Science, National Astronomical Observatory of Japan (NAOJ), Tokyo 181-8588, Japan; email: masahiro.ikoma@nao.ac.jp}
%\affil{$^2$Department of Earth and Planetary Science, The University of Tokyo, Tokyo 113-0033, Japan}
%\affil{$^3$The Graduate University for Advanced Studies (SOKENDAI), Kanagawa 240-0193, Japan}
%\affil{$^4$Department of Physics, Nagoya University, Aichi 464-8602, Japan}

%Abstract
\begin{abstract}
Gas giant planets, if present, are the most massive objects in a planetary system and play a pivotal role in shaping its overall architecture. The formation of these planets has constantly been a central issue in planetary science. Increasing evidence from spacecraft explorations of Jupiter and Saturn, as well as telescope observations of exoplanets, has provided new constraints on the formation process of gas giant planets. The classic challenge of explaining formation timescales still remains a significant issue, while new constraints on planetary interiors have introduced additional complexities. Recent shifts away from the single-size planetesimal hypothesis, nevertheless, show promise in resolving these problems. Additionally, various discoveries regarding exoplanets have led to theoretical improvements, while the discovery of numerous super-Earths and sub-Neptunes has posed new challenges in understanding gas accretion. This review synthesizes the latest theoretical advancements, discussing resolved issues and emerging challenges in giant planet formation.
\end{abstract}

%Keywords, etc.
\maketitle

%Table of Contents
\tableofcontents

%%%%%%%%%%%%%%%%
% INTRODUCTION %
%%%%%%%%%%%%%%%%
\section{Introduction}
Exploring the origin of Jupiter, one of the brightest objects in the night sky, would be a captivating scientific pursuit. While Earth is naturally the most significant planet to us as its inhabitants, from an external perspective, Jupiter, being the most massive planet in our Solar System, would likely be seen as its primary representative. The scientific significance of exploring Jupiter's origin may be heightened by the possibility that the planet has an influence on making our Earth habitable. This influence could include delivering essential building blocks for life, such as water and organic molecules, and shielding our planet from catastrophic meteorite impacts. However, the exact role that Jupiter has played in these processes remains uncertain. This uncertainty is partly due to unresolved questions regarding the timing and location of Jupiter’s formation.

In the Solar System, the giant planets--—Jupiter, Saturn, Uranus, and Neptune--—orbit far from the Sun, with semi-major axes of 5.2 astronomical units (au) or more and orbital periods of at least 12 years (or about 4400 days). Their relatively low mean densities suggest that these planets are enveloped by massive atmospheres composed primarily of hydrogen (H) and helium (He). However, their relatively high overall densities indicate the presence of heavier elements within their interiors. The proportion of hydrogen and helium compared to heavier elements varies among these planets. Jupiter and Saturn have relatively low densities (1.3~g/cm$^3$ and 0.8~g/cm$^3$, respectively), implying a higher proportion of hydrogen and helium. In contrast, Uranus and Neptune have higher densities (around 2.5~g/cm$^3$), suggesting a greater concentration of heavier elements.
\begin{marginnote}
    \entry{Mean density}{the quantity defined as the mass divided by the volume}
\end{marginnote}

The debate on the origin of giant planets began in earnest in the 1970s. Jupiter and Saturn, fluid objects dominated by hydrogen and helium (H/He), bear similarities to the Sun and other stars. This resemblance led to the development of a top-down model, suggesting that these planets formed through the gravitational fragmentation of a disk-like nebula (known as the solar nebula or proto-solar disk) surrounding the primordial Sun \citep[the \textit{disk instability} hypothesis; e.g.,][]{Kuiper51,Cameron78,Boss97}. Alternatively, due to the significant presence of heavy elements in the giant planets, a bottom-up model was proposed, extending the formation process of terrestrial planets like Earth. In this model, solids first accumulated to form a central core, after which gas was accreted onto it \citep[the \textit{core accretion} hypothesis; e.g.,][]{Perri+74,Mizuno+78,Mizuno80}. Both hypotheses have their theoretical strengths and weaknesses. However, the core accretion hypothesis has gained broader support, as it provides a straightforward explanation for why the giant planets contain more heavy elements relative to solar abundances, and why the amount of these elements varies from planet to planet.

Beyond the Solar System, most of the giant exoplanets identified so far are also thought to have formed through the core accretion process. 
Although detailed observational data for exoplanets is not as readily available as it is for Solar System planets, the large number of detected exoplanets allows for statistical analyses (see Section~\ref{sec:constraints:exoplanets} for details). 
For example, a noticeable difference exists in the occurrence-mass relationship between planets and brown dwarfs. 
In the planetary mass range ($\lesssim 13 \MJupiter$, where $\MJupiter$ is the mass of Jupiter), occurrence decreases as mass increases; 
in contrast, in the brown dwarf mass range ($\gtrsim 13~\MJupiter$), occurrence increases with mass \citep[e.g.,][]{Udry+02}. 
This suggests a fundamental difference in the formation mechanisms of brown dwarfs and gas giant planets.
Furthermore, a positive correlation between the metallicity of host stars and the occurrence of giant planets has been observed \citep[][]{Santos+04, Fischer+05}. Additionally, many giant exoplanets with measured mean densities, such as the hot Jupiter HD149026~b \citep[][]{Sato+05,Ikoma+06,Fortney+06}, exhibit higher metallicities compared to their host stars, similar to Jupiter and Saturn, suggestive of additional accretion of solids. 
While the disk instability hypothesis remains a possibility, this article primarily focuses on the formation of giant planets via the core accretion hypothesis, unless otherwise noted.

Recent advancements in observational techniques in spectroscopy and imaging have provided valuable insights into the formation of giant planets: The Atacama Large Millimeter/submillimeter Array (ALMA) and the James Webb Space Telescope (JWST) have been contributing to our understanding of protoplanetary disks and giant exoplanets. Despite this progress, however, there are still several unanswered questions. For instance, the timescale for core accretion, the role of planetary migration, and whether the cores of giant planets form primarily through the accretion of planetesimals or pebbles are still actively debated. Furthermore, the composition and internal structure of giant planets, particularly in relation to their heavy element content, require further investigation. Theoretical models must tackle these challenges and enhance our understanding of giant planet formation. In this article, we review both the observational constraints and the physical processes involved in giant planet formation, discuss how current models align with observational data, and consider the prospects for future research.

The rest of this article is organized as follows. In Section~\ref{sec:constraints}, we provide an overview of the observational evidence that constrains the formation of giant planets, both in the Solar System and beyond. In Section~\ref{sec:physics}, we delve into the physical processes involved in giant planet formation, with a focus on recent advancements in the field. Section~\ref{sec:discussion} synthesizes the observational data and theoretical understanding presented in the previous sections, discussing how well current models explain the observed characteristics of giant planets. Finally, we outline future directions and open questions in the study of giant planet formation, highlighting the areas where further research is needed.

%%%%%%%%%%%%%%%
% OBSERVATION %
%%%%%%%%%%%%%%%
\section{Observational Constraints} \label{sec:constraints}
In 1995, three epoch-making events occurred, significantly altering our understanding of subsolar objects: First, in October, the first giant exoplanet 51~Pegasi~b was discovered \citep{Mayor+95}. In November the discovery of a brown dwarf was reported \citep{Nakajima+95}. Meanwhile, in December, the probe released by NASA’s \mission{Galileo} spacecraft entered Jupiter’s atmosphere, conducting in-situ measurements of the elemental abundances\footnote{https://science.nasa.gov/mission/galileo-jupiter-atmospheric-probe/}. 
Over the subsequent three decades, there have been significant advancements in planetary exploration within our Solar System and observations of extrasolar systems. These advances have increased our understanding of how gas giant planets form and evolve. In this section, we give an overview of this progress. Note that we do not delve into the recent observations from the JWST, as these observations are still ongoing. A more detailed discussion of these findings will be reserved for future review articles.

\subsection{Solar system}
%
% SOLAR-SYSTEM: INTERIORS
%
\subsubsection{Interior of the giant planets} \label{sec:solar_interior}

\begin{figure}
    \centering
    \includegraphics[width=\textwidth]{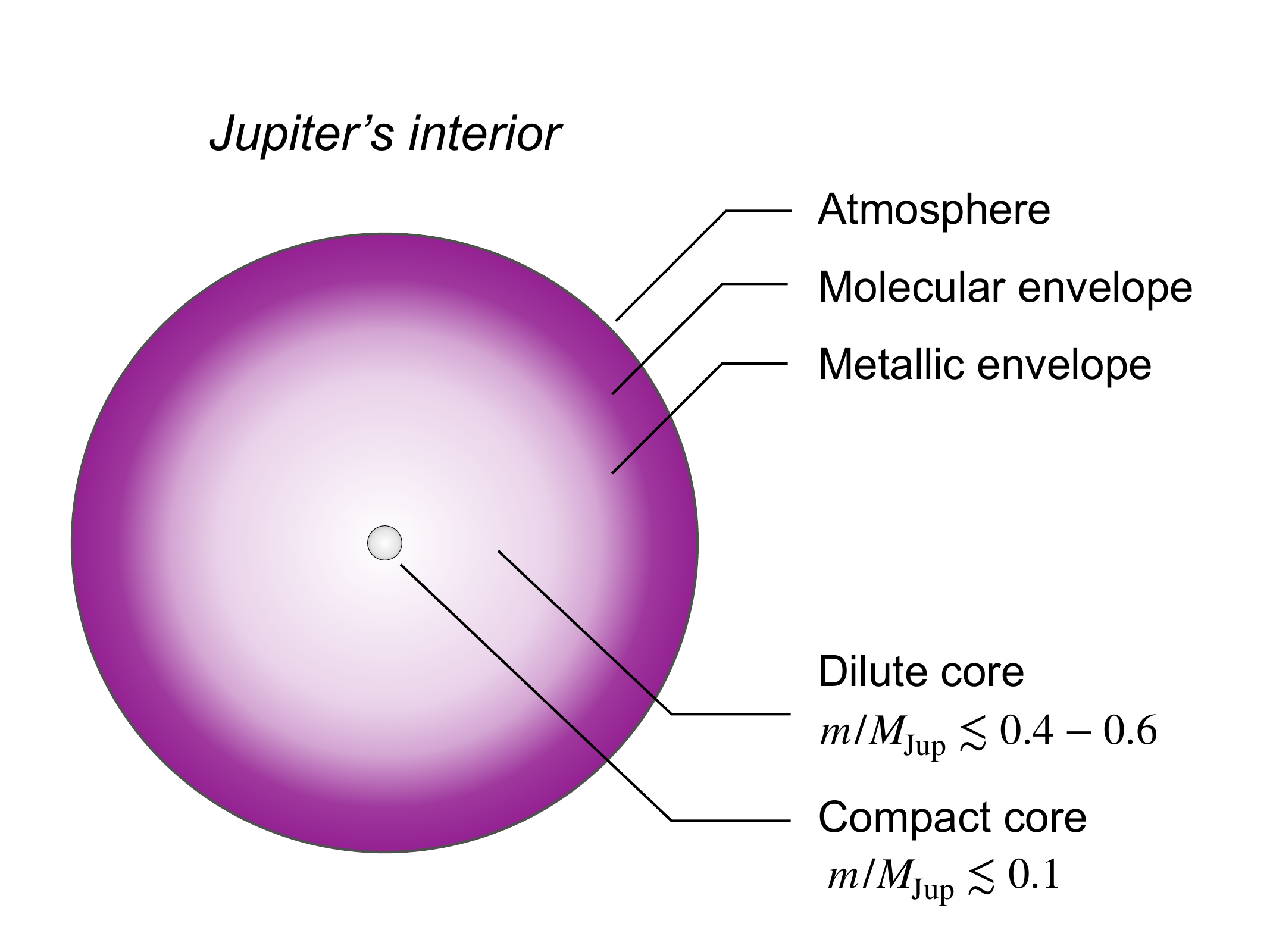}
    \caption{Schematic illustration of Jupiter's interior based on recent models that match \mission{Juno}'s gravity and wind measurements \citep[][]{Wahl+17,Miguel+22,Militzer+22}.
    }
    \label{Fig: Jupiter's interior}
\end{figure}

In the context of the core accretion hypothesis, knowledge of the amount and distribution of heavy elements in the planetary interior is crucial for understanding how the giant planets were formed. To determine the interior structure, we primarily rely on the observed gravity field around the planet and the equation of state (EOS) for hydrogen-helium mixtures at high pressures ($\gtrsim$~100~GPa). Both of these areas have advanced significantly in the last few decades, which has deepened our understanding of the interiors of the giant planets in the Solar System.

% Classic picture
Early interior models estimated that the four giant planets had cores similarly of $\sim$10 to 20~$\MEarth$, despite having different total masses, namely different masses of the H/He-dominated envelopes \citep[e.g.,][]{Slattery77, Hubbard+80}. This aligned with the concept of the so-called critical core mass (or crossover mass), beyond which runaway gas accretion occurs \citep[][see \S~\ref{sec:critical} for the detail]{Mizuno80, Bodenheimer+86}. Additionally, the similarity in core mass led to a theoretical finding of the insensitivity of the critical core mass to the outer boundary conditions for the envelope \citep{Mizuno80, Stevenson82}. 

% Progress in EOS
However, the estimated core mass depends on EOS for H-He mixtures used in interior modeling. \cite{Saumon+04}, for example, presented models of Jupiter's and Saturn's structures based on several different EOSs. They found that Jupiter's core mass is smaller than previously estimated, $\lesssim 10 \MEarth$ or zero, and its envelope contains up to $\sim 40 \MEarth$ of heavy elements. 
Regarding Saturn, they also demonstrated that the same EOSs predict a more massive core ($\sim$10-25 $\MEarth$) with the amount of heavy elements in the envelope estimated to be $0-10 \MEarth$. 
Thus, Jupiter was estimated to contain more heavy elements in total but with a smaller core than Saturn, which posed a challenge to the core accretion model for giant planet formation, as will be discussed later. 
\begin{marginnote}
    \entry{heavy elements}{refer collectively to elements with atomic numbers larger than hydrogen and helium}
\end{marginnote}

% After Juno & Cassini
Recently, the space missions \mission{Juno} and \mission{Cassini} brought new observation data, which have led to modifying understanding of the interiors of Jupiter \citep{Iess+18} and Saturn \citep{Iess+19}, respectively. The \mission{Juno} measurements yielded precise data of the gravitational harmonics of Jupiter: In particular, the measured $J_4$ and $J_6$ are found to be lower in absolute value than predicted by conventional three-layer (dense core + metallic hydrogen layer + molecular hydrogen layer) models for Jupiter. This discrepancy has led to ad hoc models with reduced densities or increased entropies, either partially or entirely in the envelope \citep[e.g.,][]{Wahl+17, Debras+18, Debras+21, Miguel+22}.

Alternatively, a new idea has recently emerged suggesting that Jupiter's interior has compositional gradients with deep regions containing more heavy elements. Many interior models accounting for \mission{Juno}'s gravity-field data include a transition layer of mixed H/He and heavy elements (referred to as a dilute core) between a compact core and a H/He-dominated envelope \citep[e.g.,][and also see Fig.~\ref{Fig: Jupiter's interior} for schematic illustration]{Wahl+17}. Interestingly, the dilute core extends to as large as tens of percent of the planet's radius and mass. The mass of the compact core, which places a crucial constraint on planet formation, is $\lesssim 6-7\MEarth$ or possibly zero, although the estimate is still uncertain and model-dependent \citep[][]{Militzer+22,Miguel+22}. The estimated total amount of heavies ranges from $\sim 20 \MEarth$ (adiabatic) to $60 \MEarth$ (non-adiabatic) \citep[][and references therein]{Helled+24}, which is not inconsistent with previous estimates by \cite{Saumon+04}.

For Saturn, the gravitational harmonics have not been measured as accurately as those for Jupiter. Instead, \mission{Cassini} conducted stellar occultation observation for Saturn's C-ring and detected some density waves generated by low-order normal-mode oscillations within the planet \citep{Hedman+13}, which was predicted by \cite{Marley91}. It turned out that the pulsation spectrum was enriched with internal gravity waves (g modes) \citep{Fuller+14}, which are restored by buoyancy, implying that part of Saturn's interior is stabilized against convection. From their joint gravity-seismology fits, \cite{Mankovich+21} placed tight constraints on the distribution of heavy elements within Saturn: They derived the total mass of heavy elements in Saturn's interior is $19.1\pm1.0\MEarth$, which aligns with previous estimates from three-layer models \citep[e.g.,][]{Saumon+04}. The core was estimated to contain $17.4\pm1.2\MEarth$ of heavy elements, also consistent with estimates ranging from 10 to 25$\MEarth$ based on three-layer models \citep{Saumon+04}. However, a significant difference is that the core contains a substantial amount of H/He, resulting in a total core mass of as much as $55.1\pm1.7\MEarth$, indicating there is a dilute core in Saturn too.

In conclusion, despite precise measurements, the updated bulk compositions of the two planets fall within the range of those from previous three-layer models. Both interiors, however, are unlikely to be three-layered but have a dilute core with compositional gradients extending to approximately 60~\% of the planetary radius and mass. A comparison between recent interior models of Jupiter and Saturn reveals that Saturn's core is more enriched with heavy elements (or less diluted with H/He) than Jupiter’s. Those similarities and differences are valuable constraints on planet formation.

\subsubsection{Atmosphere} \label{sec:atmosphere_solar}

\begin{figure}
    \centering
    \includegraphics[width=\textwidth]{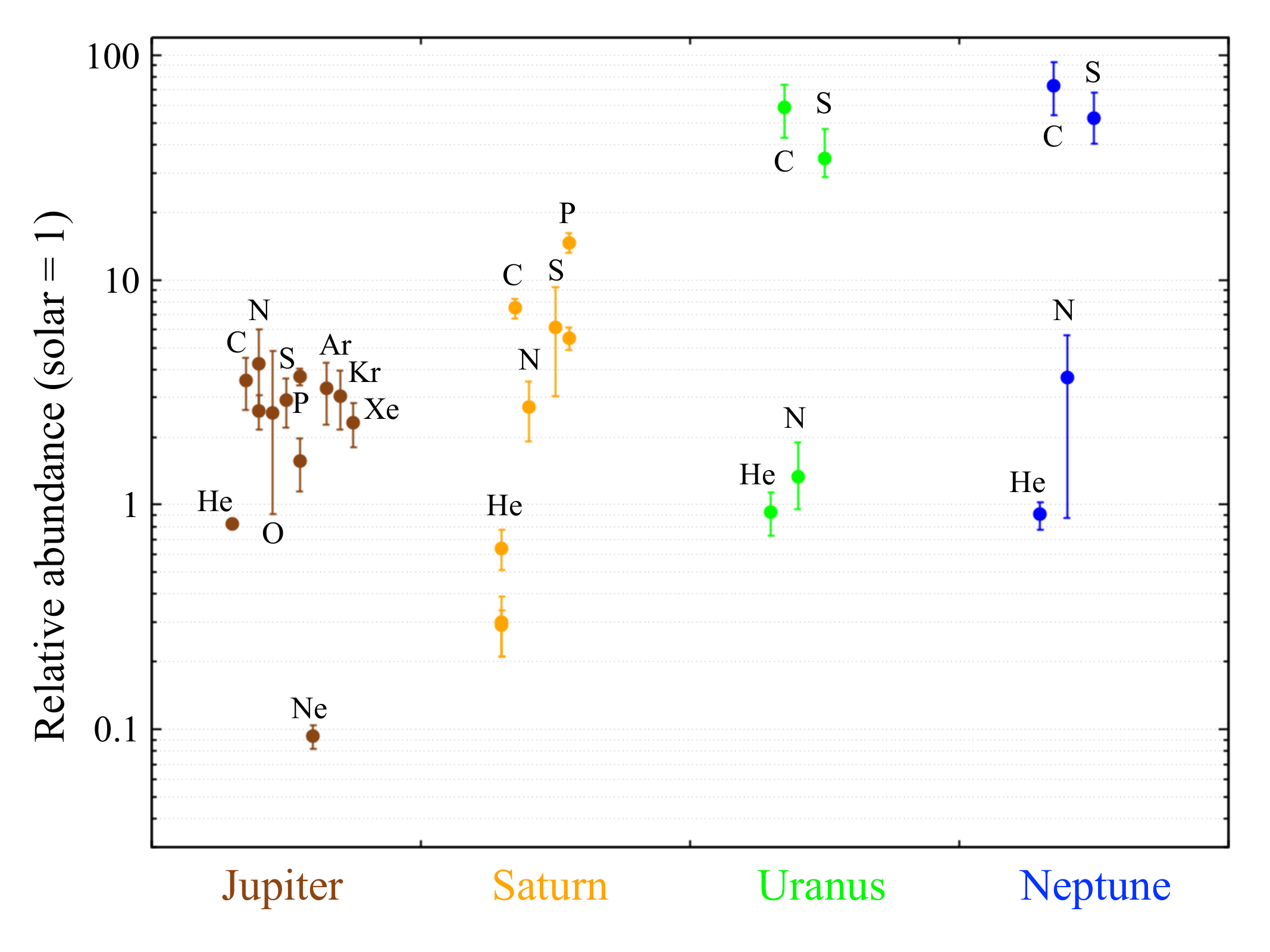}
    \caption{Measured elemental abundances of the atmospheres of the giant planets in the Solar System. Data taken from Table~2 of \cite{Guillot+23}.}
    \label{Fig: atmosphere_elements}
\end{figure}

Atmospheric remote sensing and in-situ measurements have so far shown that the atmospheres of the gas giant planets are enriched with heavy elements, as well as their interiors \citep[see Fig.~\ref{Fig: atmosphere_elements}; data from][]{Guillot+23}. 
In both Jupiter's and Saturn's atmospheres, carbon (C), nitrogen (N), phosphorus (P), sulfur (S), arsenic (As),  and germanium (Ge) have been detected in molecular forms (including condensates) of CH$_4$, NH$_3$ and NH$_4$SH, PH$_3$, H$_2$S and NH$_4$SH, AsH$_3$, and GeH$_4$, respectively. 
For Jupiter, the \mission{Galileo} probe detected noble gases such as argon (Ar), krypton (Kr), and xenon (Xe). 
While the \mission{Galileo}'s non-detection of H$_2$O caused much debate, \mission{Juno}'s microwave radiometer detected H$_2$O at tens of bars and measured its abundance \citep[e.g.,][]{Li+20}. The debate on the presence or absence of water exemplified the difficulty in determining the bulk contents of condensable molecules such as H$_2$O. 

For Jupiter and Saturn, the measured CH$_4$ abundance is expected to represent its bulk abundance because the atmospheric temperatures are too high for CH$_4$ to condense. 
That would be the case with noble gases except neon (Ne), which dissolves into He droplets in metallic hydrogen \citep[e.g.,][]{Wilson+10}. 
For Jupiter, the C/H ratio inferred from the measured CH$_4$ abundance is $(1.19 \pm 0.28) \times 10^{-3}$, which is higher by a factor of $3.56 \pm 0.92$ relative to the proto-solar value \citep{Wong+04}.
For noble gases, the inferred ratios Ar/H, Kr/H, and Xe/H are $3.28 \pm 1.00$, $3.05 \pm 0.90$, and $2.32 \pm 0.52$ times the proto-solar values, respectively \citep{Mahaffy+00}. 
For Saturn, the C/H ratio inferred from the measured CH$_4$ abundance is $(2.50 \pm 0.11) \times 10^{-3}$, which is higher by a factor of $7.50 \pm 0.76$ relative to the proto-solar value \citep{Fletcher+09}.

Phosphorus and sulfur are also notable elements in Jupiter's and Saturn's atmospheres, where those elements are observed primarily in the form of phosphine (PH$_3$) and hydrogen sulfide (H$_2$S), respectively. Phosphine is thought to originate in the deep atmospheres, where the high temperatures and pressures favor its formation. Despite being thermodynamically unstable at higher altitudes, PH$_3$ is still observed as it is transported upwards by convective mixing \citep[][]{Fegley+94, Visscher+06}. Similarly, sulfur is detected as H$_2$S with vertical mixing playing a role in its distribution \citep[][]{Niemann+98,Visscher+06}. The detection of these compounds offers valuable insights into the chemical processes occurring in the deep atmospheres and importantly implies the delivery of those elements in sold forms during the formation of Jupiter and Saturn.

Finally, some interior models for Jupiter show that the outer molecular envelope is less enriched than in the atmosphere \citep[][]{Debras+19,Guillot+23}. 
For Saturn, the amount of heavy elements in the envelope is relatively limited to a metallicity of $0.041\pm0.009$ \citep{Mankovich+21}, which {is approximately 3$\times$solar and} appears to be discrepant with the observed high metallicity of Saturn's atmosphere. 
Although other solutions accounting for the observational constraints have been proposed \citep[][]{Nettelmann+21,Militzer+22}, it is worth exploring whether planet formation processes can lead to atmospheres enriched with heavy elements relative to the interior.

\subsubsection{Cosmochemical constratints} \label{sec:cosmochemistry}
Meteorites come to Earth from the asteroid belt primarily due to Jupiter’s gravitational perturbations. They provide valuable constraints on the planet’s origin, as if Jupiter offers us clues about its own formation history. By studying the isotopic compositions and ages of meteorites, we gain crucial insights into the formation and evolution of giant planets. For a detailed review of these topics, the reader is referred to \cite{Kleine+20} and \cite{Kruijer+20}, while we focus only on the key aspects in this article.
{In summary, recent meteoritic research indicates two reservoirs isolated from each other early in the solar system's history for several million years. This places important constraints on the timing of Jupiter's formation.}

Progress in experimental analysis techniques in the 21st century has enabled the detection of nucleosynthesis isotopic anomalies in meteorites---variations in the isotopic ratios of specific elements that originate from different nucleosynthesis processes, such as the proton-capture ($p$-) process and the slow ($s$-) and rapid ($r$-) neutron-capture processes, occurring in different stellar environments before the formation of the Solar System. Such anomalies are thought to represent the heterogeneous distribution of pre-solar matter derived from multiple nucleosynthesis sources. \citet{Trinquier+07} first identified that at least two nebular reservoirs existed in the early Solar System, distinguishing between differentiated bodies and carbonaceous chondritic bodies based on $^{54}$Cr anomalies. These two reservoirs are now commonly referred to as the non-carbonaceous (NC) and carbonaceous (CC) meteorite groups \citep[e.g.,][who similarly identified two distinct reservoirs for elements such as titanium, nickel, and oxygen]{Warren11}.

\begin{marginnote}
    \entry{$\varepsilon$-unit notation for isotopic variations}{Parts-per-ten-thousand deviations from terrestrial standard values: For example, $\varepsilon^i$Mo represents the deviation of the $^i$Mo/$^{96}$Mo ratio from the terrestrial standard expressed in units of 0.01~\%. }
\end{marginnote}
The dichotomy between NC and CC meteorites (NC-CC dichotomy) is clearly defined by molybdenum (Mo) isotopic anomalies. There are seven stable Mo isotopes, produced through different nucleosynthesis processes, with $^{94}$Mo formed by the $s$- and $p$-processes and $^{95}$Mo by the $s$- and $r$-processes. \cite{Budde+16} demonstrated that in a plot of $\varepsilon^{95}$Mo vs. $\varepsilon^{94}$Mo, NC and CC meteorites align along two distinct, parallel lines, with the CC meteorites having larger $\varepsilon^{95}$Mo values. These isotopic variations along the lines reflect the heterogeneous distribution of $s$-process material, while the offset between the two lines indicates that the CC reservoir is more enriched in $r$-process material compared to the NC reservoir. Subsequent studies confirmed that all known asteroidal meteorites belong to either the NC or the CC group, and the NC-CC dichotomy extends other siderophile elements, including tungsten, ruthenium, and nickel \citep[see][references therein]{Kleine+20}. As such, the NC and CC reservoirs represent fundamentally different materials with almost no mixing between them. 

The hafnium-tungsten (Hf-W) chronology is a common radiometric dating method used to determine the timing of early solar system processes, based on the decay of the short-lived radionuclide $^{182}$Hf into $^{182}$W with a half-life of about 8.9 million years. Since Hf is lithophile (rock-loving) and W is siderophile (metal-loving), their relative abundances in different planetary materials provide insights into the timing of core formation, accretion, and differentiation events in the early solar system. By measuring the $^{182}$W/$^{184}$W ratio in a sample and comparing it to the initial ratio estimated from Calcium-Aluminum-rich Inclusions (CAIs), one can determine the age of metal-silicate separation or core formation of a planetary body.

Using the Hf-W dating method combined with thermal modeling of parent bodies heated internally by $^{26}$Al decay, \cite{Kruijer+17} estimated that core formation began in the NC and CC parent bodies at approximately 0.5~Myr and 1~Myr after CAI formation, respectively. This suggests that the NC and CC reservoirs were already separated by this time. The distinction between the NC and CC reservoirs is shaped by the addition of presolar material enriched in $r$-process nuclides to the region of the solar nebula where the CC meteorites originated. Importantly the NC-CC dichotomy is observed not only in iron meteorites but also in chondrites, indicating that the two reservoirs remained isolated from each other until the parent body accretion in both reservoirs ceased. Chondrite parent bodies accreted at around 2~Myr after CAI formation in the NC reservoir and continued until approximately 3-4~Myr in the CC reservoir \citep[see, e.g., Fig.~4 of][]{Kleine+20}. This means that the NC and CC reservoirs must have remained isolated from each other from $<$1~Myr up to around 3-4~Myr after CAI formation. Such an early but prolonged separation between the two reservoirs places constraints on the timing of Jupiter's formation \citep{Kruijer+17}. 
\subsection{Beyond the Solar system} \label{sec:constraints:exoplanets}

\subsubsection{Occurrence}
Occurrence rates of giant planets and their dependence on orbital distance and stellar properties provide valuable insights into planet formation. A well-known piece of evidence supporting the core accretion hypothesis is the correlation between giant planet occurrence and stellar metallicity. \cite{Santos+04} and \cite{Fischer+05} showed that more close-in giant planets (or hot Jupiters) are detected around Sun-like stars with higher photospheric metallicity. This tendency has also been observed to apply to giant planets with longer orbital periods \citep[$\sim$1-5~au; see][]{Fulton+21}.

Many hot Jupiters{, whose orbital periods and semi-major axes are $\lesssim$~10~days and $\lesssim$~0.1~au, respectively,} have been discovered to date. This is mainly because they are the easiest planets to detect via both radial velocity (RV) and transit methods. Exoplanet demographics, however, indicate that hot Jupiters are relatively uncommon. Radial velocity data analysis estimates the occurrence rate of hot Jupiters to be approximately 1\% for FGK-dwarfs (or solar-type stars) \citep[][and references therein]{Wright+12}. 
{Using the transit method, the occurrence rate of hot Jupiters from the \mission{Kepler} sample}
was estimated to be 0.4$-$0.6\% \citep[e.g.,][]{Howard+12,Fressin+13,Dawson+18}. The difference between the RV surveys and \mission{Kepler}'s transit survey is still a matter of debate. It is possible that the former includes stellar binaries: Indeed, \citet{Beleznay+22} estimated that the occurrence rate of hot Jupiters orbiting single stars in the \mission{Kepler} sample is 0.98~$\pm$~0.36\% for G-dwarfs, which aligns with estimates from RV surveys.

Whereas the \mission{Kepler} survey was designed to focus on FGK-dwarfs and observed a limited number of higher-mass or lower-mass stars, the Transiting Exoplanet Survey Satellite (\mission{TESS}) has provided data useful for exploring planetary occurrence both for A- and M-dwarfs. Following \citet{Zhou+19}'s early work, \citet{Beleznay+22} present their estimates for the hot-Jupiter occurrence based on data from the TESS Prime Mission, taking 97 hot Jupiters orbiting about 200,000 AFG-dwarfs. 
Their estimates are 0.55~$\pm$~0.14\% for G-dwarfs (0.8$-$1.05~$M_\odot$), 0.36~$\pm$~0.06\% for F-dwarfs (1.05$-$1.4~$M_\odot$), and 0.29~$\pm$~0.05\% for A-dwarfs (1.4$-$2.3$M_\odot$). 
This shows a negative correlation between the hot Jupiter occurrence and stellar mass for AFG-dwarfs.
{Note that this is not the case with relatively long-period giant planets (or cold Jupiters) at 1-5~au. Data from the California Legacy Survey suggest that for stars with masses $M_\ast > 1 M_\odot$, the correlation of occurrence with stellar mass is either flat or shows a slight increase \citep[][]{Fulton+21}, indicating the relative occurrence of hot Jupiters to cold Jupiters decreases with stellar mass \citep[][]{Gan+24}.}

As for M-dwarfs, \citet{Gan+23} also used the \mission{TESS} data to estimate the occurrence rate of hot Jupiters orbiting early M-dwarfs (0.45$-$0.65~$M_\odot$) at 0.27~$\pm$~0.09\%. This is lower than the occurrence rate for G-dwarfs described above. \citet{Bryant+23} have recently extended \citet{Gan+23}'s work to late M-dwarfs and conducted a systematic search for transiting giant planets in the \mission{TESS} sample. The estimated occurrence rates are 0.108~$\pm$~0.083\% and 0.137~$\pm$~0.097\% for M~dwarfs of 0.26$-$0.42~$M_\odot$ and 
0.088$-$0.26~$M_\odot$, respectively, which are lower than that for early M-dwarfs. 
As such, the occurrence rate of hot Jupiters depends on the type of host stars and decreases as stellar mass decreases for stars less massive than Sun-like stars. It should be emphasized that giant planets do exist even around late M-dwarfs, albeit low frequency \citep[][]{Kagetani+23,Hartman+23}.

How the occurrence varies with the distance from the host star is an important constraint on planet formation. \citet{Cumming+08} were the first to conduct a broad measurement of giant planet occurrence ($M \sin i \geq 0.1 M_\mathrm{J}$), using eight years of RVs from Keck-HIRES. They found an increase in the abundance of giant planets for orbit near the present-day water-ice line and found that about 10~\% of Sun-like stars have giant planets with semimajor axes of $<$~3~au. Later, \citet{Fulton+21} determined the occurrence of giant planets with longer orbital periods and confirmed the previous result \citep[e.g.,][]{Cumming+08} that the occurrence is enhanced by a factor of four beyond 1~au compared to within 1~au. They also showed that giant planets are more prevalent at 1$-$10~au compared to orbits interior or exterior of that range. They estimated the giant planet rate to be as high as 14.1$^{+2.0}_{-1.8}$~\% with semimajor axes of 2$-$8~au and 8.9$^{+3.0}_{-2.4}$~\% in 8$-$32~au, showing a fall-off at around 8~au \citep[see also][who claimed such a broken power law distribution]{Fernandes+19}. \cite{Fulton+21} also found a qualitatively similar occurrence enhancement around 1$-$10~au for both the sub-Jovian planets of 0.1$-$1$\MJupiter$.

Finally, in the \mission{Kepler} sample, gas giants are quite rare compared to smaller planets such as super-Earths and sub-Neptunes at orbital periods of less than 100 days. \mission{Kepler} data have shown that only about 1\% of Sun-like stars host a hot/warm Jupiter, whereas the occurrence rates of super-Earths and sub-Neptunes are approximately 30–50\% \citep{Howard+12,Fressin+13}. This demographic distinction is crucial for shaping our understanding of planetary formation and migration processes. Additionally, the \mission{Kepler} data reveal there is a clear gap in the planet size distribution between two populations typically around 1.8~$R_\oplus$, which is sometimes called the radius valley \citep{Fulton+17}. This feature indicates a division in planet formation and evolutionary processes: Planets below this gap tend to be rocky super-Earths, while planets above it retain significant hydrogen-helium atmospheres, classifying them as sub-Neptunes. 

\subsubsection{Internal composition} \label{sec:extrasolar_interior}
The bulk composition of gas giant planets often differs significantly from that of their host stars. This difference is thought to be due to the different processes through which solids and gas accumulate in the core accretion model, often leading to gas giants that have much higher heavy element contents than their host stars. Those gas giants are sometimes referred to as metal-rich gas giants. 

An extreme example of metal-rich gas giant is HD~149026b, a planet discovered using the Subaru Telescope \citep{Sato+05}. This planet is notable for having an extraordinarily high bulk density compared to its host star. Interior models of HD~149026b estimate that its interior contains an immense amount of heavy elements, with $\gtrsim 50$$M_\oplus$ of heavy elements out of a total planet mass of 110$M_\oplus$ \citep{Ikoma+06,Fortney+06}. This implies that nearly half of the planet’s total mass is composed of elements heavier than hydrogen and helium. Such a high heavy element content strongly supports the core accretion hypothesis. However, it is important to note that core accretion models do not always predict such massive central cores for gas giants \citep[][]{Ikoma+06}, implying that HD~149026b likely has a heavy-element-enriched envelope as well. This points to the significance of continued solid accretion during and possibly even after the phase of gas accretion, contributing to the enrichment of the planet’s envelope in heavy elements.

Further evidence for the prevalence of metal-rich gas giants comes from statistical studies: \cite{Guillot+06} and \cite{Miller+11} initially pointed out a positive correlation between planetary heavy element content and the stellar metallicity [Fe/H], using relatively small samples of exoplanets. They observed that metal-rich stars tend to form planets with higher heavy element contents. 
However, \cite{Thorngren+16}, using a larger sample, did not robustly confirm this correlation. 
Meanwhile, \cite{Thorngren+16} found a clear correlation between the planetary heavy element content $M_Z$ and the total planet mass $M_p$, which is approximately given by $M_Z \propto \sqrt{M_p}$. This indicates that more massive planets tend to have larger absolute amounts of heavy elements, but the bulk metallicity, defined as the ratio of heavy element mass to total mass ($Z_p = M_Z/M_p$), decreases with increasing planet mass. Specifically, $Z_p$ follows an inverse square root dependence on the planet mass, $Z_p \propto M_p^{-1/2}$, suggesting that smaller gas giants tend to have higher bulk metallicities relative to their total mass, while larger planets are more dominated by hydrogen and helium.
Note that \cite{Muller+20} confirmed this mass-metallicity trend but found that the correlation disappears when removing a 20~$M_\oplus$ core from the planets in their models. They highlighted that theoretical uncertainties, such as core mass assumptions and opacity models, significantly affect the interpretation of planetary composition, suggesting that the heavy element content may be more complex than previously thought.
\color{black}

\subsubsection{Atmospheres} \label{sec:atmosphere_extrasolar}
Recent advancements in space-based observatories such as the HST and the JWST have provided deeper insights into the atmospheric characteristics of giant exoplanets. 
In contrast to the giant planets in the Solar System, whose atmospheres are cold enough for water to condense, water vapor can thermodynamically exist and may be detectable in the atmospheres of close-in gas giants \citep[e.g.,][]{Tinetti+07}. Transmission spectra obtained with the Wide Field Camera 3 (WFC3) onboard the HST have provided a wealth of data (around the wavelength of 1.4~$\mu$m) on water vapor in exoplanet atmospheres \citep[][]{Sing+16, Pinhas+19}. It turned out that water features are significantly diverse, suggesting that water contents may differ from planet to planet, or clouds and hazes may affect the spectral features. 

The improved resolution and extended wavelength coverage of the JWST allow for more refined analyses of atmospheric spectra of transiting exoplanets. 
Its Early Release Science (ERS) observations have provided unprecedented insights into the atmospheric composition of gas giants, starting with the detection of key molecules in the atmosphere of WASP-39b \citep[][]{Ahrer+23}. The identification of CO$_2$ is significant because it indicates that the atmosphere is rich in heavy elements (or metals) and therefore impacts the understanding of the planet formation process. In addition to CO$_2$, JWST identified other molecules, including SO$_2$, Na, and H$_2$O, in WASP-39b’s transmission spectra \citep[][]{Alderson+23,Rustamkulov+23} {\citep[also see][for SO$_2$ detection in WASP-107b's atmosphere]{Dyrek+24}}. The presence of SO$_2$ is particularly important, as it suggests active photochemical processes in the atmosphere {\citep[][]{Polman+23}}. 

The atmospheric metallicity of hot Jupiters provides a useful constraint on planet formation. \cite{Bean+23} recently estimated the atmospheric metallicity of the Saturn-mass hot Jupiter HD~149026b based on the CO$_2$ and H$_2$O features in the thermal emission spectrum measured by the JWST, and found that its atmosphere is quite enriched in heavy elements, far exceeding expectations based on the planet mass. 
{These results from the JWST align with its inferred high bulk metallicity based on measured mass and radius.}
This challenges the conventional trend in the solar system, where more massive planets tend to have lower atmospheric metallicity. Their analysis, which includes data from WASP-39b, WASP-77Ab, and the Solar System giants, highlights the diversity in the compositions of giant planet atmospheres, suggesting a more complex relationship between planetary mass and atmospheric composition than previously thought. 
Notably, \cite{Bean+23} found that atmospheric metallicity is more strongly correlated with bulk metallicity than with planet mass. 
This indicates that factors other than mass, such as the overall heavy element content, could play a crucial role in shaping a planet's atmosphere. These findings underscore the importance of taking into account the complete composition of a planet when researching its atmospheric characteristics, thus adding to our comprehensive knowledge of the formation of giant planets.
JWST observations are still in the early stages, and as more data is collected and our understanding continues to evolve, deeper insights will emerge. Therefore, this article will refrain from delving too deeply into the results from JWST, leaving a more detailed analysis for future review papers.

%\subsubsection{Protoplanetary disks}

%%%%%%%%%%%
% PHYSICS %
%%%%%%%%%%%
\section{Physics Relevant to Giant Planet Formation} \label{sec:physics}

\color{black}

\subsection{Proto-envelope structure}
\subsubsection{Envelope retention conditions}
Planets can maintain an atmosphere (or an envelope) when their gravity is strong enough to overcome the thermal motion of gas. 
The condition for atmospheric retention is mathematically expressed as
\begin{equation}
    1 \lesssim \frac{GM_p}{R_p c^2} \equiv \lambda,
\end{equation}
where 
$M_p$ and $R_p$ are the mass and radius of the planet, respectively, 
$G$ is the gravitational constant, and 
$c$ is the thermal velocity of the gas. 
The non-dimensional quantity $\lambda$ is termed the escape parameter, which is estimated to be
\begin{marginnote}
    \entry{atmosphere and envelope}{For growing planets, the terms "atmosphere" and "envelope" are not distinguished from each other, unless otherwise stated, unlike for the present-day giant planets.}
\end{marginnote}
\begin{equation}
    \lambda = 27 
    \left(\frac{M_p}{0.1\MEarth}\right)^{2/3}
    \left(\frac{T}{100~{\rm K}}\right)^{-1}
    \left(\frac{\mu}{2.3}\right)^{}
    \left(\frac{\bar{\rho}}{2 \rm g \, cm^{-3}}\right)^{1/3},
\end{equation}
where $T$ and $\mu$ are the temperature and mean molecular weight of the gas, respectively, and $\bar{\rho}$ is the planet's mean density. 
This indicates that 10\% of the Earth's mass (i.e., Mars' mass) is sufficient for a planet to possess a H-He envelope in the absence of external influences. 

During the early stages of planet formation, planets are embedded in a protoplanetary disk. 
The outer layer of the planet's envelope seamlessly merges with the surrounding disk, and there is no distinct boundary between the two.
The boundary can be physically defined as follows: 
For $\lambda = 1$, the Bondi radius is defined as \citep[][]{Bondi52}
\begin{equation}
    R_\mathrm{B} = \frac{GM_p}{c^2}.
\end{equation}
In addition, tidal forces can present the gravitational binding of its envelope, as the planet orbits its host star. 
This is expressed by the Hill radius as \citep[][]{Hill1878}
\begin{equation}
    R_\mathrm{H} = a \left(\frac{M_p}{3M_\ast}\right)^{1/3},
\end{equation}
where $a$ is the semi-major axis and $M_\ast$ is the host star mass. 
These two radii depend on planetary mass in different ways:
In low-mass regimes, the Bondi radius is less than the Hill radius, while the opposite is true in high-mass regimes.
For instance, $M_p \simeq$~20~$\MEarth$ for $R_\mathrm{B} = R_\mathrm{H}$, if $T$~=~100~K, $\mu$~=~2.3, $a$~=~5~au, and $M_\ast$~=~1~$M_\odot$. 

Most models of growing gas giants assume that the envelope connects smoothly to the surrounding disk; specifically, the pressure and temperature at the outer boundary of the envelope are presumed to be the same as those of the disk.
Three-dimensional hydrodynamic simulations of the flow around a protoplanet, however, demonstrate that some of the accretion flow is not gravitationally bound by the planet and therefore returns to the surrounding disk \citep[e.g.,][]{Lissauer+09, D'Angelo+13}. Further discussion is presented in \S~\ref{sec:recycling}.

\subsubsection{Physical description of envelope structure}
It is safe to assume that the deep envelope is in hydrostatic equilibrium and has a spherically symmetric structure.
The structure can be mathematically described by the following two equations:
\begin{equation}
    \frac{\partial P}{\partial r} = - \frac{GM_r}{r^2} \rho,
\label{eq:hydrostatic}
\end{equation}
\begin{equation}
    \frac{\partial M_r}{\partial r} = 4 \pi r^2 \rho,
\label{eq:Mr}
\end{equation}
where $M_r$ is the total mass contained inside the sphere of radius $r$, $P$ the pressure, and $\rho$ the mass density.

To solve the two equations above, one needs the relationship between $P$ and $\rho$, namely, an equation of state (EOS). 
Since the accreting envelope is hot enough and relatively less dense (especially below the critical mass; see \S~\ref{sec:critical}), the ideal-gas approximation causes no large errors \citep[e.g.,][]{Bodenheimer+86}:
\begin{equation}
    P = \frac{k_\mathrm{B}}{m_\mathrm{u}} \frac{\rho T}{\mu (\rho, T)},
\end{equation}
where 
$k_\mathrm{B}$ is the Boltzmann constant and $m_\mathrm{u}$ is the atomic mass unit. 
In accreting envelopes, various chemical reactions such as the dissociation of H$_2$ occur and, thereby, the mean molecular weight and heat capacity vary significantly in the envelope, which greatly affect the envelope structure \citep[][]{Mizuno+78,Hori+11,Venturini+15}. 

Temperature is closely related to energy transfer. 
During the stages of planet growth, a substantial amount of accretion energy is brought into the envelope, leading to vigorous convection in the deep envelope.
In the shallow regions of the envelope where the density is lower, radiation dominates energy transfer rather than inefficient convection.
The energy transfer equation is often expressed in the form
\begin{equation}
    \frac{\partial T}{\partial r} = 
    \nabla \frac{T}{P} \frac{\partial P}{\partial r}
\end{equation}
with $\nabla$ being dependent on energy transfer mechanisms. In dry convective regions, the adiabatic is a dry one, $\nabla_\mathrm{dry}$:
{For a perfect gas, for example, $\nabla_\mathrm{dry} = 1 - \gamma^{-1}$, with the specific heat ratio $\gamma$, and is constant, while $\nabla_\mathrm{dry}$ varies in real envelopes due to chemical reactions \citep[e.g.,][]{Hori+11}.} 
When the envelope contains condensable species such as H$_2$O, the \textit{wet} adiabat is given by \citep[see, e.g.,][]{Kurosaki+17}
\begin{equation}
    \nabla = \nabla_\mathrm{dry}
    \frac{1+\sum_i \displaystyle{\frac{x_i}{1-x_i}\nabla_{\mathrm{vap},i}}}
    {1+\displaystyle{\frac{R_g}{C_p}\sum_i \frac{x_i}{1-x_i}\nabla_{\mathrm{vap},i}}}
    \,\,\, \equiv \nabla_\mathrm{wet},
\end{equation}
where $x_i$ is the mole fraction of the condensable species $i$, 
$R_g$ the gas constant (= $k_\mathrm{B}/m_\mathrm{u}$), 
$C_p$ the specific heat capacity at constant pressure, 
and $\nabla_{\mathrm{vap},i} \equiv d \ln P^\ast / d \ln T$ with the vapor pressure $P^\ast$.
In optically thick radiative regions, the so-called radiation-conduction approximation is valid and $\nabla$ is given as
\begin{equation}
    \nabla = \frac{3}{64 \pi \sigma_\mathrm{SB}G}
            \frac{\kappa L_r}{M_r} \frac{P}{T^4}
            \,\,\, \equiv \nabla_\mathrm{rad},
\label{eq:radiation conduction}
\end{equation} 
where $\sigma_\mathrm{SB}$ is the Stefan-Boltzmann constant, $\kappa$ the Rosseland-mean opacity, and $L_r$ the total energy flux passing through a sphere of radius $r$. 

The total energy flux passing through a sphere of radius $r$, which is often called the luminosity, is given by 
\begin{equation}
    \frac{\partial L_r}{\partial M_r} = \epsilon - T \frac{ds}{dt}, 
\label{eq:entropy}
\end{equation}
where $\epsilon$ is the energy generation rate per unit mass {and $s$ the specific entropy,} and the second term is the rate of generation/extinction of heat. 
In the early stages of planetary formation, heavy bombardments of core building blocks (or solids) provide a significant amount of energy, with the first term dominating over the second.
If all incoming solids reach the surface of the core with a mass accretion rate of $\dot{M}_\mathrm{sol}$, then the luminosity can be approximated by 
\begin{equation}
    L_r % \approx \int_{R_\mathrm{core}}^r \epsilon \, dM_r 
    \approx \frac{GM_\mathrm{core}}{R_\mathrm{core}} \dot{M}_\mathrm{sol},
\label{eq:luminosity}
\end{equation}
where $M_\mathrm{core}$ and $R_\mathrm{core}$ are the mass and radius of the core, respectively. 

\subsubsection{Nature of embedded envelopes}
Approximate solutions obtained analytically provide insights into the properties and behavior of envelopes that connect to the surrounding disk.
Under the assumption that the envelope mass is negligibly small relative to the core mass (i.e., $M_r = M_\mathrm{core}$), no energy generation occurs within the envelope (i.e., $L_r =$ constant $\equiv L$), and the quantities $\mu$, $\gamma$, and $\kappa$ are also constant throughout the envelope, the integration of 
the equations above yields the density profile in deep parts of the wholly radiative envelope as \citep[][]{Mizuno80, Stevenson82}, 
\begin{equation}
    \rho \approx \frac{\pi \sigma_\mathrm{SB}}{12 \kappa L} \frac{\Phi^4}{r^3},
\label{eq:density_rad}
\end{equation}
where $\Phi$ is a function of the core mass and mean molecular weight, namely,
\begin{equation}
    \Phi \, (M_\mathrm{core}, \mu) \equiv \frac{GM_\mathrm{core}\mu m_\mathrm{u}}{k_\mathrm{B}}.
\label{eq:Phi}
\end{equation}
The equation shows that mass density $\rho$ increases linearly as luminosity $L$ or opacity $\kappa$ decreases and does not explicitly depend on the outer boundary conditions \citep[see the Appendix of][for the exact integration under these assumptions]{Inaba+03AA}.
In fully convective envelopes, the density profile is given by \citep[][]{Wuchterl93, Inaba+03AA}
\begin{equation}
    \rho = \frac{\rho_0/T_0^n}{(1+n)^{n}}
    \frac{\Phi^n}{r^n},
\label{eq:density_con}
\end{equation}
where $n \equiv 1/(\gamma-1)$, and $\rho_0$ and $T_0$ are the density and temperature, respectively, at the outer boundary of the envelope. 
The structure of the deep convective envelope depends on the outer boundary conditions, unlike radiative envelopes. Usually, embedded envelopes mainly made up of H and He have a broad radiative layer where the impact of the outer boundary conditions diminishes quickly as one moves inward, except in extreme situations such as dense or cold protoplanetary disks \citep[][]{Ikoma+01}.

The temperature is given using a similar analytical approach and simply expressed as
\begin{equation}
    T \approx \frac{1}{1+n^\ast} \frac{\Phi}{r},
\end{equation}
where $n^\ast = 3$ for radiative envelopes and $n^\ast = n$ for convective envelopes.
It is interesting to note that even for radiative envelopes, the temperature in deep regions is independent of luminosity and opacity, but determined only by the core mass and mean molecular weight. This feature is obviously different from that of isolated atmospheres. 
While isolated atmospheres change temperature due to variations in luminosity or opacity, embedded atmospheres adjust by altering their mass (i.e., optical depth).
The temperature is not affected by the outer boundary conditions for convective envelopes as well as radiative ones.

\subsubsection{Critical/crossover mass for the onset of runaway gas accretion} \label{sec:critical}
The envelope mass $M_\mathrm{env}$ is obtained by integrating the density profile from the core surface ($r = R_\mathrm{core}$) to the outer boundary ($r = R_0)$. 
Here we focus on radiative envelopes: Using Eq.~(\ref{eq:density_rad}), we obtain
\begin{equation}
    M_\mathrm{env} \approx 
    \frac{\pi^2 \sigma_\mathrm{SB} }{3}
    \frac{\Phi^4}{\kappa L}, %\ln \left( \frac{R_0}{R_\mathrm{core}}\right)
\label{eq:envelope_mass}
\end{equation}
where the factor $\ln (R_0/R_\mathrm{core})$ has been omitted. 
This demonstrates that the envelope mass increases rapidly with the core mass (see Eq.~(\ref{eq:Phi})). 
The mass at which $M_\mathrm{env}$ = $M_\mathrm{core}$ is called the \textit{crossover mass}. 
Using Eqs.~(\ref{eq:Phi}) and (\ref{eq:envelope_mass}), the crossover mass is given by 
\begin{equation}
    M_\mathrm{core}^\dagger \approx \Theta_L \frac{L^{1/3} \kappa^{1/3}}{\mu^{4/3}},
\label{eq:crossover mass}    
\end{equation}
where $\Theta_L^{3} \equiv (3/\pi^2\sigma_\mathrm{SB}) (k_\mathrm{B}/Gm_\mathrm{u})^4$, or using the solid accretion rate $\dot{M}_\mathrm{sol}$ (see Eq.~(\ref{eq:luminosity})), 
\begin{equation}
    M_\mathrm{core}^\dagger \approx \Theta_M \frac{\dot{M}_\mathrm{sol}^{3/7}\kappa^{3/7}}{\mu^{12/7}},
\label{eq:crossover mass 2}
\end{equation}
where $\Theta_M^7 \equiv \Theta_L^9 G^3 4 \pi \bar{\rho}/3$.
For $\dot{M}_\mathrm{sol}= 1 \times 10^{-6} \MEarth/{\rm yr}$, $\kappa = 0.01 {\rm cm^2/g}$, $\mu =2.3$, $\rho = 4 {\rm g/cm^3}$, for example, Eq.~{(\ref{eq:crossover mass 2})} yields $M_\mathrm{core}^\dagger = 28 \MEarth$ 
\citep[See][for numerical calculations]{Ikoma+00, Hori+10}

Numerical integration shows that no purely hydrostatic solutions exist beyond a critical mass of the core for a given value of $L$ or $\dot{M}_\mathrm{sol}$ \citep[e.g.,][]{Perri+74,Mizuno80}. The critical core mass $M_\mathrm{core}^\mathrm{crit}$ is similar in value 
to 
the crossover mass \citep[e.g.,][]{Bodenheimer+86}.
It was previously thought that once the critical core mass was surpassed, the envelope would undergo gravitational ``collapse'' \citep[e.g.,][]{Perri+74, Mizuno80}. However, what actually occurs is quasi-static gravitational ``contraction'', similar to what takes place in stellar evolution \citep[][]{Bodenheimer+86,Ikoma+00}.
Once envelope mass becomes similar to core mass, both the deep envelope as well as the core contribute to gravity, pulling the outer envelope inward. 
At this point, the energy provided by solid accretion is not enough, causing the envelope to contract and release potential energy to maintain hydrostatic equilibrium. 
However, the contraction increases the envelope's self-gravity, leading to further contraction. 
This causes the envelope to rapidly accumulate in a process known as runaway gas accretion. 

As Eq.~(\ref{eq:crossover mass 2}) shows, the crossover mass ($M_\mathrm{core}^\dagger$) is positively linked to the solid rate ($\dot{M}_\mathrm{sol}$) associated with the energy source and the opacity ($\kappa$) related to the envelope's blanketing effect. 
While the power index for $\dot{M}_\mathrm{sold}$ and $\kappa$ is 3/7 in the analytical expression above, the actual dependence is slightly weaker due to convection at deeper levels. Detailed numerical calculations estimate that the index is approximately 0.2-0.3 \citep{Ikoma+00}. 
Note that such a weak dependence of the crossover mass on the solid accretion rate suggests that the rapid growth of cores leads to large cores. 
Detailed discussion on the solid accretion rate $\dot{M}_\mathrm{sol}$ is deferred to Section~\ref{sec:core accretion}.

Opacity{, especially in the infra-red,} is an important factor that controls the envelope accumulation. 
In protoplanetary envelopes, opacity is caused by dust grains in low-temperature regions ($\lesssim$~1500~K) and by gas molecules in high-temperature regions. 
As an extreme case, dust-free envelopes with solar abundances can trigger runaway accretion at core masses as small as $\sim$1 to 2~$\MEarth$ \citep{Hori+10}.
Despite its significance, many studies have traditionally taken the interstellar medium (ISM) opacity of $\sim$1~cm$^2$/g as the reference and simply multiplied it by an arbitrary depletion factor \citep[e.g.,][]{Mizuno80}.
\cite{Podolak03} was the first to conduct direct calculations of dust opacity in protoplanetary envelopes by solving the \cite{Smoluchowski16} coagulation equation with grain microphysics such as settling, collisional growth, and destruction via vaporization \citep[also see][]{Movshovitz+08}. 
They demonstrated that the grain size distribution in the envelopes is quite different from that in the ISM and the dust opacity is much lower than the ISM value.
Later, \cite{Movshovitz+10} incorporated their dust opacity calculations into the numerical code of \cite{Pollack+96} and simulated giant planet growth in a self-consistent way, demonstrating that the formation time as well as the final core mass are greatly reduced. 
More extensive investigations were conducted for dust opacity in protoplanetary envelopes using semi-analytical \citep{Mordasini14} or simplified microphysical models \citep{Ormel14}. 
They show that dust grains grow through the differential settling of grains of different sizes {in the envelope}, independent of the dust injection rate, rendering the radiative region devoid of dust grains.

\subsubsection{Polluted envelope}

\begin{figure}
    \centering
    \includegraphics[width=\linewidth]{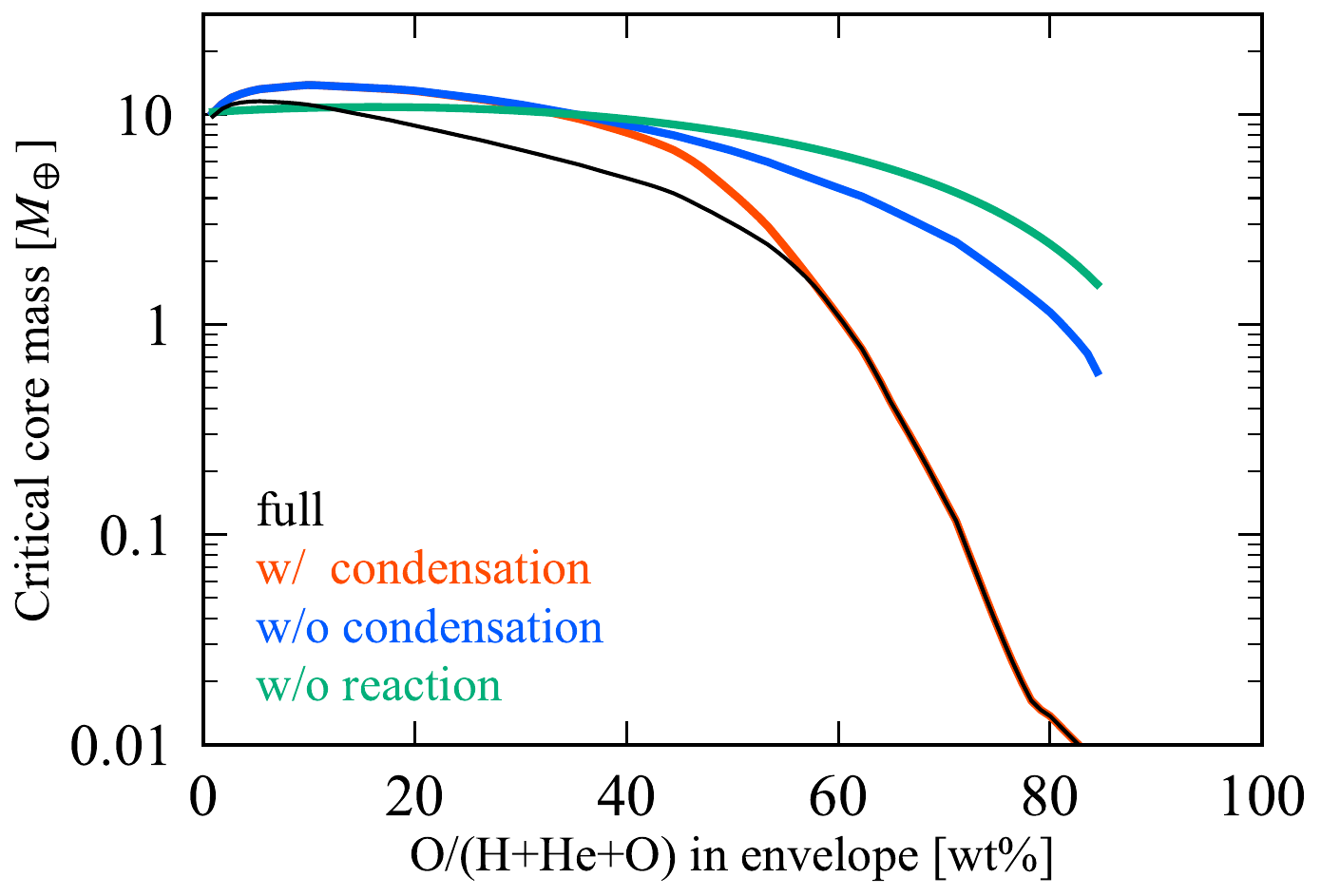}
    \caption{The critical core mass for polluted envelopes is shown. The black curve represents the full-model results with grain opacity calculated based on \cite{Ormel14}, while the other curves assume interstellar opacity from \cite{Semenov+03}. The red curve considers water condensation, the blue curve does not, and the green curve does not include any reactions relevant to H$_2$O. The solid accretion rate is set at $1 \times 10^{-5} \MEarth/{\rm yr}$. The disk pressure and temperature are assumed to be 0.01~Pa and 150~K, respectively. The planet is assumed to orbit at 5.2~au from a Sun-like star. These calculations were conducted using the code developed by \cite{Kimura+20}.}
    \label{fig:critical_polluted}
\end{figure}

The above discussion assumes that protoplanetary envelopes are primarily made of hydrogen and helium, for simplicity. However, many giant planets, both in and outside the Solar System, are known to contain significant amounts of heavy elements, as discussed in Section~\ref{sec:constraints}.
In addition, recent core accretion models suggest that small solid particles contribute to core growth, as will be discussed in Section~\ref{sec:core accretion}. 
Those small particles are expected to readily evaporate on the way to the core, polluting the envelope with heavy elements \citep[][]{Podolak+88,Valletta+19}.

The pollution increases the molecular weight of the envelope gas, leading to a reduction in the envelope's pressure scale height. This causes the envelope to contract and gain additional mass (see Eq.~[\ref{eq:envelope_mass}]). Consequently, the increased concentration of heavy elements reduces the critical core mass (or the crossover mass; see Eqs.~[\ref{eq:crossover mass}] and [\ref{eq:crossover mass 2}]). This was initially proposed analytically by \cite{Stevenson82} and further explored numerically by \cite{Hori+11}.

It is often stated in the literature that an increase in the average molecular weight leads to a decrease in critical core mass. However, this explanation may not be completely accurate, as other effects play more significant roles. 
{Figure~\ref{fig:critical_polluted} shows the dependence of the critical core mass on the heavy element abundance in the envelope:}
\cite{Hori+11} considered molecules containing carbon and oxygen (such as H$_2$O, CH$_4$, CO, CO$_2$) in addition to hydrogen and helium, and integrated the envelope structure simultaneously solving for chemical equilibrium. Their findings reveal that the decrease in critical core mass is much more significant than what is predicted based solely on the effect of mean molecular weight{, demonstrating} the importance of the increase in the effective specific heat associated with chemical reactions on envelope structure and critical core mass (see the green and blue curves of Fig.~\ref{fig:critical_polluted}). Additionally, if volatile components condense within the envelope, this should be treated as a separate chemical reaction. \cite{Venturini+15} carried out similar computations, considering the effects of water vapor sublimation/condensation{, and} found that the impact of latent heat is significant, resulting in a critical core mass that is orders of magnitude smaller compared to when dry convection is assumed (see the red curve of Fig.~\ref{fig:critical_polluted}).

To determine whether pollution actually speeds up the formation of the envelope, several factors need to be considered. First, chemical reactions do not occur evenly throughout the envelope but take place within specific temperature ranges. Condensation specifically occurs in the cold outer envelope, while other chemical reactions happen at temperatures of several hundred kelvins. Therefore, it is crucial to accurately account for the processes of evaporation, pollution, and subsequent mixing. This is particularly important during the runaway gas accretion phase, where the mixing process between the fresh H/He gas and the pre-existing polluted gas should be properly addressed. \cite{Venturini+16} demonstrated that runaway gas accretion proceeds after the critical core mass is attained without slowing down due to dilution by fresh H/He gas, assuming complete mixing in the envelope, which remains to be verified.

\subsection{Gas accretion} \label{sec:gas_accretion}
When the critical core mass or crossover mass is reached, the envelope’s self-gravity becomes too strong to be ignored relative to the core’s gravity. As a result, the envelope’s gravitational pull begins to dominate gas accretion. The envelope’s contraction and mass growth further enhance its self-gravity, leading to continued growth of the envelope mass, a process known as runaway gas accretion. Initially, gas accretion is thought to proceed in a roughly spherically symmetric fashion. Recent 2D and 3D hydrodynamic simulations of the accretion flow near the growing protoplanet, however, suggest that early-phase accretion may be hindered or delayed by recycling flows. In the later phases, the supply of disk gas becomes insufficient to meet the demand of the contracting envelope, and gas accretion becomes more limited, occurring fully in a 2D or 3D manner governed by the disk’s evolving structure. In the following subsections, we discuss each phase of gas accretion in more detail, examining the physical processes that govern the early and late phases.

\subsubsection{Early-phase accretion} \label{sec:1D_gas_accretion}

\begin{figure}
    \centering
    \includegraphics[width=1.2\linewidth]{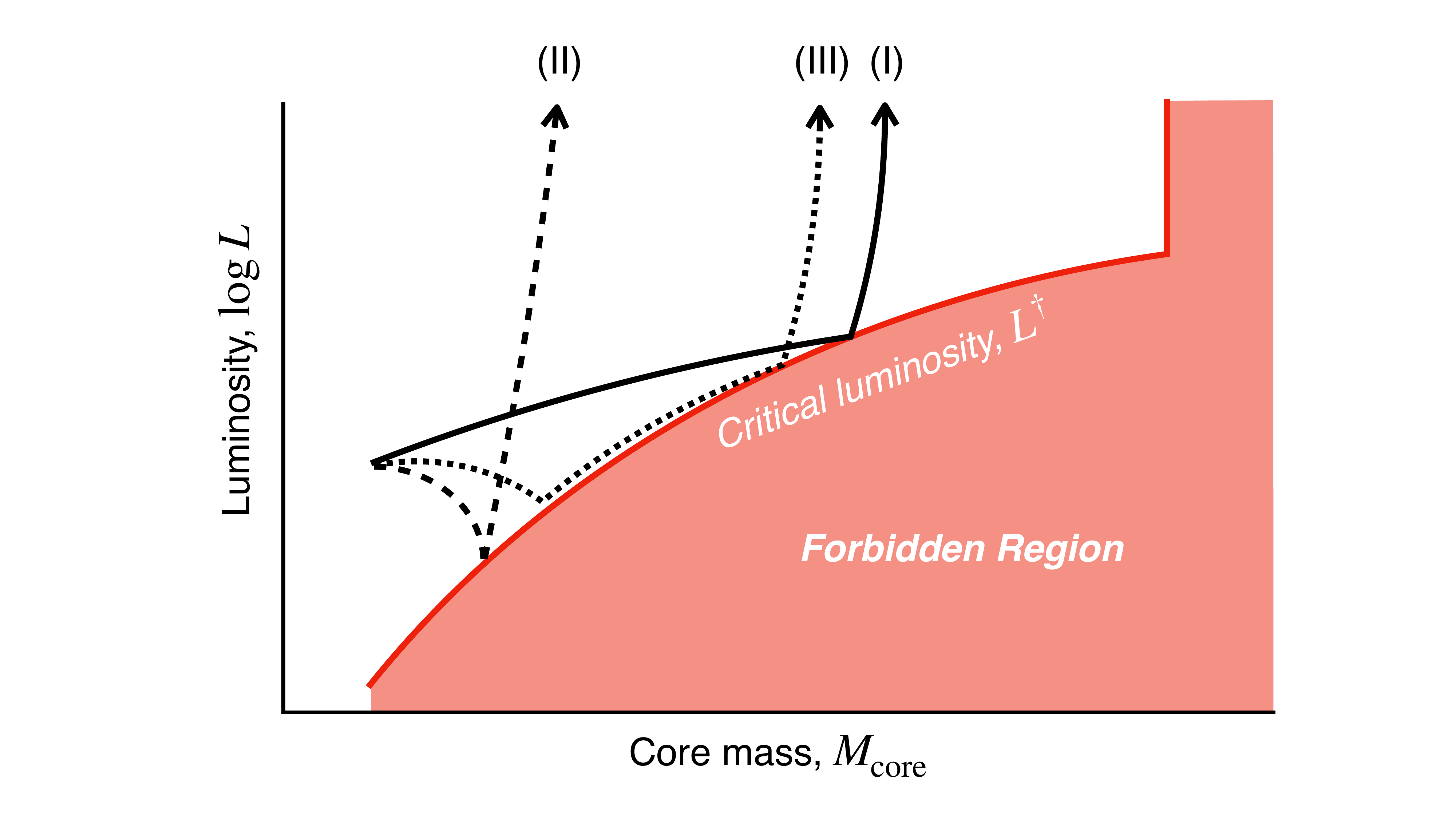}
    \caption{Schematic illustration explaining how the luminosity of protoplanetary envelopes changes with the core mass. {In the ``forbidden region'', the envelope is unable to be in the hydrostatic equilibrium \citep[][]{Ikoma+00}.} Path~(I) represents the core growing to a critical mass, leading to the runaway gas accretion \citep[e.g.,][]{Bodenheimer+86}. Path~(II) shows the core becoming isolated, followed by the runaway gas accretion \citep[e.g,][]{Ikoma+00}. Path~(III) corresponds to Phase~2, where gas and solid accretion balance each other out \citep[][]{Pollack+96}.}
    \label{fig:critical_luminosity}
\end{figure}

The early-phase gas accretion process has been well studied. It is driven by the quasi-hydrostatic contraction of a self-gravitating envelope in a similar way to the stellar pre-main-sequence evolution, although the detailed evolution differs between planetary and stellar evolution due to different interior temperatures. 
The quasi-static evolution is defined as the interior that remains in hydrostatic equilibrium but not thermal equilibrium; in other words, the evolution is dictated by a change in entropy (see Eq.~[\ref{eq:entropy}]).
Its characteristic timescale, often referred to as the Kelvin-Helmholtz timescale, can be expressed for the primordial envelope as \citep[][]{Ikoma+00}
\begin{equation}
    t_\mathrm{KH} \approx \frac{GM_\mathrm{core}M_\mathrm{env}}{R_\mathrm{core}L}.
\end{equation}
This equation implies that understanding the relationship between the core mass $M_\mathrm{core}$, the envelope mass $M_\mathrm{env}$, and the luminosity $L$ is crucial to determining the timescale of gas accretion. At the critical core mass, $M_\mathrm{env}^\dagger \simeq M_\mathrm{core}^\dagger$; the approximate analytic solution for radiative envelopes (Eq.~[\ref{eq:crossover mass}]) gives the corresponding luminosity (or the critical luminosity), ending up with
\begin{equation}
    t_\mathrm{KH} \approx
    \left(\frac{4\pi\bar{\rho}}{3}G^3\right)^{1/3}
    \Theta_L^3 \mu^{-4} \kappa M_\mathrm{core}^{\dagger -4/3}.
\label{eq:KH_time}
\end{equation}
This formula indicates that the larger the critical core mass, the shorter the Kelvin-Helmholtz timescale is specifically for the early phase of gas accretion. The detailed simulations show that the dependence of $t_\mathrm{KH}$ on $M_\mathrm{core}^\dagger$ is {rather strong} and its index is about $-3$ \citep[][]{Ikoma+00,Ikoma+Genda06}.

{In theory, various pathways toward the onset of runaway gas accretion are possible, as schematically illustrated in Fig.~\ref{fig:critical_luminosity}. In the ``forbidden region'', the envelope is unable to achieve hydrostatic equilibrium \citep[][]{Ikoma+00}. Path~(I) is the conventional pathway with constant core accretion rate, as first demonstrated by \citet{Bodenheimer+86}.}
Modern core growth models show that once a core reaches a sufficient mass, it starts to gravitationally influence the surrounding remnant bodies or disk gas, preventing further solid accumulation, regardless of planetesimal accretion \citep[e.g.,][]{Kokubo+98} or pebble accretion \citep[e.g.,][]{Lambrechts+2014a} (see Section~\ref{sec:core accretion} for the state-of-the-art understanding). This phenomenon is called isolation and the mass at that time is called the isolation mass. 
The isolation, by definition, corresponds to $\dot{M}_\mathrm{sol} \rightarrow 0$ or $L \rightarrow 0$; thus, as shown in equation~(19), the isolated mass corresponds to the critical core mass. 
In reality, however, the luminosity never reaches zero but instead decreases down to the critical luminosity \citep[e.g., Path~(II) in Fig.~\ref{fig:critical_luminosity};][]{Ikoma+00,Hubickyj+05}. The gas accretion timescale after isolation is determined by the critical luminosity, as discussed above. 
Note that in Phase~2 found by \cite{Pollack+96}, planetesimal accretion continues after isolation, and the luminosity is maintained just above the critical luminosity (Path~(III); see Section~\ref{sec:formation_timescale_problem} for a discussion of theoretical issues in their model).

%%%%%%%%%%%%%%%%%%%%%%%%%%%%%%%%%%%

\subsubsection{Envelope recycling} \label{sec:recycling}
The discussion above assumes that the primordial envelope has a spherically symmetric hydrostatic structure connected directly to the surrounding gas disk. In reality, however, the primordial envelope is never a closed, spherically symmetric structure. 
The issue was acknowledged in early studies, but the non-1D effects mainly impact the outer part of the envelope and do not significantly affect the deep structure and, therefore, the critical core mass of the primordial envelope in hydrostatic equilibrium \citep[][]{Mizuno80,Stevenson82}, except under extreme conditions \citep[][]{Ikoma+01}. However, it needs to be confirmed by hydrodynamic calculations whether the disk gas flowing into the planet's Roche lobe can settle into hydrodynamic equilibrium.

Following the pioneering work by \cite{Ayliffe+09}, \cite{D'Angelo+13} directly performed global radiative-hydrodynamic simulations based on grid-based methods for protoplanets growing at 5 and 10~au from a solar-mass central star, assuming the formation of Jupiter and Saturn. They focused on relatively low-mass protoplanets, neglecting self-gravity. Their simulations revealed that disk gas flows much further inwards than the Hill radius, and even further inwards than the Bondi radius. Despite this, the 3D calculations, like the 1D ones, resulted in a hydrostatic structure that is nearly spherically symmetric at relatively low masses ($\lesssim$~0.1$\MJupiter$). It is worth noting that the size of the closed envelope is roughly half that of the Bondi radius. The resulting envelope mass and gas accretion rate only differed by a factor of $\sim$2, and little difference in results was observed between 5~au and 10~au. In conclusion, the envelope structure within roughly half of the Bondi radius is nearly isolated from and not significantly affected by the 3D flow. Note {\cite{Ayliffe+09} point out the significance of the equation of state.}

This issue was revisited by \cite{Ormal+15a} and \cite{Ormel+15b} in the context of the formation of short-period super-Earths/sub-Neptunes. 
As mentioned in Section~\ref{sec:constraints:exoplanets}, the \mission{Kepler} statistics show that those planets are much more common than gas giant planets at orbital periods shorter than 100~days. Their observed mass-radius relationships indicate that many of them retain H-He envelopes, albeit to the mass fraction of less than 10\%. 
\cite{Ormel+15b} and \cite{Popovas+18} conducted 3D hydrodynamic calculations under isothermal and adiabatic assumptions, respectively. They found continuous gas exchange between the protoplanet's envelope and the surrounding gas disk (referred to as ``recycling''), preventing runaway gas accretion. This contradicted the findings of \cite{D'Angelo+13} who had found almost complete isolation between the envelope and gas disk. This was thought to arise from the difference in distance from the central star, although there are differences in the treatment of energy transfer. To verify that, \cite{Cimerman+17} later performed 3D hydrodynamic calculations considering radiative transfer, demonstrating that radiative transport plays a decisive role and no recycling occurs {at 0.1~au}, supporting \cite{D'Angelo+13}. The gas flowing from the disk into the envelope has higher entropy than the deep envelope and needs to overcome the buoyancy barrier to mix with it. Under isothermal or adiabatic assumptions, there is no buoyancy barrier, making mixing and recycling easily achievable \citep[also see][]{Kurokawa+18}.

However, the issue may not be fully resolved, as the buoyancy barrier is not completely unbreakable. Recently, Moldenhauer et al. (2022) conducted a systematic investigation of envelope recycling using high-resolution 3D radiation-hydrodynamic calculations (Moldenhauer et al., 2021). Unlike previous 3D radiation-hydrodynamic calculations, their findings demonstrate that the entropy loss in the deep envelope is offset by the entropy injection from the disk gas, leading to effective recycling and the successful crossing of the buoyancy barrier. This outcome may be attributed to the high resolution of their calculations. 
{Additionally, choice of boundary conditions may affect the results of \cite{Cimerman+17}.}

\subsubsection{Late-phase accretion and gap formation} 
\label{sec:2D/3D_gas_accretion}

Following earlier 2D hydrodynamic simulations with low spatial resolutions by \cite{Miki82} and \cite{Lubow+99}, \cite{Tanigawa+02} performed high-resolution 2D simulations to study flows near the orbit of a proto-gas giant. Two major shockwaves were discovered around the planet: bow shocks that extend beyond the planet's Hill sphere and spiral shocks that wind inward within the Hill sphere. These shockwaves play crucial roles in directing gas accretion toward the planet. It was also found that gas accretes into the planet’s Hill sphere through the Lagrangian points only from narrow bands, which are located $\sim$2 to 3 $r_\mathrm{H}$ away from the plant, on both sides of the planetary orbit. Based on their simulations, \cite{Tanigawa+02} derived an empirical formula for the gas accretion rate as 
\begin{equation}
    \frac{d M_{\rm p}}{dt} = D \Sigma_\mathrm{d},
    \label{eq:tanigawa02}
\end{equation}
where $\Sigma_\mathrm{d}$ is the gas surface density in the accretion bands, 
\begin{equation}
    D = 0.29 \left( \frac{M_{\rm p}}{M_*}\right)^{4/3} 
        \left( \frac{h_{\rm g}}{a}\right)^{-2} a^2 \Omega,
\end{equation}
$M_\ast$ is the mass of the central star, and $h_\mathrm{g}$ and $\Omega$ are, respectively, the gas disk scale-height and Kepler frequency at the planet's semi-major axis. By comparing the 2D accretion rate with the 1D rate from \cite{Tajima+97}, they found that the 2D accretion begins to limit planetary growth at a planet mass of around 100~$\MEarth$.

Three-dimensional simulations show that the accretion flow is more complex inside the Hill sphere. Unlike the 2D picture, where the flow entering through the Lagrangian points spirals down toward the central planet, the 3D flow settles from high altitudes onto the mid-plane. This was suggested by \cite{D'Angelo+03} and \cite{Bate+03} with low-resolution simulations and confirmed by \cite{Tanigawa+12}. \cite{Tanigawa+12} also discovered outward flows in the circumplanetary disk, indicating that gas does not simply flow inward but also circulates outward. 
The detailed structure of the flow, including the formation of shock waves and circumplanetary disks, is influenced by the choice of the equation of state used in the simulations \citep[][]{Szulagyi+16,Takasao+21}. 
Many papers have been published focusing on the formation of circumplanetary disks, which falls outside the scope of this review.

%%%
When a planet becomes massive enough, the tidal torque it exerts on the nearby gas disk dominates over viscous diffusion, leading to the formation of a gap near the planet's orbit in the protoplanetary disk \citep[][]{Papaloizou+84}. This gap can affect the gas accretion rate, as the accretion bands mentioned above can be located inside the gap. The gas surface density profile inside the gap is influenced by the gas flow crossing the gap, which has been shown to exist in hydrodynamic simulations \citep[e.g.,][]{Duffell+13}. The steady-state density profile in the gap has been obtained through hydrodynamic simulations, and the surface density in the gap is empirically given by \cite{Kanagawa+17}
\begin{equation}
    \Sigma_{\rm gap} = \frac{\Sigma_{\rm out}}{1+0.04 K}, 
    \label{eq:sigma_gap}
\end{equation}
where $\Sigma_{\rm out}$ is the surface density outside the gap and $K$ is the dimensionless parameter given by 
\begin{equation}
    K = \alpha^{-1} \left( \frac{M_{\rm p}}{M_*}\right)^2 
        \left( \frac{h_{\rm g}}{a}\right)^{-5}, 
\end{equation}
$\alpha$ is the viscosity parameter in the disk.  
Using Eq.~(\ref{eq:sigma_gap}) in Eq.~(\ref{eq:tanigawa02}), one can calculate the gas supply rate to the planet's Hill sphere. The accretion rate is given by the smaller of $\dot M_{\rm p}$ given in Eq.~(\ref{eq:tanigawa02}) and $M_{\rm p} / t_{\rm KH}$ given in Eq.~(\ref{eq:KH_time}). Note that recent MHD simulations show a deeper gap compared to that expressed by Eq.~(\ref{eq:sigma_gap}) \citep[][]{Aoyama+23}.

\subsection{Core accretion} \label{sec:core accretion}
\color{black}

As explained above, cores must exceed the critical core mass or the crossover mass to initiate sufficient gas accretion in protoplanetary disks. 
The formation timescale of a core with the crossover mass is, therefore, shorter than the disk lifetime. 
Moreover, cores migrate inward due to gravitational interaction with the surrounding protoplanetary disk. This migration is referred to as type I migration, which is faster than type II migration, because gaps are not opened by cores. 
The orbital decay timescale of a core with a semimajor axis $a$ for the type I migration is estimated to be 
\citep[e.g.,][]{Paardekooper+23}
\begin{eqnarray}
 t_{\rm mig,core} &=& {\tilde \Gamma}^{-1} \left(\frac{\Sigma_{\rm g} a^2}{M_*}\right)^{-1} 
  \left(\frac{M_{\rm core}}{M_*}\right)^{-1} \left(\frac{h_{\rm g}}{a}\right)^{-2} \Omega^{-1} 
  \nonumber
  \\
&=& 1.3 \times 10^5 \left(\frac{{\tilde \Gamma}}{4}\right)^{-1} 
\left(\frac{M_{\rm core}}{10 M_\oplus}\right)^{-1}
\left(\frac{\Sigma_{\rm g}}{100\,{\rm g \,cm}^{-2}}\right)^{-1}
\left(\frac{h_{\rm g}}{0.05a}\right)^2
\nonumber
\\ &&\times 
\left(\frac{a}{5\,{\rm au}}\right)^{-1/2}
\left(\frac{M_*}{M_\odot}\right)^{3/2} {\rm yr}, \label{eq:mig1} 
\end{eqnarray}
where ${\tilde \Gamma}$ is the dimensionless migration coefficient, $h_{\rm g}$ is the scale height of the disk and $\Sigma_{\rm g}$ is the surface density of the gas disk. In the isothermal disk, ${\tilde \Gamma} \approx 4$ \citep{Tanaka+02}. The formation of cores with the crossover mass prior to their orbital decay requires $t_{\rm grow,core} \la t_{\rm mig,core}$, where $t_{\rm grow,core}$ is the growth timescale of a core. 
Once a core reaches the crossover mass, the runaway gas accretion onto the core forms a gas giant planet (see Section 3.1). The gas accretion opens up the gap around its orbit. The migration timescale is then much longer than the estimate in Eq.~(\ref{eq:mig1}) because of the onset of type II migration. Therefore, the formation timescale of a core with the crossover mass is required to be comparable to or shorter than the type I migration timescale. 

\subsubsection{Planetesimal accretion}
\label{sc:planetesimal_accretion}
In the classical scenario, cores are formed via the collisional coagulation in a swarm of planetesimals. 
Runaway growth of planetesimals forms planetary embryos, which further grow to cores via collisions with remnant planetesimals. The orbital separations of planetary embryos are determined by their mutual Hill radii, $2^{1/3} R_{\rm H}$, multiplied by a dimensionless factor $\tilde b \approx 10$ obtained from $N$-body simulations of planetesimal accretion \citep{Kokubo+98}. Therefore, the growth of cores is limited by the amount of planetesimals; $M_{\rm core} < 2^{4/3} \pi \Sigma_{\rm s} a \tilde b R_{\rm H}$, where $\Sigma_{\rm s}$ is the surface density of solid bodies. 
\begin{equation}
    \Sigma_{\rm s} > 6.3 \left( \frac{M_{\rm core}}{10\,M_\oplus}\right)^{2/3} \left( \frac{M_*}{M_\odot} \right)^{1/3} \left( \frac{a}{7 \, {\rm au}}\right)^{-2} \left(\frac{\tilde b}{10} \right)^{3/2} \, {\rm g/cm}^2,
\end{equation}
where $M_\odot$ is the solar mass. Therefore, a certain amount of planetesimals is required for core formation via planetesimal accretion. 

The collision rate depends on the vertical distribution of planetesimals. 
If the planetesimal disk is geometrically thick, 
the growth rate of a core is then given by 
\begin{equation}
    \frac{dM_{\rm core}}{dt} \sim \sigma_{\rm col} \Sigma_{\rm s} \Omega, 
%    2 \pi G M_{\rm core} R_{\rm core} \Sigma_{\rm s} \Omega_{\rm K} / v_{\rm rel}^2
\label{eq:collision_rate_3d}
\end{equation}
where $\sigma_{\rm col}$ is the collisional cross-section between the core and planetesimals,  
$\Sigma_{\rm s}$ is the surface denstiy of solid bodies, and  
$\Omega$ is the Keplerian frequency. 
In a protoplanetary disk, the mean relative velocity of planetesimals to the core is slower than the surface escape velocity of the core, the gravitational focusing is dominant, 
and then 
\begin{equation}
\sigma_{\rm col} \approx \frac{G M_{\rm core} R_{\rm e}}{v_{\rm rel}^2},     
% 2 \pi
\end{equation}
$v_{\rm rel}$ is the mean relative velocity between the core and planetesimals and $R_{\rm e}$ is the effective collisional radius of the core. 
For a solid core with $M_\mathrm{core} \sim M_\mathrm{core}^\dagger$, the envelope of the core enhances the collisional radius of the core. The collisional radius $R_{\rm e}$ is then larger than $R_{\rm core}$ \citep{Inaba+03AA}.

The relative velocity is determined by the random velocity of planetesimals, which increases by the stirring by the core and decreases by the gas drag. 
{The relative velocity is assumed to be determined by the equilibrium between gas drag and the stirring by the core \citep[][]{Kokubo+98}}
\begin{equation}
    v_{\rm rel} \sim \left( \frac{m}{\rho_{\rm neb} d^2 a}\right)^{1/5}  R_{\rm H} \Omega,
\end{equation}
{where $d$ is the planetesimal diameter.}
For $d \ga 1\,$km, the assumption for derivation is valid. Thus, $v_{\rm rel}$ is several times larger than $R_{\rm H} \Omega_{\rm K}$, which is called the Hill velocity. 

The growth timescale 
$t_{\rm grow,core} = M_{\rm core} / \dot M_{\rm core}$
is then estimated to be
\begin{equation}
    t_{\rm grow,core} \sim \left( \frac{M_*}{\Sigma_{\rm s} a^2} \right) \left( \frac{R_{\rm H}^2}{R_{\rm e} a}\right) \left( \frac{m}{\rho_{\rm neb} d^2 a}\right)^{2/5} \Omega^{-1}. 
    \label{eq:rough_tg}
\end{equation}
{If $R_{\rm e} = R_{\rm core}$, Eq.~(\ref{eq:rough_tg}) gives $t_{\rm grow,core} \sim 10^7$\, years for $a \approx 5$\,au and $d \approx 10$\,km. }
However, the collisional radius of a core is enhanced by its envelope, resulting in $R_{\rm e} > R_{\rm core}$ and shorter $t_{\rm grow,core}$. \citet{Inaba+03AA} derived the analytic formula of $R_{\rm e}$, based on the hydrostatic density profile of the envelope.
According to \citet{Inaba+03AA}, we here estimate $R_{\rm e}$. 
When a planetesimal passes around the distance $R_{\rm e}$ from the core with velocity $v_{\rm pass}$, gas drag decreases from $v_{\rm pass}$ to $v_{\rm pass} - \Delta v_{\rm pass}$, which results in the decrease of energy, $\Delta E_{\rm pass} \approx m v_{\rm pass} \Delta v_{\rm pass}$.  We estimate $\Delta v_{\rm pass}$ via the gas drag force $\sim d^2 \rho v_{\rm pass}^2$ and the passing time $\sim R_{\rm e} /v_{\rm pass}$; $\Delta v_{\rm pass} \sim d^2 \rho v_{\rm pass} R_{\rm e} /m$.
The passing velocity is roughly determined by the gravitational energy of the core; $v_{\rm pass}^2 \sim G M_{\rm core} / R_{\rm e}$. Therefore, $\Delta E \sim d^2 \rho G M_{\rm core}$. 
On the other hand, the planetesimal is captured in the Hill sphere of the core, which are eventually accreted onto the core. The condition gives $\Delta E \sim G m M_{\rm c} / R_{\rm H}$; 
\begin{equation}
    \pi \rho d^2 R_{\rm H} \approx m. 
\end{equation}
For the wholly radiative envelope, $\rho$ is given by Eq.~(\ref{eq:density_rad}) and then 
\begin{eqnarray}    
    \frac{R_{\rm e}}{R_{\rm core}} &\approx& \left[ \frac{\pi^2 \sigma_{\rm SB} \Phi^4 d^2 R_{\rm H}}{12 \kappa L m} \right]^{1/3} \frac{1}{R_{\rm core}},
    \nonumber
    \\
    &\approx& 6.7 \left(\frac{d}{10\,{\rm km}} \right)^{-1/3}
        \left( \frac{\kappa}{0.01\, {\rm cm}^2 {\rm g}^{-1}}\right)^{1/3}
        \left( \frac{M_{\rm core}}{10 M_\oplus}\right)^{8/9}
        \left( \frac{M_{\rm core}/\dot M_{\rm sol}}{10^5 {\rm yr}}\right)^{8/9},
        \label{eq:enhancement_radius}
\end{eqnarray}
where Eq.~(\ref{eq:luminosity}) was used for $L$. As the estimate above, the radius enhancement due to envelope is significant. The enhancement shortens the growth timescale, which is still longer than the type I migration timescale for $d = 10$\,km. However, {$t_{\rm grow,core} \propto d^{11/15}$ according to Eqs.~(\ref{eq:collision_rate_3d})
 and (\ref{eq:enhancement_radius}). Therefore, for $d \la 1$\,km at a = 5\,au, $t_{\rm grow,core}$ can be comparable to or shorter than the type I migration timescale. }

For smaller planetesimals of $d \la 100$\,m, the vertical distribution is not wider than $R_{\rm H}$. The accretion rate given by Eq.~(\ref{eq:collision_rate_3d}) is invalid. Instead, the accretion rate is estimated from the accretion length $\sim \sqrt{\sigma_{\rm col}}$ and the approach velocity $R_{\rm H} \Omega$. 
The relative velocity is then approximated to the approach velocity, $v_{\rm rel} \sim R_{\rm H} \Omega$. 
The accretion rate is given by (Ida \& Nakazawa 1989)
\begin{equation}
    \frac{dM_{\rm core}}{dt} \sim \Sigma_{\rm s} \sqrt{\sigma_{\rm col}} R_{\rm H} \Omega \approx \sqrt{6 \pi} \Sigma_{\rm s} R_{\rm e}^{1/2} R_{\rm H}^{3/2} \Omega.  
    \label{eq:planetesimal_acc_2d}
\end{equation}
{Equation~(\ref{eq:planetesimal_acc_2d}) gives $t_{\rm grow,core} \propto d^{1/6}$. }
Even for very small planetesimals, $R_{\rm e}$ cannot exceed $R_{\rm H}$. 
Orbital simulations in the gas disk perturbed by the core show $R_{\rm e} \la R_{\rm H}/8$ \citep{Okamura+21}. 

{
The comparison between $t_{\rm core,grow}$ and $t_{\rm core, mig}$ suggests that
the core formation with small planetesimals with $d \la 1\,$km can occur prior to type I migration. However, such small planetesimals are strongly influenced by radial drift and mutual collisions. The simulations that account for the collisional evolution and the radial drift show the crossover-mass core cannot form prior to type I migration in planetesimal disks under the planetesimal hypothesis, because of the loss of small planetesimals due to radial drift \citep[e.g.,][]{Inaba+03Icar, Kobayashi+11}. Therefore core formation is difficult in a simple planetesimal disk. }

{
It should be noted that the core growth timescale is short enough via small planetesimal accretion. If small planetesimals are sufficiently supplied, the core formation via planetesimal accretion is possible. 
}

\subsubsection{Pebble accretion}

Recently, pebble accretion is alternatively proposed for core formation {citep{Ormel+10,Johansen+17,Drazkowska+23}. In the theory of planet formation, pebbles are characterized by the stopping time due to gas drag $t_{\rm st}$, which is roughly estimated as \citep{Adachi+1976}
\begin{equation}
    t_{\rm st} \Omega = \frac{3 m}{2 d^2 \Sigma_{\rm g}} \approx 
    4 \times 10^{-2} \left( \frac{d}{1\,{\rm cm}}\right) \left( \frac{\rho_{\rm mat}}{1\,{\rm g/cm}^2} \right)
    \left(\frac{\Sigma_{\rm g}}{100\,{\rm g/cm}^2} \right)^{-1}, 
    \label{eq:stopping_time}
\end{equation}
where $\rho_{\rm mat}$ is the density of pebbles. 
Here, bodies with $t_{\rm st} \Omega < 1$ are simply referred to as pebbles, while those with $t_{\rm st} \Omega > 1$ are called planetesimals. Bodies with $t_{\rm st} \sim 1$ drift inward on short timescales unless disks have pressure bumps. Although the pile-up of bodies at pressure bumps influences planet formation, the origins and lifetimes of pressure bumps in the absence of pre-existing planets remain debated \citep{Bae+23}. We here focus on planet formation in protoplanetary disks without pressure bumps. }

We here estimate the pebble accretion rate. As discussed in Section~\ref{sc:planetesimal_accretion},  the growth timescale can be much shorter than the migration timescale. However, the surface density of small bodies decreases because of their radial drift due to gas drag.
Therefore, we estimate the core growth timescale via pebble accretion differently.  

For pebble accretion, the existence of cores roughly larger than Mars is assumed in the inner disk. Cores grow via collisions with pebbles drifting from the outer disk.  
The growth rate of a planetary core via pebble accretion is given by 
\begin{equation}
 \frac{dM_{\rm core}}{dt} = \epsilon \frac{d M_{\rm F}}{dt}, \label{eq:accretion_rate_pebble} 
\end{equation}
where $d M_{\rm F} / dt$ is the mass flux of pebbles across the orbit of the core and 
$\epsilon$ is the accretion efficiency of drifting pebbles. 

Pebbles are formed from collional growth of dust in the protoplanetary disks.
The pebble formation timescale is given by 
\begin{equation}
    t_{\rm pebble} \sim \left( \frac{\Sigma_{\rm g}}{\Sigma_{\rm s}} \right) \Omega^{-1}.  
\end{equation}
Dust growth occurs from the inner disk, which progresses outwards. The growth-front radius $a_{\rm gf}$ is estimated via $t_{\rm pebble} \sim t_{\rm age}$ as, 
\begin{equation}
    a_{\rm gf} \sim  \left( \frac{\Sigma_{\rm s}}{\Sigma_{\rm g}} \right)^{2/3} \left( G M_* t_{\rm age}^2 \right)^{1/3}, 
    \label{eq:gf_radii}
\end{equation}
where $t_{\rm age}$ is the age of the host star.
{The growth front may be seen as a ring structure for the millimeter wavelength observation and the growth-front radii given by Eq. (\ref{eq:gf_radii}) are consistent with outermost ring radii around young protostars with the ages of $\ll 10^6$ years \citep{Ohashi+21}.} {Note that the mechanism of growth fronts is one of several possibilities for explaining the observed dust rings with ALMA and that this mechanism works for very young disks. }
Assuming the drift of pebbles occurs immediately after their formation, the drifting pebble flux is obtained as
\citep[e.g.,][]{Lambrechts+2014a}
\begin{equation}
    \dot M_{\rm F} \sim \frac{a_{\rm gf}^2 \Sigma_{\rm s}}{t_{\rm age}}. 
\end{equation}
This equation is valid until $a_{\rm gf}$ reaches the outer edge of disk $\sim 100$\,au. The growth front reaches the outer edge at $t_{\rm out} \sim 10^5$\,years \citep{Ohashi+21}. We estimate $a_{\rm gf}^2 \Sigma_{\rm s} \sim M_{\rm solid,disk}$ at $t_{\rm age} = t_{\rm out}$, where $M_{\rm solid,disk}$ is the total mass of solid bodies in the disk. 
Using Eq.~(\ref{eq:accretion_rate_pebble}), we estimate the core growth timescale via pebble accretion $t_{\rm grow,pe}$ as 
\begin{eqnarray}    
    t_{\rm grow,pe} &=& \frac{M_{\rm core}}{\epsilon \dot M_{\rm F}} \sim \frac{M_{\rm core} t_{\rm out}}{\epsilon M_{\rm solid,disk}}\\
    &\sim& 10^5 \left(\frac{\epsilon}{0.1}\right)^{-1}
        \left( \frac{M_{\rm core}}{10 M_\oplus} \right)
      \left(\frac{M_{\rm solid,disk}}{100\,M_\oplus} \right)^{-1}
    \left(\frac{t_{\rm out}}{10^5{\rm yr}}\right)^{-1/3} 
    {\rm yr} \label{eq:tgpe} 
\end{eqnarray}    
The growth timescale $t_{\rm grow,pe}$ is therefore comparable to the migration timescale. The core formation  occurs with significant migration. 

However, because of $\epsilon \ll 1$, the large total mass of pebbles is required for the formation of $M_{\rm core} = M_{\rm core}^\dagger$. The required pebble mass, $M_{\rm min,pebble}$, is estimated from the integration of $\dot M_{\rm F} = \dot M_{\rm core} /\epsilon$ with $\epsilon \propto M_{\rm core}^{2/3}$ \citep{Okamura+21}; 
\begin{equation}
    M_{\rm min,pebble} = \int \dot{M_{\rm F}} dt \approx \int_0^{M_{\rm core}^\dagger} \epsilon^{-1} d M_{\rm core} = \frac{3M_{\rm core}^\dagger}{\epsilon^\dagger},  
    \label{eq:min_mass}
\end{equation}
where $\epsilon^\dagger$ is $\epsilon$ for $M_{\rm core} = M_{\rm core}^\dagger$. 

{The envelope around the core is important to determine $\epsilon^\dagger$. The headwind of the core in the gas disk induces the gas flow to enter the envelope. If the headwind velocity ($\sim \Omega h_{\rm g}^2/a$)
is smaller than the Keplerian shear velocity at the envelope radius ($\sim R_{\rm B} \Omega$), the isolated envelope is formed around a core with \citep{Kuwahara+20}, 
\begin{equation}
    M_{\rm core} \ga \frac{4}{15} \left( \frac{h_{\rm g}}{a }\right)^4 M_*. 
\end{equation}
Therefore, a crossover mass core with $M_{\rm core}^\dagger \sim 10 M_\oplus$ has an isolated envelope if $h_{\rm g} \la 0.1 a$. }

{If an isolated envelope forms around a core, the gas flow surrounding the envelope significantly influences pebble accretion. The core's gravity drives a downflow from the high-altitude disk to the envelope, inducing the outflow from the outer edge of the envelope around the disk midplane.} The small pebbles with $t_{\rm st} \Omega \la 0.2 (R_H /h_{\rm g})^6$ are carried away by the outflow and cannot enter the isolated envelope \citep{Okamura+21}. Consequently, the pebble accretion onto the core is possible only for
\begin{equation}
    t_{\rm st} \Omega \ga 0.03 \left( \frac{a}{5\,{\rm au}}\right)^{-8/7}, 
    \label{eq:pebble_limit}
\end{equation}
where the disk temperature is assumed for the optical thick disk, given by $T_{\rm neb} \approx 100 (a /1\,{\rm au})^{-3/7}$\,K.

Therefore, the pebble accretion is effective only for $t_{\rm st} \Omega \ga 0.1$. The isolated envelope has the radius comparable to $R_{\rm B}$. 
Pebble accretion occurs via entering the isolated envelope during passing around the envelope. The accretion rate is thus estimated to $2 \Sigma_{\rm s} G M_{\rm core} t_{\rm st} /R_{\rm B}$ \citep{Okamura+21}. 
The pebble flux is given by $2 \pi \Sigma_{\rm s} c_{\rm s}^2 / \Omega$, where the drift velocity is approximated to be $(c_{\rm s} / a \Omega)^2 a \Omega$ \citep{Adachi+1976}. Then, $\epsilon$ is given by the ratio of the accretion rate to the flux as
\begin{equation}
    \epsilon^\dagger \approx \frac{t_{\rm st} \Omega}{\pi}. 
    \label{eq:epsilon_envelope}
\end{equation}
Therefore, $\epsilon^\dagger \la 0.1$ for $t_{\rm st} \Omega \la 0.3$ and $\epsilon^\dagger$ is significantly smaller for $t_{\rm st} \Omega \ll 0.1$ (see Eq.~(\ref{eq:pebble_limit})). 

\begin{figure}[t]
\includegraphics[width=\textwidth]{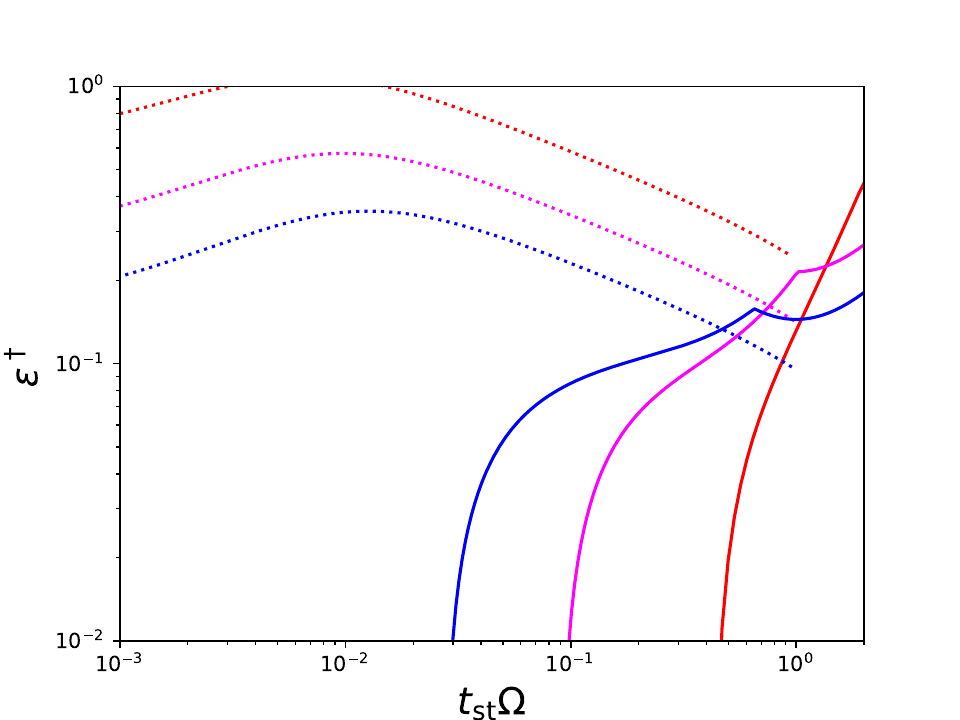}
\caption{Pebble accretion efficiency $\epsilon^\dagger$ at $a = 2$\,au (red), 5\,au (magenta), and 10\,au (blue) as a function of $t_{\rm st} \Omega$ of pebbles. {Corresponding sizes of pebbles are calculated from Eq.~(\ref{eq:stopping_time}).} For reference, the dotted curves show $\epsilon^\dagger$ obtained for $\alpha = 10^{-3}$ under the assumption of unperturbed circular gas motion \citep{Ormel+18}. }
\label{fig:epsi_ik}
\end{figure}

\citet{Okamura+21} analytically derived the accretion rate as a function of $t_{\rm st} \Omega$, which well reproduces the results of orbital simulations in the disk perturbed by the core. 
Figure \ref{fig:epsi_ik} shows $\epsilon^\dagger$ for $M_{\rm core}^\dagger = 10 M_\oplus$ calculated from the accretion rate formula by \citet{Okamura+21} and the drift rate in the disk with $\Sigma_{\rm g} = 480 (a/1\,{\rm au})^{-1} \, {\rm g/cm}^2$, $T_{\rm neb} = 100 (a/1\,{\rm au})^{-3/7}$\,K and the densities of the core and pebbles of $1.4\,{\rm g/cm}^3$. 
For small pebbles ($t_{\rm st} \Omega \la 0.01$), $\epsilon^\dagger \ll 0.01$ because pebbles cannot enter the envelope due to the flow around the envelope. The relatively large pebbles satisfying Eq.~(\ref{eq:pebble_limit}) have $\epsilon^\dagger$ comparable to or smaller than Eq.~(\ref{eq:epsilon_envelope}), because the derivation of Eq.~(\ref{eq:epsilon_envelope}) do not include the reduction by the outflow. 
It should be noted that the gas flow perturbed by the core plays a significant role in determining $\epsilon^\dagger$. Assuming an unperturbed disk leads to a substantial overestimation of $\epsilon^\dagger$, especially for small $t_{\rm st} \Omega$ (see Fig.~\ref{fig:epsi_ik}). Additionally, the transition between gas drag regimes has a strong influence on the accretion rate \citep{Okamura+21}. Consequently, $\epsilon^\dagger$ is significantly smaller than 0.1 for $t_{\rm st} \Omega \lesssim 1$.

Using Eqs.~(\ref{eq:min_mass}), we can estimate the required solid mass for pebble accretion as 
\begin{equation}
    M_{\rm min,pebble} = 300 \left( \frac{M_{\rm core}^\dagger}{10\,M_\oplus} \right) \left( \frac{\epsilon^\dagger}{0.1}\right)^{-1} M_\oplus. 
\end{equation}
The value of $\epsilon^\dagger = 0.1$ is achieved only for $t_{\rm st} \Omega \sim 1$ and $\epsilon^\dagger \ll 0.1$ for small pebbles (see Fig.~\ref{fig:epsi_ik} and Eq.~\ref{eq:epsilon_envelope}). Therefore, $M_{\rm min,pebble}$ is much larger than $300 \, M_\oplus$.
Even for Class 0 objects, the disks with total solid masses $\geq 300\,M_\oplus$ are too rare to explain the occurrence rate of gas giant planets \citep{Tychoniec20,Mulders+21}.  

{Therefore, the core formation via pebble accretion is also difficult, because of too small $\varepsilon^\dagger$. Planetesimals with $t_{\rm st} \Omega > 1$ exhibit much larger $\varepsilon^\dagger$, enabling core formation without significant mass loss.  However, pebbles play a critical role of transporting solid bodies from the outer to inner disks, facilitating core formation in inner disks. Despite the separate treatment of pebbles and planetesimals, a seamless collisional evolution of solid bodies may be essential for gas-giant core formation. }

\subsection{Migration}

Gas giants undergo orbital migration due to their interaction with the surrounding protoplanetary disk. As discussed earlier, type I migration constrains the formation timescale of planetary cores. However, once a planet grows to a mass of several $10 M_\oplus$ through gas accretion, it creates a gap in the disk near its orbit. The migration of such gap-opening planets is referred to as type II migration.

The density profile within the gap plays a critical role in determining the rate of type II migration, as well as influencing the gas accretion rate onto the planet. The type II migration timescale depends on the surface density within the gap, $\Sigma_{\rm gap}$ \citep[e.g., see][]{Paardekooper+23}, rather than the unperturbed gas surface density, $\Sigma_{\rm g}$, as used for type I migration (see Eq.~[\ref{eq:mig1}]). $\Sigma_{\rm gap}$ is itself determined by the structure of the gap, as described by Eq.~(\ref{eq:sigma_gap}). Consequently, both type I and type II migration can be consistently described using the surface density profile.

The migration rate $\dot{a}$ and the gas accretion rate $\dot{M}_p$ during type II migration are both proportional to $\Sigma_{\rm gap}$. The evolutionary track of a planet on the $M_p$-$a$ plane is governed by the ratio $\dot{M}_p / \dot{a}$, which leads to the cancellation of $\Sigma_{\rm gap}$. This implies that the evolutionary track is independent of the overall disk evolution \citep{Tanaka+20}.
When type II migration is considered alongside gas accretion, its significance diminishes for planets with masses comparable to or smaller than Jupiter’s \citep{Kanagawa+18}. This finding aligns with the observed orbital mass distribution of giant exoplanets \citep[see also Section \ref{sec:hot_warm_jupiter};][]{Tanaka+20}.

%%%%%%%%%%
% ISSUES %
%%%%%%%%%%
\section{Formation of Giant Planets in and outside the Solar System} \label{sec:discussion}
Putting together our current understanding of relevant physical processes, we discuss how to resolve known issues regarding giant planet formation and what we need to study more. The following are broadly recognized as theoretical challenges.
\begin{enumerate}
    \item \textbf{Early formation of massive cores}---Can primordial cores grow massive enough to trigger rapid gas accretion before the depletion of disk gas and solid materials and the orbital migration of planetary cores? Does the debate about planetesimals vs. pebbles truly capture the essence of the issue?
    \item \textbf{Enriched interiors with diluted cores}---What caused the high bulk metallicities of Jupiter, Saturn, and some gas giant exoplanets? Are diluted cores predicted theoretically within the core accretion hypothesis?    
    \item \textbf{Enriched atmospheres}---When and how did Jupiter and Saturn's atmospheres become enriched with heavy elements, including highly volatile ones (i.e., noble gases), as well as solid-loving ones (i.e., phosphorus)?
    Did similar or different processes lead to the enrichment of close-in giant exoplanet atmospheres?
    \item \textbf{Existence of hot and warm Jupiters}---What causes a small fraction of gas giant planets to orbit close to their host star?
    \item \textbf{Occurrence vs. stellar distance and other properties}---What causes the uneven distribution of giant exoplanets in their orbits? Do the correlations between giant planet occurrence and stellar properties match theoretical predictions? 
    \item \textbf{Existence of many failed cores with short orbital periods}---How do massive cores avoid runaway gas accretion and remain as super-Earths or sub-Neptunes? Are their origins different from those of Uranus and Neptune?
    \item \textbf{Long-period gas giants}---Long-period gas giants with orbital distances $>$~30~au have been detected, and their thermal emissions have been observed. Is the existence of long-period gas giants consistent with current formation theories? What constraints do thermal emissions impose on gas giant planet formation?
    \item \textbf{Architecture of the Solar System}---How did Jupiter and Saturn contribute to the formation of the present-day architecture of the Solar System? Especially, did a great voyage of the giant planets take place and play a vital role?
\end{enumerate}

\subsection{Early formation of massive cores} \label{sec:formation_timescale_problem}

\color{black}

As discussed in Section~\ref{sec:core accretion}, the accretion of planetesimals of 10\,km or larger takes too long for core formation. In contrast, pebble accretion leads to rapid core growth; however, it requires an excessive total mass of pebbles. Furthermore, to fulfill the assumption of pre-existing cores, additional mechanisms for core formation are necessary in conjunction with pebble accretion. In discussions of planetesimal and pebble accretion, it is often assumed that the size of bodies accreting onto cores is fixed. However, the core growth rate actually depends on the size distribution of the small bodies that accrete onto it. These small bodies grow through mutual collisional evolution. The total collisional evolution from dust grains to planets may help reconcile these issues.

\begin{figure}[t]
\includegraphics[width=\textwidth]{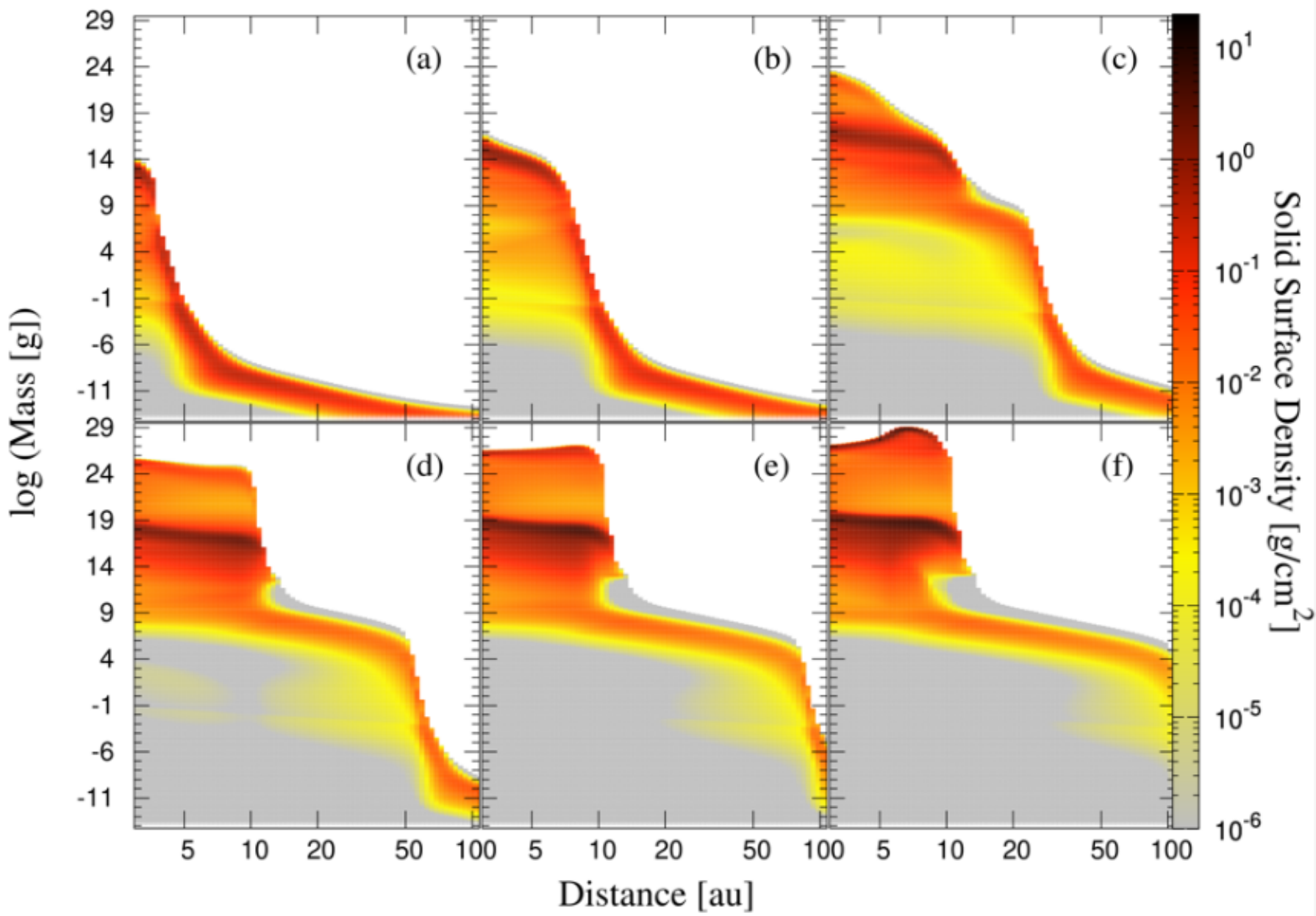}
\caption{The snapshots of the mass distribution of bodies obtained from the simulation for collisional evolution from dust to core at (a) $5.6\times 10^2$, (b) $2.1 \times 10^3$, (c) $1.5 \times 10^4$, (d) $5.6 \times 10^4$, (e) $1.2 \times 10^5$, and (f) $2.1 \times 10^5$ years {in the protoplanetary disk with the solid mass of $200\,M_\oplus$ around the solar-mass host star.}The color represents the solid surface densities of bodies with masses on the vertical axis and orbital distances in the horizontal axis.  {The initial sizes of solid bodies are given by a single size population with $d = 0.2\,\mu$m corresponding to $m \sim 10^{-16}\,$g. Bodies with Moon mass ($7\times 10^{25}$\,g) or Mars mass ($6\times 10^{26}$\,g) appear at $\la 10$\,au in $\sim 10^5$ years (d,e). Bodies with the crossover mass ($\sim 10 M_\oplus = 6 \times 10^{28}$\,g) are formed at $\approx 7$\,au in $2 \times 10^5$ years (f). }
This figure is from Fig.~3 of \cite{Kobayashi+21}.
}
\label{fig:dust_to_planet_simulation}
\end{figure}

\citet{Kobayashi+21} performed the simulations that treat solid bodies from dust grains to cores in the whole disk. All the processes dependent on the size of solid bodies introduced in Section \ref{sec:core accretion} are included in the simulations. 
Collisional evolution between similar-sized dust aggregates forms fluffy aggregates \citep{Suyama+12}, {while pebbles and small planetesimals are compressed by ram pressure and self-gravity \citep{Kataoka+13}.  
The porosity of bodies in the simulations was modeled based on such collisions and compression.} The simulation result is shown in Fig.~\ref{fig:dust_to_planet_simulation}. Dust growth occurs from the inner disk (Fig.~\ref{fig:dust_to_planet_simulation}a). The growth front of pebbles propagates beyond 10\,au at $t \ga 2 \times 10^3$ years. Pebbles grow up to planetesimals in the disk ($a \la 10$\,au, Fig.~\ref{fig:dust_to_planet_simulation}b), because pebbles with high porosity grow prior to their significant drift.   
In the inner disk ($a \la 10$\,au), 
the characteristic mass of planetesimals appears at $m \sim 10^{18}$\,g, {corresponding to $d \sim 10$\,km }(Fig.~\ref{fig:dust_to_planet_simulation}c,d,e).
This is caused by the onset of runaway growth of planetesimals because the gravitational focusing is effective \citep{Kobayashi+16}. 
On the other hand, pebbles are formed and drift inward from the outer disk, where the growth timescale of pebbles is longer than the drift timescale. The pebbles drift into the inner disk ($a \la 10\,$au) and then grow to planetesimals, {because the growth timescale becomes shorter than the drift timescale due to the transition of gas drag regimes caused by high porosity.} The pebble supply and planetesimal formation result in the high surface density of planetesimals in the inner disk.  In addition, such planetesimal formation continuously occurs until the growth front of pebbles reaches the outer edge of the disk. Small planetesimals are sustainably supplied. The core growth is accelerated by the large amount of planetesimals and the sustainable supply of small planetesimals. A core with mass of $10\,M_\oplus$ is formed in $2\times 10^5$ years (Fig.~\ref{fig:dust_to_planet_simulation}e). In addition, even when the core reaches the crossover mass, the surface density of planetesimals remain large, which can assist the pollution of the envelope and atmosphere with heavy elements (see Section \ref{sec:enrichment}).

The mechanism for such rapid core formation is planetesimal formation via pebble growth in the inner disk, which is caused by the high porosity of pebbles. However, the millimeter-wavelength polarized light of protoplanetary disks imply that the porosity of large dust aggregates is not so high \citep{Tazaki+19}. In addition, the high mass ratio collisions between dust aggregates lead to low porosity dust aggregates \citep{Tanaka+23}.  \citet{Kobayashi+23} thus investigated the dust-to-core simulations for compact pebbles. 
Planetesimals form in the inner disk via collisional growth even for non-porous pebbles, which is determined by the disk temperature. 
{If the temperature is lower than $80 (M_* / M_\odot)^{-1}$\,K at 10\,au in protoplanetary disks, planetesimals are formed in $a \la 7-15$\,au. }As a result, cores are formed at $a \approx 3-5\,$au in several $10^5$ years via similar way to porous pebbles. 

It should be noted that catastrophic disruption due to collisions of pebbles prohibits collisional growth from dust to cores. 
Realistic laboratory studies are still difficult for high-velocity collisions between icy aggregates in conditions equivalent to protoplanetary disks {\citep{Blum+08}}. However, collisional simulations of icy dust aggregates investigated collisional outcomes, which depend on monomer sizes of aggregates. For sub-micron-sized monomers, collisional fragmentation is insignificant unless the relative velocities are larger than 50--80\,m/s \citep{Wada+13}. In addition, even relatively high-velocity collisions result in mass transfer between target aggregates to projectile aggregates, which are quite different from catastrophic disruption \citep{Hasegawa+21,Hasegawa+23}. 
Dust aggregates composed of sub-micron-sized monomers account for optical and near-infrared polarimetric observations of protoplanetary disks \citep{Tazaki+22,Tazaki+23}. Therefore, critical collisional fragmentation does not disturb during collisional growth. 

Once cores exceed the crossover mass, gas accretion is significant. The gas accretion strongly depends on the solid accretion rate, corresponding to the core growth rate discussed in Section \ref{sec:core accretion}. \citet{Pollack+96} investigated the gas accretion of cores, assuming the solid accretion rates with low orbital inclinations of planetesimals. The overestimate of the solid accretion leads to the slow gas accretion phase {called ``Phase 2'' in Path~(III) of Fig.~\ref{fig:critical_luminosity}. The core formation with realistic solid accretion rates results in prompt runaway gas accretion without the slow gas accretion phase (see Path (II) of Figure \ref{fig:critical_luminosity}). Therefore, the solid accretion rates are important for the gas accretion of cores as well as core formation. }

\subsection{Enriched interior with diluted cores}
\label{sec:enrichment}
\color{black}
As mentioned in Section~\ref{sec:constraints}, gravity measurements indicate that Jupiter and Saturn contain substantial amounts of heavy elements distributed throughout their interiors. Both planets may have a dilute core rather than a massive compact core, partially mixed with the surrounding H-He envelope. This suggests that their internal structure is more complex than the simple core-envelope one that was traditionally thought. 
Meanwhile, many giant exoplanets with relatively short orbital periods (warm Jupiters) exhibit even more significant heavy-element enrichment compared to Jupiter and Saturn. These findings emphasize the variety in giant planet composition, which may indicate diverse formation and evolutionary histories resulting in the inclusion of heavy elements.

First, the existence of dilute cores is qualitatively consistent with recent core-accretion models for giant planet formation. While traditional core-accretion models involve the formation of a distinct solid core followed by rapid gas accretion \citep[e.g.,][]{Pollack+96}, recent theories have demonstrated that small-sized solids, such as small planetesimals \citep[e.g.,][]{Kobayashi+21} or pebbles \citep[e.g.,][]{Johansen+17}, significantly contribute to planetary growth. 
Such small solids evaporate in the hot envelope without reaching the core surface, polluting the envelope with heavy elements \citep[e.g.,][]{Valletta+19,Brouwers+18}. 
Envelope pollution leads to a significant reduction in the critical core mass, allowing small cores to accrete disk gas more efficiently and grow more rapidly \citep[][]{Hori+11,Venturini+15,Venturini+16}. Thus, a small compact core surrounded by a polluted envelope can be formed. 

The critical core mass is, however, much smaller than the total mass of heavy elements inferred in Jupiter, Saturn, and other metal-rich warm Jupiters, indicating that solid accretion continues after the initial core formation. Theories suggest that further accretion of solids becomes increasingly difficult due to the phenomenon known as isolation: For planetesimals, a growing planet clears its feeding zone, limiting the supply of planetesimals \citep[e.g.,][]{Kokubo+98,Shibata+19}; for pebbles, the planet creates a pressure bump in the gas disk, preventing the inward drift of pebbles and halting pebble accretion \citep[][]{Lambrechts+2014b}. 
Given the limitations imposed by the isolation, alternative pathways must be considered to explain the substantial enrichment of heavy elements observed in the envelopes of gas giant planets.

\begin{figure}
    \centering
    \includegraphics[width=\textwidth]{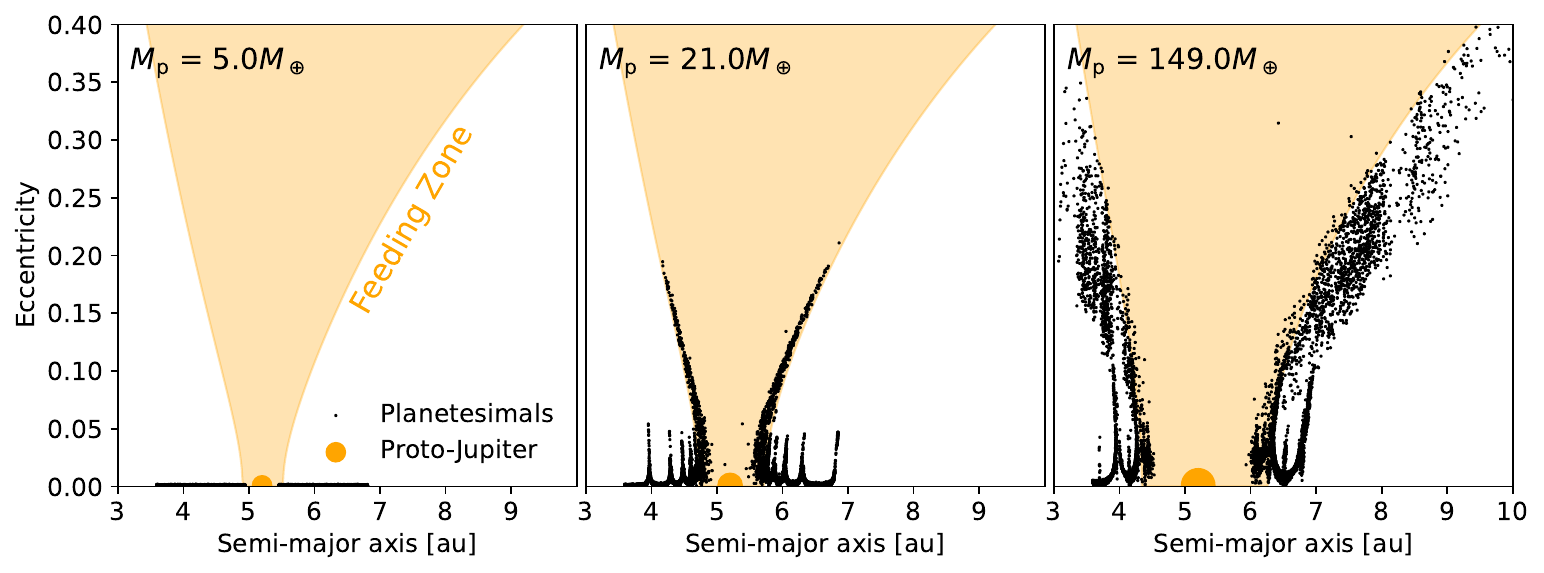}
    \caption{Dynamics of planetesimals around a proto-Jupiter {growing from 5~$\MEarth$ to 149~$\MEarth$. The planetesimals are represented by black dots, while proto-Jupiter is shown as a filled orange circle. The shaded area represents the planet's feeding zone, which expands as the planet's mass increases. The simulations have been done according to \cite{Shibata+19}.} 
    Here it is assumed that the planet growth rate is $1 \times 10^{-2} M_\oplus/{\rm yr}$ and the planetesimal radius is $1 \times 10^5$~cm.
    }
    \label{Fig: Shibata+19_Fig2}
\end{figure}

\begin{marginnote}
    \entry{Feeding zone}{Circumstellar region around the planet only inside which solid particles can encounter the planet.}
\end{marginnote}
One potential way for a planet to accumulate more heavy elements is by expanding its feeding zone as it grows due to rapid gas accretion{; An example of numerical simulation is given in Fig.~\ref{Fig: Shibata+19_Fig2}.} This would allow the planet to capture more planetesimals or pebbles that drift into the feeding zone. However, this process is not very efficient due to the planet's strong gravitational scattering \citep[{see Fig.~\ref{Fig: Shibata+19_Fig2};}][]{Shiraishi+08,Shibata+19}. Planetesimals can only enter the feeding zone through specific channels near the mean motion resonances with the planet \citep[][]{Shibata+19}. 
The sufficient accumulation of heavy elements into the planet is possible in  a large surface density of planetesimals, which is achieved in the dust-to-core simulations \citep{Shibata+23}. 
Note that pebbles or small planetesimals are too small to have their eccentricities enhanced at the mean-motion resonances and to enter the feeding zone. Since the drift timescale is shorter than the scattering timescale, small bodies $s \la 100$\,m are not disturbed to enter the feeding zone \citep{Kobayashi+10Icar}.

One alternative pathway involves the engulfment of planetesimals by migrating giant planets. \cite{Shibata+20} studied how migrating giant planets can accumulate heavy elements within their interiors by capturing planetesimals ahead of themselves. They assumed that as a giant planet migrates inward, it encounters swarms of planetesimals. However, the encounter does not always lead to significant enrichment: From the dynamical perspective, there are two shepherding processes, the resonant shepherding and the aerodynamic shepherding, which act as barriers for planetesimals to be captured by a migrating giant planet. 
Their dynamical simulations, nevertheless, identified an optimal ``sweet spot'' during planetary migration where the efficient capture of planetesimals occurs. 
The migrating planet pushes planetesimals ahead of itself forward like a bulldozer by trapping the planetesimals in its mean motion resonances. The trapped planetesimals are dynamically excited and, thereby, collide with the planet. However, as the planet moves inward, disk gas density becomes so high that aerodynamic damping dominates over the dynamic excitation \citep[see][for the detailed analysis]{Shibata+22}. 
Note that the initial concept was proposed by \cite{Alibert+05}, who suggested that Jupiter and Saturn were formed farther from their current locations and subsequently migrated. However, their treatment of the collisional cross-section was oversimplified and did not fully consider the complex interactions between migrating planets and surrounding solids.

Based on the pebble accretion hypothesis, \cite{Schneider+21} have recently examined the role of accreting enriched gas from the protoplanetary disk in the context of the formation of metal-rich warm Jupiters {\citep[see also][]{Kalyaan+21,Kalyaan+23,Mah+23}}. Disk gas inside the snowline becomes enriched with heavy elements due to the evaporation or ablation of inward-drifting pebbles. If a core captures this enriched gas in situ, the resulting envelope {and atmosphere} also become enriched. Simulating the 1D thermal and chemical evolution of protoplanetary disks with drifting and evaporating pebbles, they demonstrated that this process can significantly enrich close-in giant planets. \cite{Schneider+21} also highlighted that the timing of planet formation and the location within the disk are crucial factors that determine the extent of this enrichment. Additionally, their models could reproduce the observed positive correlation between planet metallicity and planet mass, suggesting that more massive planets are capable of accreting a larger amount of metal-rich gas and solids, thereby enhancing their overall heavy element content.

\color{black}
\subsection{Enriched atmospheres}

The atmospheres of Jupiter and Saturn are enriched in heavy elements, including noble gases such as argon, krypton, and xenon, and phosphorus, as well as carbon and nitrogen at abundances several times higher than those found in the Sun’s photosphere, as described in Section~\ref{sec:atmosphere_solar}. This uniform enhancement of heavy elements presents a challenge to planet formation theories.
Phosphorus (P) has been detected in the form of phosphine (PH$_3$) in those atmospheres. In the solar nebula, the temperature in the region where Jupiter and Saturn formed was likely lower than the condensation temperature of PH$_3$ (approximately 140K). This suggests that phosphorus was likely present in the form of solid phosphorus-containing minerals within solids rather than in a gaseous form. 
Conversely, noble gases are highly volatile and typically exist in a gaseous state. For Ar to be incorporated into solids, temperatures must have been very low: around 30~K or less for adsorption onto water ice \citep[e.g.,][]{Bar-Nun+85} and 40-60~K for clathrate hydrates \citep[e.g.,][]{Hersant+04}. 

\cite{Owen+99} first proposed that planetesimals formed at low temperatures in the outer solar nebula trapped noble gases and other volatiles in amorphous ice or clathrate hydrates. 
Later, \cite{Mousis+12} explored the depletion of water in the protoplanetary disk,  
suggesting that icy planetesimals formed in water-poor regions could have different volatile profiles {\citep[see also][]{van_Dishoeck+14,van_Dishoeck+21}}. 
\cite{Oberg+19} discuss the delivery mechanisms of nitrogen and other heavy elements via icy planetesimals. 
\cite{Bosman+19} examined how volatiles, including \rev{nitrogen and} noble gases, were trapped in icy planetesimals depending on their formation location in the protoplanetary disk, using isotopic ratios to trace these origins. 
\cite{Ohno+21} highlight the role of shadows cast by forming planets or other bodies in the protoplanetary disk, creating localized cold regions. In these shadowed regions, temperatures could drop sufficiently low to allow noble gases to become trapped in water ice or clathrate hydrates. These cold traps could lead to the formation of icy grains and planetesimals enriched in noble gases, which were later accreted by Jupiter, enriching its atmosphere. 
{In summary, it remains unclear what caused the observed element abundances in Jupiter's atmosphere.}

The possibility that Jupiter’s atmosphere might be more enriched than its interior also poses a challenge to planet formation theories. As of writing, there has been no study that specifically addresses this issue. Most studies have focused on explaining the overall enrichment of Jupiter’s {envelope and} atmosphere and have generally assumed that the heavy element distribution should be relatively uniform or that the deep interior should have a higher concentration due to the core accretion process.
The data from missions like Juno, as well as advanced modeling of Jupiter’s formation and interior processes, may eventually provide insights into this possibility. 

Finally, recent observations of hot Jupiters using JWST have revealed that their atmospheres are also enriched in heavy elements compared to their host stars. However, the origin of hot Jupiters remains a matter of debate with multiple formation theories such as in-situ formation, disk migration, and high-eccentricity migration, all suggesting different processes and histories. Because of this uncertainty, the causality behind the metal enrichment observed in the atmospheres of hot Jupiters is not well understood. {The processes described in Section~\ref{sec:enrichment} are also applicable to hot Jupiters.} While many studies have recently attempted to predict the chemical compositions of hot Jupiters' atmospheres, such predictions may be premature, given our limited knowledge of how hot Jupiters formed and how various molecules are distributed in protoplanetary disks. More observations are needed to provide a clearer understanding of the atmospheric characteristics and formation histories of hot Jupiters.

\subsection{Existence of hot/warm Jupiters}
\label{sec:hot_warm_jupiter}

\begin{figure}
    \centering
    \includegraphics[width=1.0\linewidth]{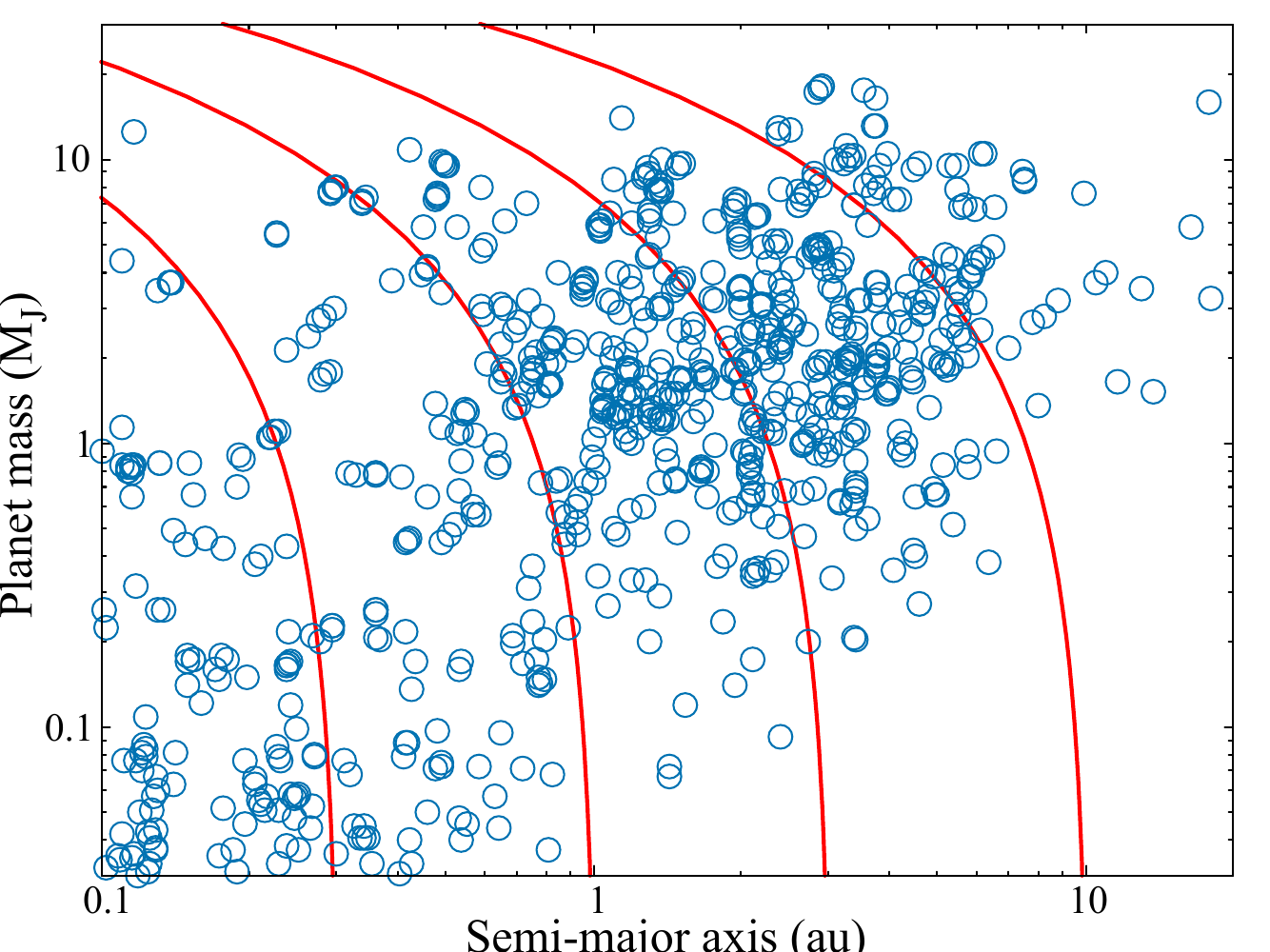}
    \caption{Predicted evolution paths of planets that concurrently grow and migrate (red curves). The open circles show the observed (minimum-)masses and semi-major axes of exoplanets detected through radial-velocity measurements around G dwarfs (0.84-1.15 $M_\odot$; data from NASA Exoplanet Archive). Modified from Figure 1 of Tanaka et al. (2020).
}
    \label{fig:tanaka+20}
\end{figure}

The discovery of hot and warm Jupiters, although less common compared to other types of exoplanets, poses a challenge to our understanding of planet formation and migration. These giant planets, found in close orbits around their host stars, are believed to have formed beyond the snowline before migrating inward. Two primary migration pathways have been proposed to explain their current locations: disk-assisted migration and dynamical migration. Both theories provide insights but also pose challenges in fully explaining the observed characteristics of these planets. Since a comprehensive review of the origin of hot and warm Jupiters has been published by \citet{Dawson+18}, this section will provide an overview of the current understanding and highlight any recent studies that have further refined these theories.

Regarding the formation of hot/warm Jupiters via disk-assisted migration, building on \cite{Tanigawa+07} and \cite{Tanigawa_Tanaka16}, \cite{Tanaka+20} recently developed a new model for the concurrent growth and migration of gas giant planets by incorporating the effects of photo-evaporation of disk gas and an updated model for Type~II planetary migration (see Fig.~\ref{fig:tanaka+20}). 
Their model suggests that when planets of $<$ a few $\MJupiter$ undergo runaway gas accretion, they {initially} hardly experience significant radial migration due to their rapid growth. The study emphasizes that the final mass of giant planets of these masses is mainly determined by the timing of gas accretion and the disk's photo-evaporation rate, and gas accretion is nearly completed in situ by collecting the total amount of gas available in the disk. 
In contrast, more massive planets (greater than a few Jupiter masses) allow some gas to flow past the gap, reducing accretion efficiency. 
As illustrated in Fig.~\ref{fig:tanaka+20}, if their model is accurate and no other migration mechanisms work, hot/warm Jupiters of $\lesssim 10 \MJupiter$ and $a \lesssim 1~{\rm au}$ must have initiated runaway gas accretion at their present-day locations, which again pose a challenge to core accretion theories.

On the other hand, dynamical migration involves gravitational interactions, such as planet-planet scattering or Kozai-Lidov cycles induced by stellar companions. These interactions can increase a planet’s orbital eccentricity, placing it on a highly elliptical orbit that gradually circularizes due to tidal forces \citep[][]{Rasio+96,Noaz+11}. Observational evidence, such as the presence of distant stellar companions around hot Jupiter hosts \citep[e.g.,][]{Knutson+14} and the detection of warm Jupiters with high eccentricities, supports this migration pathway. However, this model requires specific initial conditions and is less effective at explaining low-eccentricity warm Jupiters \citep[][]{Dawson+18}.

Emerging theories propose hybrid scenarios combining the disk-assisted and dynamical migration. These models assume that a planet might first migrate inward through its natal disk and later undergo dynamical interactions that modify its orbit. Such hybrid models are gaining attraction, especially for warm Jupiters, which exhibit a wide range of orbital eccentricities and semi-major axes. Additionally, the controversial in situ formation hypothesis posits that some hot Jupiters may form near their current locations, challenging the traditional view of giant planet formation exclusively beyond the snowline \citep[][]{Batygin+16, Huang+16}. 

\subsection{Correlation with stellar properties}
Observationally, several correlations between the occurrence rates of giant planets and stellar properties have been identified, as mentioned in Section~\ref{sec:constraints:exoplanets}. However, from a theoretical perspective, the process of giant planet formation remains highly complex and not yet fully understood. As a result, quantitative comparisons between observations and theoretical models are premature at this stage. Nevertheless, the state-of-the-art core accretion models align qualitatively with observed correlations between giant planet occurrence rates and stellar properties. 

Core accretion models suggest that gas giant planets form more efficiently around metal-rich stars, whether through planetesimal accretion \citep[e.g.,][]{Ida+04b,Mordasini+12,Emsenhuber+21} or pebble accretion \citep[e.g.,][]{Brugger+18,Ndugu+18}. This aligns with observations indicating a strong correlation between stellar metallicity and the frequency of giant planets \citep{Santos+04,Fischer+05}. The reason for this is that cores large enough to trigger runaway gas accretion form more easily in disks that are richer in metals.

Models also predict that giant planets form predominantly beyond the snow line, where solid materials are abundant. This aligns with the observed distribution of cold Jupiters located several au away from Sun-like stars \citep[e.g.,][]{Fulton+21}. Surveys also indicate that approximately 10\% of solar-type stars host gas giants beyond 1~au, which is reasonably consistent with the core accretion models. 
In Class~0 objects, about 10\% of the disks contain more than 100 Earth masses of solid material \citep{Tychoniec20,Mulders+21}. These disks can produce cores with $\sim 10$ Earth masses within several $10^5$ years, as demonstrated by dust-to-planet simulations \citep{Kobayashi+21,Kobayashi+23}. 

The known dependences on stellar mass can also be qualitatively explained through core accretion models.
Observations show that the occurrence rate of hot Jupiters is highest at Sun-like (G dwarf) stars: Specifically, the frequency of these planets increases with stellar mass from M dwarfs to G dwarfs, whereas it decreases from G dwarfs to A dwarfs \citep[e.g.,][]{Bryant+23}. In contrast, giant planets with relatively long orbits (1-5~au) show no such trends \citep[][]{Fulton+21,Gan+24}. 
Additionally, recent ALMA surveys of nearby star-forming regions indicate that the fraction of structured disks, especially transition disks, is dependent on stellar mass, suggesting long-period giant planets are more common around more massive stars \citep[][]{van_der_Marel+21}. 
These observational facts may be understood in a manner that gas giant formation is less efficient in disks with limited solid material around sub-solar mass stars \citep[e.g.,][]{Ida+05}, while early disk dissipation for super-solar-mass (intermediate-mass) stars inhibits gas giant formation or orbital migration \citep[e.g.,][]{Currie+09,Yasui+14}.

However, there are still quantitative challenges, particularly regarding the origins of hot/warm Jupiters and failed cores, which have yet to be clarified.

\subsection{Existence of many failed cores with short orbital periods} \label{sec:failed_core}
The \mission{Kepler}'s finding that super-Earths and sub-Neptunes are common while hot Jupiters are rare is of particular interest from the perspective of core accretion theory, which predicts that those objects are massive enough to undergo rapid gas accretion and evolve into gas giants (see \S~\ref{sec:critical}). Understanding the origin of these ``failed cores'' is also crucial for comprehending the formation of gas giant planets.

The discovery of the first low-density super-Earths orbiting Kepler-11 \citep[][]{Lissauer+11} inspired theorists to consider the formation of failed cores at short orbital periods. \cite{Ikoma+12} first proposed that low-mass protoplanets without atmospheres migrate from afar in a protoplanetary disk. As the disk starts to dissipate, the dynamical system becomes destabilized, causing the protoplanets to collide with each other. These collisions allow them to become massive enough to collect surrounding disk gas. However, runaway gas accretion stops once the disk gas is depleted. Additionally, \cite{Lee+14} suggested the possibility of dusty envelopes avoiding runaway growth.

Envelope recycling is another possibility as discussed in Section~\ref{sec:recycling}. In this process, gas around the planetary core moves back into the protoplanetary disk instead of being fully retained \citep[e.g.,][references therein]{Moldenhauer+22}. This recycling of the gas prevents the envelope from cooling effectively, which delays the onset of runaway accretion. Recent 3D hydrodynamic simulations have shown that envelope recycling is especially important for planets with short orbital periods that are surrounded by dense protoplanetary disks. In these situations, the recycling flows restrict the net accumulation of gas. This mechanism helps to explain why many short-period planets gather moderate envelopes but do not transform into gas giants.

Several ideas have been proposed to explain the formation of failed cores, suggesting that many planets can gather some gas without undergoing runaway accretion, ultimately resulting in super-Earths or sub-Neptunes. However, our understanding of the building blocks that form cores in regions interior to the snowline remains insufficient. To fully understand failed core formation, it is essential to consider these planets within the broader framework of planetary system formation, which includes the formation of long-period giant planets. Integrating both short- and long-period planetary formation processes will provide a more comprehensive picture of how planetary systems evolve.
\subsection{Long-period gas giants}
Direct imaging has become a key technique in exoplanet research, offering valuable insights into planetary systems, particularly for gas giants with long orbital periods that are challenging to detect using radial velocity or transit methods. By targeting young, massive planets still emitting radiation that originates from their accretion heat, direct imaging that observes their intrinsic luminosities not only complements other detection techniques but also offers critical constraints on planetary formation and evolution. Discoveries such as four planets orbiting $>$~15~au from HR~8799 of age $\sim$30~Myr \citep[][]{Marios+08,Marios+10} and Beta Pictoris~b orbiting $\sim$10 au from its host star of age $\sim$20~Myr \citep[][]{Lagrange+10} demonstrated the capability of direct imaging to detect long-period gas giants and probe their thermal states. 

Surveys suggest that long-period gas giants are relatively rare \citep[e.g.,][]{Nielsen+19,Fulton+21}.
As discussed in Section~\ref{sec:core accretion}, theoretically, early core formation is feasible at $a \la 10$\,au via the accretion of small planetesimals or pebbles. However, core formation becomes significantly more challenging at $a \ga 30$\,au. One alternative mechanism for the formation of gas giants is the gravitational instability of gas disks. However, in these unstable disks, newborn gas giants undergo rapid orbital decay \citep{Machida+11}, which should be addressed in future studies. In addition, the planet-hosting stars of HR~8799 and Beta Pictris are accompanied by debris disks, which implies planetesimals exist in the systems \citep{Marios+10,Smith+84}. The origin of planetesimal belts should be addressed consistently with gas giant formation.   

There is ongoing debate about the thermal evolution of gas giants based on observed thermal emissions, specifically regarding the “hot start” with high initial entropy and the “cold start” with low initial entropy.
\cite{Marley+07} first presented a model for an extremely cold start and showed that a gas giant evolving from the cold start emits intrinsic luminosity that is significantly lower---by a few orders of magnitude---at ages of 10-100~Myr compared to a hot start with an extremely high initial entropy. The former and latter evolutions are often referred to as those for gas giants formed via core accretion and disk instability, respectively. There is still insufficient understanding of the entropy that gas giants acquire during their accretion phase. For the cold start model, \cite{Marley+07} adopted one-dimensional core-accretion models from \cite{Hubickyj+05}, which utilized isothermal accretion shock models that brought low-entropy gas into the envelope \citep[][]{Bodenheimer+00}; in contrast, the hot start model is based on 1D thermal evolution models for brown dwarfs with arbitrarily high initial entropies.

Recently \cite{Marleau+19} carried out one-dimensional radiation-hydrodynamic simulations to investigate post-shock entropies \citep[see also][]{Marleau+17}. Their findings revealed significantly high post-shock entropies, suggesting warm or hot starts. Their conclusion aligns with the observed thermal emissions of gas giants, Beta Pictris b and c, 51 Eri b, and HR 8799 e, that have measured dynamical masses \citep[][respectively]{Snellen+18,Dupuy+22,Brandt+21}. 
Nevertheless more studies will be needed: Gas accretion occurs in three dimensions, not just one, as demonstrated by hydrodynamic simulations (see \S~\ref{sec:2D/3D_gas_accretion}). Accreting gas giants have been detected with H$\alpha$ observations \citep[PDS70 b and c][]{Wagner+18,Haffert+19}. Radiation-hydrodynamic models that account for the observed H$\alpha$ emissions show that accretion does not occur across the entire surface of the planet, but rather in narrow regions of its surface \citep[][]{Aoyama+19}. It remains unclear how the accreting gas mixes with the interior and how entropy is deposited.
\color{black}

\subsection{Architecture of the Solar System}
Our Solar System contains at least two gas giant planets: Saturn and Jupiter. The core of Saturn has been thought to be relatively larger than that of Jupiter (see \S~\ref{sec:solar_interior}) which raises interesting questions about the differences in the formation processes of these two planets. 
The standard core accretion model predicts that the critical core mass for Saturn should be smaller than that for Jupiter due to Saturn’s lower accretion rate (see \S~\ref{sec:critical}). This difference suggests that additional mechanisms may be necessary to achieve high solid accretion rates for Saturn.
\cite{Kobayashi+12} addressed this issue by demonstrating that the pressure bump caused by Jupiter could effectively accelerate the accumulation of Saturn’s core via planetesimal accretion. Alternatively, based on the pebble accretion model, \cite{Lambrechts+2014a} demonstrated that Saturn’s core grows rapidly by capturing pebble-sized particles drifting through the protoplanetary disk, reaching a larger pebble isolation mass than at Jupiter's location. 
However, as mentioned in Section~\ref{sec:solar_interior}, recent interior models do not always infer a larger core for Saturn, making this aspect of Saturn’s formation uncertain, highlighting the need for further understanding of both interior structure and planet formation.

Uranus and Neptune did not undergo significant gas accretion, {as indicated by} their relatively thin envelopes. Their different semimajor axes suggest it is unlikely that their formation coincided with the depletion of the gas. They formed after the substantial depletion of the solar nebula, with planetesimals remaining after the gas depletion. Uranus and Neptune may have formed via planetesimal accretion. The simple planetesimal accretion takes too long as discussed in Section \ref{sec:core accretion}. Planetesimals are stirred by protoplanets, leading to destructive collisions between them, which produce {numerous} fragments. If {frequent} collisions between fragments result in significant collisional damping, the formation timescale of Uranus and Neptune becomes sufficiently short due to the accretion of {fragments with low eccentricies and inclinations} \citep{Goldreich04}. However, the simulations for planetesimal accretion after the gas depletion show that massive solid planets such as Uranus and Neptune are not formed because of insignificance of collisional damping in the collisional cascade of planetesimals \citep{Kobayashi+14}. Therefore, the formation process of Uranus and Neptune remains uncertain.

The non-carbonaceous (NC) and carbonaceous (CC) dichotomy observed in meteorites is thought to have been established during Jupiter’s growth. This dichotomy suggests that Jupiter formed relatively early (0.5-1 Myr after the CAI formation), acting as a barrier that maintained a prolonged (up to 4 Myr) separation between the inner NC and outer CC reservoirs (see §\ref{sec:cosmochemistry}). As a result, the materials in the inner and outer Solar System evolved independently for an extended period, which is why we observe distinct differences in the isotopic compositions of meteorites today. As Jupiter began to accrete material, its growing mass created a gap in the protoplanetary disk, likely inhibiting the inward drift of dust grains or pebbles from the outer regions once Jupiter’s core reached about 20~$M_\oplus$ \citep[][]{Morbidelli+16,Alibert+18}. This separation prevented the mixing of materials between the inner and outer disk, leading to the formation of distinct reservoirs and resulting in the observed NC-CC dichotomy. Furthermore, the gravitational stirring by Jupiter began mixing NC and CC materials when the planet’s mass reached approximately 50~$M_\oplus$ \citep[][]{Alibert+18}. \cite{Brasser+20} proposed another idea that the initial separation of the NC and CC reservoirs was not caused by Jupiter itself but rather by a pressure maximum in the disk near the location where Jupiter would later form. Nevertheless, the efficiency of Jupiter’s barrier in maintaining this dichotomy depends on the size of dust grains and pebbles, as smaller particles might still cross the gap. Further studies will be necessary to understand these dynamics in greater detail.

The mixing of the reservoirs might have been caused by the migration of the gas giant planets. The Grand Tack model \citep{Walsh+11} has been one of the most influential theories explaining the present-day architecture of the Solar System. According to this model, Jupiter initially migrated inward before reversing its direction due to the influence of Saturn, resulting in the planets’ current positions. This migration significantly shaped the distribution of inner Solar System planetesimals, explaining the observed differences between the terrestrial planets and the structure of the asteroid belt. However, recent theoretical advancements in gas giant migration, such as the work by \cite{Tanaka+20}, suggest that such large-scale migrations may not be as common as once thought. These newer models, which effectively account for the scarcity of hot Jupiters around other stars, cast doubt on the necessity of the Grand Tack scenario to explain the current configuration of our Solar System.

%%%%%%%%%%%
% OUTLOOK %
%%%%%%%%%%%
%\section{Summary and Future Perspectives} \label{sec:outlook}

% Summary Points
\color{black}
\begin{summary}[SUMMARY POINTS]
\noindent
{This article is summarized as follows: The first three points are observational facts, while the next four points outline necessary actions to address them.}
\begin{enumerate}
\item 
{Jupiter's and Saturn's envelopes and atmospheres contain significant amounts of heavy elements and their cores may not be compact but rather diluted.} 
\item 
{Exoplanet observations suggest that close-in gas giants also have interiors and atmospheres rich in heavy elements.}
\item 
{Hot Jupiters are rather rare, while cold Jupiters at several~au are more common. At shorter orbital periods, failed cores such as super-Earths and sub-Neptunes are more prevalent, being significantly more abundant than hot Jupiters.}
\item 
{Understanding of early-stage envelope accumulation has advanced significantly, particularly regarding the impact of heavy element pollution. However, the challenge of recycling flows between the disk and planet remains unresolved.}
\item Instead of viewing core formation as a debate between planetesimal and pebble accretion, the recent unified model provides a more realistic understanding, which tracks the size distribution of solid particles and its evolution over a wide range, offering a comprehensive view of how gas giant cores form. This has resolved {how gas giants can form so quickly.} 
\item To account for the excess of heavy elements in both the interior and atmosphere, additional steps for heavy element acquisition are necessary {beyond the primary accretion that contributes to core growth}. While various ideas have been proposed, each has its own advantages and limitations, and the issue remains unresolved.
\item 
{The migration of giant planets through disk-assisted processes occurs more slowly than once believed, aligning with the rarity of hot Jupiters. While dynamical migration can explain hot Jupiters, it leaves the formation of warm Jupiters unresolved.}
\end{enumerate}
\end{summary}

% Future Issues
\begin{issues}[FUTURE ISSUES]
\begin{enumerate}
\item \textbf{Theoretical modeling:} It is crucial to develop self-consistent models integrating solid accretion, envelope pollution, and the subsequent interior evolution of gas giant planets with compositional gradients. These processes are interconnected and have significant implications for the overall structure and composition of these planets. 
\item \textbf{Tracking the size distribution of solids:} Laboratory experiments focusing on the collision, sticking, and fragmentation of particles are crucial for understanding how solid materials evolve within protoplanetary disks. These experiments will refine models of solid growth and accretion, which are essential for understanding the formation of planetary cores and their subsequent growth into gas giants.
\item \textbf{Solar-system gas giants:} Obtaining a detailed understanding of Jupiter and Saturn's interiors will be greatly aided by seismology, namely the observation of surface oscillations. This approach offers critical insights into the distribution of heavy elements within their interiors. This understanding is key to unraveling the formation and evolution of Jupiter and Saturn.
\item \textbf{Exoplanet Atmospheres}: In the next decade, we expect significant advancements in our understanding of the atmospheric compositions of exoplanets. The JWST and Ariel will play pivotal roles in this progress. These observations will allow for more precise determinations of elemental abundances, {leading to the validation of predictions from models that can be tested by observations.}
\item \textbf{Exoplanet demographics:} The {microlensing survey with the} Roman Space Telescope will expand the search for exoplanets, allowing us to study planetary systems that are more widely distributed, resembling our Solar System. It will also help us understand the distribution of planetary cores and transition objects ($\lesssim$ 30~$M_\oplus$) near the snowline and beyond. These observations will provide critical details about the processes of planetary growth and migration, giving us important insights into the formation of gas giant planets. 
\item \textbf{Habitable planet formation:} The formation of habitable planets has traditionally been investigated separately from the formation of gas giant planets. However, the latter can significantly affect the former by shaping the overall architecture of planetary systems. It would be particularly important to understand how volatile-rich solids, which failed to become the cores of gas giants, are incorporated into rocky planets in the inner regions of protoplanetary disks. This process plays a crucial role in determining the composition and potential habitability of these planets.
\end{enumerate}
\end{issues}

\section*{DISCLOSURE STATEMENT}
If the authors have noting to disclose, the following statement will be used: The authors are not aware of any affiliations, memberships, funding, or financial holdings that
might be perceived as affecting the objectivity of this review. 

% Acknowledgements
\section*{ACKNOWLEDGMENTS}
We would like to extend our sincere gratitude to Drs.~Ewine van Dishoeck and Jonathan Fortney for their support and the opportunity to write this review article. 
We appreciate our colleagues for fruitful discussions, in particular, the contributions of Drs.~Tadahiro Kimura and Sho Shibata for providing their calculation data.
% References
%
% Margin notes within bibliography
\section*{LITERATURE\ CITED}

To download the appropriate bibliography style file, please see \url{http://www.annualreviews.org/page/authors/author-instructions/preparing/latex}.

\noindent
Please see the Style Guide document for instructions on preparing your Literature Cited.

The citations should be listed in alphabetical order, with no titles. For example:x

\newcommand{\aap}{\textit{Astron. Astrophys.}}
\newcommand{\aj}{\textit{Astron.~J.}}
\newcommand{\apj}{\textit{Astrophys.~J.}}
\newcommand{\apjl}{\textit{Astrophys.~J.~Lett.}}
\newcommand{\apjs}{\textit{Astrophys.~J.~Supp.}}
\newcommand{\araa}{\textit{Ann.~Rev.~Astron.~Astrophys.}}
\newcommand{\aaps}{\textit{Astron.~Astrophys.~Supp.}}
\newcommand{\nat}{\textit{Nature}}
\newcommand{\icarus}{\textit{Icarus}}
\newcommand{\pasj}{\textit{Publ.~Astron.~Soc.~Japan}}
\newcommand{\planss}{\textit{Planet.~Space~Sci.}}
\newcommand{\ssr}{\textit{Space~Sci.~Rev.}}
\newcommand{\grl}{\textit{Geophys.~Res.~Lett.}}
\newcommand{\jgr}{\textit{J.~Geophys.~Res.}}
\newcommand{\mnras}{\textit{Mon. Not. Royal Astron. Soc.}}
\newcommand{\pasp}{\textit{Publ.~Astron.~Soc.~Pac.}}
\newcommand{\prl}{\textit{Phys.~Rev.~Lett.}}
\newcommand{\psj}{\textit{Planet.~Sci.~J.}}

\bibliography{refs}

\begin{thebibliography}{}
\expandafter\ifx\csname natexlab\endcsname\relax\def\natexlab#1{#1}\fi

\bibitem[{{Adachi}, {Hayashi} \& {Nakazawa}(1976)}]{Adachi+1976}
{Adachi} I, {Hayashi} C, {Nakazawa} K. 1976.
\textit{Progress of Theoretical Physics} 56:1756--1771

\bibitem[{{Alderson} et~al.(2023){Alderson}, {Wakeford}, {Alam}, {Batalha}, {Lothringer} et~al.}]{Alderson+23}
{Alderson} L, {Wakeford} HR, {Alam} MK, {Batalha} NE, {Lothringer} JD, et~al. 2023.
\textit{\nat} 614:664--669

\bibitem[{{Alibert} et~al.(2005){Alibert}, {Mordasini}, {Benz} \& {Winisdoerffer}}]{Alibert+05}
{Alibert} Y, {Mordasini} C, {Benz} W, {Winisdoerffer} C. 2005.
\textit{\aap} 434:343--353

\bibitem[{{Alibert} et~al.(2018){Alibert}, {Venturini}, {Helled}, {Ataiee}, {Burn} et~al.}]{Alibert+18}
{Alibert} Y, {Venturini} J, {Helled} R, {Ataiee} S, {Burn} R, et~al. 2018.
\textit{Nature Astronomy} 2:873--877

\bibitem[{{Aoyama} \& {Bai}(2023)}]{Aoyama+23}
{Aoyama} Y, {Bai} XN. 2023.
\textit{\apj} 946:5

\bibitem[{{Aoyama} \& {Ikoma}(2019)}]{Aoyama+19}
{Aoyama} Y, {Ikoma} M. 2019.
\textit{\apjl} 885:L29

\bibitem[{{Ayliffe} \& {Bate}(2009)}]{Ayliffe+09}
{Ayliffe} BA, {Bate} MR. 2009.
\textit{\mnras} 393:49--64

\bibitem[{{Bae} et~al.(2023){Bae}, {Isella}, {Zhu}, {Martin}, {Okuzumi} \& {Suriano}}]{Bae+23}
{Bae} J, {Isella} A, {Zhu} Z, {Martin} R, {Okuzumi} S, {Suriano} S. 2023.
\textit{{Structured Distributions of Gas and Solids in Protoplanetary Disks}}. In \textit{Protostars and Planets VII}, eds. S~{Inutsuka}, Y~{Aikawa}, T~{Muto}, K~{Tomida}, M~{Tamura}, vol. 534 of \textit{Astronomical Society of the Pacific Conference Series}

\bibitem[{{Bar-Nun} et~al.(1985){Bar-Nun}, {Herman}, {Laufer} \& {Rappaport}}]{Bar-Nun+85}
{Bar-Nun} A, {Herman} G, {Laufer} D, {Rappaport} ML. 1985.
\textit{\icarus} 63:317--332

\bibitem[{{Bate} et~al.(2003){Bate}, {Lubow}, {Ogilvie} \& {Miller}}]{Bate+03}
{Bate} MR, {Lubow} SH, {Ogilvie} GI, {Miller} KA. 2003.
\textit{\mnras} 341:213--229

\bibitem[{{Batygin}, {Bodenheimer} \& {Laughlin}(2016)}]{Batygin+16}
{Batygin} K, {Bodenheimer} PH, {Laughlin} GP. 2016.
\textit{\apj} 829:114

\bibitem[{{Bean} et~al.(2023){Bean}, {Xue}, {August}, {Lunine}, {Zhang} et~al.}]{Bean+23}
{Bean} JL, {Xue} Q, {August} PC, {Lunine} J, {Zhang} M, et~al. 2023.
\textit{\nat} 618:43--46

\bibitem[{{Beleznay} \& {Kunimoto}(2022)}]{Beleznay+22}
{Beleznay} M, {Kunimoto} M. 2022.
\textit{\mnras} 516:75--83

\bibitem[{{Blum} \& {Wurm}(2008)}]{Blum+08}
{Blum} J, {Wurm} G. 2008.
\textit{\araa} 46:21--56

\bibitem[{{Bodenheimer}, {Hubickyj} \& {Lissauer}(2000)}]{Bodenheimer+00}
{Bodenheimer} P, {Hubickyj} O, {Lissauer} JJ. 2000.
\textit{\icarus} 143:2--14

\bibitem[{{Bodenheimer} \& {Pollack}(1986)}]{Bodenheimer+86}
{Bodenheimer} P, {Pollack} JB. 1986.
\textit{\icarus} 67:391--408

\bibitem[{{Bondi}(1952)}]{Bondi52}
{Bondi} H. 1952.
\textit{\mnras} 112:195

\bibitem[{{Bosman}, {Cridland} \& {Miguel}(2019)}]{Bosman+19}
{Bosman} AD, {Cridland} AJ, {Miguel} Y. 2019.
\textit{\aap} 632:L11

\bibitem[{{Boss}(1997)}]{Boss97}
{Boss} AP. 1997.
\textit{Science} 276:1836--1839

\bibitem[{{Brandt} et~al.(2021){Brandt}, {Brandt}, {Dupuy}, {Michalik} \& {Marleau}}]{Brandt+21}
{Brandt} GM, {Brandt} TD, {Dupuy} TJ, {Michalik} D, {Marleau} GD. 2021.
\textit{\apjl} 915:L16

\bibitem[{{Brasser} \& {Mojzsis}(2020)}]{Brasser+20}
{Brasser} R, {Mojzsis} SJ. 2020.
\textit{Nature Astronomy} 4:492--499

\bibitem[{{Brouwers}, {Vazan} \& {Ormel}(2018)}]{Brouwers+18}
{Brouwers} MG, {Vazan} A, {Ormel} CW. 2018.
\textit{\aap} 611:A65

\bibitem[{{Br{\"u}gger} et~al.(2018){Br{\"u}gger}, {Alibert}, {Ataiee} \& {Benz}}]{Brugger+18}
{Br{\"u}gger} N, {Alibert} Y, {Ataiee} S, {Benz} W. 2018.
\textit{\aap} 619:A174

\bibitem[{{Bryant}, {Bayliss} \& {Van Eylen}(2023)}]{Bryant+23}
{Bryant} EM, {Bayliss} D, {Van Eylen} V. 2023.
\textit{\mnras} 521:3663--3681

\bibitem[{{Budde} et~al.(2016){Budde}, {Burkhardt}, {Brennecka}, {Fischer-G{\"o}dde}, {Kruijer} \& {Kleine}}]{Budde+16}
{Budde} G, {Burkhardt} C, {Brennecka} GA, {Fischer-G{\"o}dde} M, {Kruijer} TS, {Kleine} T. 2016.
\textit{Earth and Planetary Science Letters} 454:293--303

\bibitem[{{Cameron}(1978)}]{Cameron78}
{Cameron} AGW. 1978.
\textit{Moon and Planets} 18:5--40

\bibitem[{{Cimerman}, {Kuiper} \& {Ormel}(2017)}]{Cimerman+17}
{Cimerman} NP, {Kuiper} R, {Ormel} CW. 2017.
\textit{\mnras} 471:4662--4676

\bibitem[{{Cumming} et~al.(2008){Cumming}, {Butler}, {Marcy}, {Vogt}, {Wright} \& {Fischer}}]{Cumming+08}
{Cumming} A, {Butler} RP, {Marcy} GW, {Vogt} SS, {Wright} JT, {Fischer} DA. 2008.
\textit{\pasp} 120:531

\bibitem[{{Currie}(2009)}]{Currie+09}
{Currie} T. 2009.
\textit{\apjl} 694:L171--L176

\bibitem[{{D'Angelo} \& {Bodenheimer}(2013)}]{D'Angelo+13}
{D'Angelo} G, {Bodenheimer} P. 2013.
\textit{\apj} 778:77

\bibitem[{{D'Angelo}, {Kley} \& {Henning}(2003)}]{D'Angelo+03}
{D'Angelo} G, {Kley} W, {Henning} T. 2003.
\textit{\apj} 586:540--561

\bibitem[{{Dawson} \& {Johnson}(2018)}]{Dawson+18}
{Dawson} RI, {Johnson} JA. 2018.
\textit{\araa} 56:175--221

\bibitem[{{Debras} \& {Chabrier}(2018)}]{Debras+18}
{Debras} F, {Chabrier} G. 2018.
\textit{\aap} 609:A97

\bibitem[{{Debras} \& {Chabrier}(2019)}]{Debras+19}
{Debras} F, {Chabrier} G. 2019.
\textit{\apj} 872:100

\bibitem[{{Debras}, {Chabrier} \& {Stevenson}(2021)}]{Debras+21}
{Debras} F, {Chabrier} G, {Stevenson} DJ. 2021.
\textit{\apjl} 913:L21

\bibitem[{{Duffell} \& {MacFadyen}(2013)}]{Duffell+13}
{Duffell} PC, {MacFadyen} AI. 2013.
\textit{\apj} 769:41

\bibitem[{{Dupuy}, {Brandt} \& {Brandt}(2022)}]{Dupuy+22}
{Dupuy} TJ, {Brandt} GM, {Brandt} TD. 2022.
\textit{\mnras} 509:4411--4419

\bibitem[{{Dyrek} et~al.(2024){Dyrek}, {Min}, {Decin}, {Bouwman}, {Crouzet} et~al.}]{Dyrek+24}
{Dyrek} A, {Min} M, {Decin} L, {Bouwman} J, {Crouzet} N, et~al. 2024.
\textit{\nat} 625:51--54

\bibitem[{{Emsenhuber} et~al.(2021){Emsenhuber}, {Mordasini}, {Burn}, {Alibert}, {Benz} \& {Asphaug}}]{Emsenhuber+21}
{Emsenhuber} A, {Mordasini} C, {Burn} R, {Alibert} Y, {Benz} W, {Asphaug} E. 2021.
\textit{\aap} 656:A70

\bibitem[{{Fegley} \& {Lodders}(1994)}]{Fegley+94}
{Fegley} Bruce J, {Lodders} K. 1994.
\textit{\icarus} 110:117--154

\bibitem[{{Fernandes} et~al.(2019){Fernandes}, {Mulders}, {Pascucci}, {Mordasini} \& {Emsenhuber}}]{Fernandes+19}
{Fernandes} RB, {Mulders} GD, {Pascucci} I, {Mordasini} C, {Emsenhuber} A. 2019.
\textit{\apj} 874:81

\bibitem[{{Fischer} \& {Valenti}(2005)}]{Fischer+05}
{Fischer} DA, {Valenti} J. 2005.
\textit{\apj} 622:1102--1117

\bibitem[{{Fletcher} et~al.(2009){Fletcher}, {Orton}, {Teanby}, {Irwin} \& {Bjoraker}}]{Fletcher+09}
{Fletcher} LN, {Orton} GS, {Teanby} NA, {Irwin} PGJ, {Bjoraker} GL. 2009.
\textit{\icarus} 199:351--367

\bibitem[{{Fortney} et~al.(2006){Fortney}, {Saumon}, {Marley}, {Lodders} \& {Freedman}}]{Fortney+06}
{Fortney} JJ, {Saumon} D, {Marley} MS, {Lodders} K, {Freedman} RS. 2006.
\textit{\apj} 642:495--504

\bibitem[{{Fressin} et~al.(2013){Fressin}, {Torres}, {Charbonneau}, {Bryson}, {Christiansen} et~al.}]{Fressin+13}
{Fressin} F, {Torres} G, {Charbonneau} D, {Bryson} ST, {Christiansen} J, et~al. 2013.
\textit{\apj} 766:81

\bibitem[{{Fuller}(2014)}]{Fuller+14}
{Fuller} J. 2014.
\textit{\icarus} 242:283--296

\bibitem[{{Fulton} et~al.(2017){Fulton}, {Petigura}, {Howard}, {Isaacson}, {Marcy} et~al.}]{Fulton+17}
{Fulton} BJ, {Petigura} EA, {Howard} AW, {Isaacson} H, {Marcy} GW, et~al. 2017.
\textit{\aj} 154:109

\bibitem[{{Fulton} et~al.(2021){Fulton}, {Rosenthal}, {Hirsch}, {Isaacson}, {Howard} et~al.}]{Fulton+21}
{Fulton} BJ, {Rosenthal} LJ, {Hirsch} LA, {Isaacson} H, {Howard} AW, et~al. 2021.
\textit{\apjs} 255:14

\bibitem[{{Gan} et~al.(2024){Gan}, {Guo}, {Liu}, {Wang}, {Mao} et~al.}]{Gan+24}
{Gan} T, {Guo} K, {Liu} B, {Wang} SX, {Mao} S, et~al. 2024.
\textit{\apj} 967:74

\bibitem[{{Gan} et~al.(2023){Gan}, {Wang}, {Wang}, {Mao}, {Huang} et~al.}]{Gan+23}
{Gan} T, {Wang} SX, {Wang} S, {Mao} S, {Huang} CX, et~al. 2023.
\textit{\aj} 165:17

\bibitem[{{Goldreich}, {Lithwick} \& {Sari}(2004)}]{Goldreich04}
{Goldreich} P, {Lithwick} Y, {Sari} R. 2004.
\textit{\araa} 42:549--601

\bibitem[{{Guillot} et~al.(2023){Guillot}, {Fletcher}, {Helled}, {Ikoma}, {Line} \& {Paramentier}}]{Guillot+23}
{Guillot} T, {Fletcher} LN, {Helled} R, {Ikoma} M, {Line} MR, {Paramentier} V. 2023.
\textit{{Giant Planets from the Inside-Out}}. In \textit{Protostars and Planets VII}, eds. S~{Inutsuka}, Y~{Aikawa}, T~{Muto}, K~{Tomida}, M~{Tamura}, vol. 534 of \textit{Astronomical Society of the Pacific Conference Series}

\bibitem[{{Guillot} et~al.(2006){Guillot}, {Santos}, {Pont}, {Iro}, {Melo} \& {Ribas}}]{Guillot+06}
{Guillot} T, {Santos} NC, {Pont} F, {Iro} N, {Melo} C, {Ribas} I. 2006.
\textit{\aap} 453:L21--L24

\bibitem[{{Haffert} et~al.(2019){Haffert}, {Bohn}, {de Boer}, {Snellen}, {Brinchmann} et~al.}]{Haffert+19}
{Haffert} SY, {Bohn} AJ, {de Boer} J, {Snellen} IAG, {Brinchmann} J, et~al. 2019.
\textit{Nature Astronomy} 3:749--754

\bibitem[{{Hartman} et~al.(2023){Hartman}, {Bakos}, {Csubry}, {Howard}, {Isaacson} et~al.}]{Hartman+23}
{Hartman} JD, {Bakos} G{\'A}, {Csubry} Z, {Howard} AW, {Isaacson} H, et~al. 2023.
\textit{\aj} 166:163

\bibitem[{{Hasegawa} et~al.(2021){Hasegawa}, {Suzuki}, {Tanaka}, {Kobayashi} \& {Wada}}]{Hasegawa+21}
{Hasegawa} Y, {Suzuki} TK, {Tanaka} H, {Kobayashi} H, {Wada} K. 2021.
\textit{\apj} 915:22

\bibitem[{{Hasegawa} et~al.(2023){Hasegawa}, {Suzuki}, {Tanaka}, {Kobayashi} \& {Wada}}]{Hasegawa+23}
{Hasegawa} Y, {Suzuki} TK, {Tanaka} H, {Kobayashi} H, {Wada} K. 2023.
\textit{\apj} 944:38

\bibitem[{{Hedman} \& {Nicholson}(2013)}]{Hedman+13}
{Hedman} MM, {Nicholson} PD. 2013.
\textit{\aj} 146:12

\bibitem[{{Helled} \& {Stevenson}(2024)}]{Helled+24}
{Helled} R, {Stevenson} DJ. 2024.
\textit{AGU Advances} 5:e2024AV001171

\bibitem[{{Hersant}, {Gautier} \& {Lunine}(2004)}]{Hersant+04}
{Hersant} F, {Gautier} D, {Lunine} JI. 2004.
\textit{\planss} 52:623--641

\bibitem[{Hill(1878)}]{Hill1878}
Hill GW. 1878.
\textit{American Journal of Mathematics} 1:245--260

\bibitem[{{Hori} \& {Ikoma}(2010)}]{Hori+10}
{Hori} Y, {Ikoma} M. 2010.
\textit{\apj} 714:1343--1346

\bibitem[{{Hori} \& {Ikoma}(2011)}]{Hori+11}
{Hori} Y, {Ikoma} M. 2011.
\textit{\mnras} 416:1419--1429

\bibitem[{{Howard} et~al.(2012){Howard}, {Marcy}, {Bryson}, {Jenkins}, {Rowe} et~al.}]{Howard+12}
{Howard} AW, {Marcy} GW, {Bryson} ST, {Jenkins} JM, {Rowe} JF, et~al. 2012.
\textit{\apjs} 201:15

\bibitem[{{Huang}, {Wu} \& {Triaud}(2016)}]{Huang+16}
{Huang} C, {Wu} Y, {Triaud} AHMJ. 2016.
\textit{\apj} 825:98

\bibitem[{{Hubbard} \& {Macfarlane}(1980)}]{Hubbard+80}
{Hubbard} WB, {Macfarlane} JJ. 1980.
\textit{\jgr} 85:225--234

\bibitem[{{Hubickyj}, {Bodenheimer} \& {Lissauer}(2005)}]{Hubickyj+05}
{Hubickyj} O, {Bodenheimer} P, {Lissauer} JJ. 2005.
\textit{\icarus} 179:415--431

\bibitem[{{Ida} \& {Lin}(2004)}]{Ida+04b}
{Ida} S, {Lin} DNC. 2004.
\textit{\apj} 616:567--572

\bibitem[{{Ida} \& {Lin}(2005)}]{Ida+05}
{Ida} S, {Lin} DNC. 2005.
\textit{\apj} 626:1045--1060

\bibitem[{{Iess} et~al.(2018){Iess}, {Folkner}, {Durante}, {Parisi}, {Kaspi} et~al.}]{Iess+18}
{Iess} L, {Folkner} WM, {Durante} D, {Parisi} M, {Kaspi} Y, et~al. 2018.
\textit{\nat} 555:220--222

\bibitem[{{Iess} et~al.(2019){Iess}, {Militzer}, {Kaspi}, {Nicholson}, {Durante} et~al.}]{Iess+19}
{Iess} L, {Militzer} B, {Kaspi} Y, {Nicholson} P, {Durante} D, et~al. 2019.
\textit{Science} 364:aat2965

\bibitem[{{Ikoma}, {Emori} \& {Nakazawa}(2001)}]{Ikoma+01}
{Ikoma} M, {Emori} H, {Nakazawa} K. 2001.
\textit{\apj} 553:999--1005

\bibitem[{{Ikoma} \& {Genda}(2006)}]{Ikoma+Genda06}
{Ikoma} M, {Genda} H. 2006.
\textit{\apj} 648:696--706

\bibitem[{{Ikoma} et~al.(2006){Ikoma}, {Guillot}, {Genda}, {Tanigawa} \& {Ida}}]{Ikoma+06}
{Ikoma} M, {Guillot} T, {Genda} H, {Tanigawa} T, {Ida} S. 2006.
\textit{\apj} 650:1150--1159

\bibitem[{{Ikoma} \& {Hori}(2012)}]{Ikoma+12}
{Ikoma} M, {Hori} Y. 2012.
\textit{\apj} 753:66

\bibitem[{{Ikoma}, {Nakazawa} \& {Emori}(2000)}]{Ikoma+00}
{Ikoma} M, {Nakazawa} K, {Emori} H. 2000.
\textit{\apj} 537:1013--1025

\bibitem[{{Inaba} \& {Ikoma}(2003)}]{Inaba+03AA}
{Inaba} S, {Ikoma} M. 2003.
\textit{\aap} 410:711--723

\bibitem[{{Inaba}, {Wetherill} \& {Ikoma}(2003)}]{Inaba+03Icar}
{Inaba} S, {Wetherill} GW, {Ikoma} M. 2003.
\textit{\icarus} 166:46--62

\bibitem[{{Johansen} \& {Lambrechts}(2017)}]{Johansen+17}
{Johansen} A, {Lambrechts} M. 2017.
\textit{Annual Review of Earth and Planetary Sciences} 45:359--387

\bibitem[{{JWST Transiting Exoplanet Community Early Release Science Team} et~al.(2023){JWST Transiting Exoplanet Community Early Release Science Team}, {Ahrer}, {Alderson}, {Batalha}, {Batalha} et~al.}]{Ahrer+23}
{JWST Transiting Exoplanet Community Early Release Science Team}, {Ahrer} EM, {Alderson} L, {Batalha} NM, {Batalha} NE, et~al. 2023.
\textit{\nat} 614:649--652

\bibitem[{{Kagetani} et~al.(2023){Kagetani}, {Narita}, {Kimura}, {Hirano}, {Ikoma} et~al.}]{Kagetani+23}
{Kagetani} T, {Narita} N, {Kimura} T, {Hirano} T, {Ikoma} M, et~al. 2023.
\textit{\pasj}

\bibitem[{{Kalyaan} et~al.(2023){Kalyaan}, {Pinilla}, {Krijt}, {Banzatti}, {Rosotti} et~al.}]{Kalyaan+23}
{Kalyaan} A, {Pinilla} P, {Krijt} S, {Banzatti} A, {Rosotti} G, et~al. 2023.
\textit{\apj} 954:66

\bibitem[{{Kalyaan} et~al.(2021){Kalyaan}, {Pinilla}, {Krijt}, {Mulders} \& {Banzatti}}]{Kalyaan+21}
{Kalyaan} A, {Pinilla} P, {Krijt} S, {Mulders} GD, {Banzatti} A. 2021.
\textit{\apj} 921:84

\bibitem[{{Kanagawa} et~al.(2017){Kanagawa}, {Tanaka}, {Muto} \& {Tanigawa}}]{Kanagawa+17}
{Kanagawa} KD, {Tanaka} H, {Muto} T, {Tanigawa} T. 2017.
\textit{\pasj} 69:97

\bibitem[{{Kanagawa}, {Tanaka} \& {Szuszkiewicz}(2018)}]{Kanagawa+18}
{Kanagawa} KD, {Tanaka} H, {Szuszkiewicz} E. 2018.
\textit{\apj} 861:140

\bibitem[{{Kataoka} et~al.(2013){Kataoka}, {Tanaka}, {Okuzumi} \& {Wada}}]{Kataoka+13}
{Kataoka} A, {Tanaka} H, {Okuzumi} S, {Wada} K. 2013.
\textit{\aap} 557:L4

\bibitem[{{Kimura} \& {Ikoma}(2020)}]{Kimura+20}
{Kimura} T, {Ikoma} M. 2020.
\textit{\mnras} 496:3755--3766

\bibitem[{{Kleine} et~al.(2020){Kleine}, {Budde}, {Burkhardt}, {Kruijer}, {Worsham} et~al.}]{Kleine+20}
{Kleine} T, {Budde} G, {Burkhardt} C, {Kruijer} TS, {Worsham} EA, et~al. 2020.
\textit{\ssr} 216:55

\bibitem[{{Knutson} et~al.(2014){Knutson}, {Fulton}, {Montet}, {Kao}, {Ngo} et~al.}]{Knutson+14}
{Knutson} HA, {Fulton} BJ, {Montet} BT, {Kao} M, {Ngo} H, et~al. 2014.
\textit{\apj} 785:126

\bibitem[{{Kobayashi} \& {L{\"o}hne}(2014)}]{Kobayashi+14}
{Kobayashi} H, {L{\"o}hne} T. 2014.
\textit{\mnras} 442:3266--3274

\bibitem[{{Kobayashi}, {Ormel} \& {Ida}(2012)}]{Kobayashi+12}
{Kobayashi} H, {Ormel} CW, {Ida} S. 2012.
\textit{\apj} 756:70

\bibitem[{{Kobayashi} \& {Tanaka}(2021)}]{Kobayashi+21}
{Kobayashi} H, {Tanaka} H. 2021.
\textit{\apj} 922:16

\bibitem[{{Kobayashi} \& {Tanaka}(2023)}]{Kobayashi+23}
{Kobayashi} H, {Tanaka} H. 2023.
\textit{\apj} 954:158

\bibitem[{{Kobayashi}, {Tanaka} \& {Krivov}(2011)}]{Kobayashi+11}
{Kobayashi} H, {Tanaka} H, {Krivov} AV. 2011.
\textit{\apj} 738:35

\bibitem[{{Kobayashi} et~al.(2010){Kobayashi}, {Tanaka}, {Krivov} \& {Inaba}}]{Kobayashi+10Icar}
{Kobayashi} H, {Tanaka} H, {Krivov} AV, {Inaba} S. 2010.
\textit{\icarus} 209:836--847

\bibitem[{{Kobayashi}, {Tanaka} \& {Okuzumi}(2016)}]{Kobayashi+16}
{Kobayashi} H, {Tanaka} H, {Okuzumi} S. 2016.
\textit{\apj} 817:105

\bibitem[{{Kokubo} \& {Ida}(1998)}]{Kokubo+98}
{Kokubo} E, {Ida} S. 1998.
\textit{\icarus} 131:171--178

\bibitem[{{Kruijer} et~al.(2017){Kruijer}, {Burkhardt}, {Budde} \& {Kleine}}]{Kruijer+17}
{Kruijer} TS, {Burkhardt} C, {Budde} G, {Kleine} T. 2017.
\textit{Proceedings of the National Academy of Science} 114:6712--6716

\bibitem[{{Kruijer}, {Kleine} \& {Borg}(2020)}]{Kruijer+20}
{Kruijer} TS, {Kleine} T, {Borg} LE. 2020.
\textit{Nature Astronomy} 4:32--40

\bibitem[{{Kuiper}(1951)}]{Kuiper51}
{Kuiper} GP. 1951.
\textit{Proceedings of the National Academy of Science} 37:1--14

\bibitem[{{Kurokawa} \& {Tanigawa}(2018)}]{Kurokawa+18}
{Kurokawa} H, {Tanigawa} T. 2018.
\textit{\mnras} 479:635--648

\bibitem[{{Kurosaki} \& {Ikoma}(2017)}]{Kurosaki+17}
{Kurosaki} K, {Ikoma} M. 2017.
\textit{\aj} 153:260

\bibitem[{{Kuwahara} \& {Kurokawa}(2020)}]{Kuwahara+20}
{Kuwahara} A, {Kurokawa} H. 2020.
\textit{\aap} 643:A21

\bibitem[{{Lagrange} et~al.(2010){Lagrange}, {Bonnefoy}, {Chauvin}, {Apai}, {Ehrenreich} et~al.}]{Lagrange+10}
{Lagrange} AM, {Bonnefoy} M, {Chauvin} G, {Apai} D, {Ehrenreich} D, et~al. 2010.
\textit{Science} 329:57

\bibitem[{{Lambrechts} \& {Johansen}(2014)}]{Lambrechts+2014a}
{Lambrechts} M, {Johansen} A. 2014.
\textit{\aap} 572:A107

\bibitem[{{Lambrechts}, {Johansen} \& {Morbidelli}(2014)}]{Lambrechts+2014b}
{Lambrechts} M, {Johansen} A, {Morbidelli} A. 2014.
\textit{\aap} 572:A35

\bibitem[{{Lee}, {Chiang} \& {Ormel}(2014)}]{Lee+14}
{Lee} EJ, {Chiang} E, {Ormel} CW. 2014.
\textit{\apj} 797:95

\bibitem[{{Li} et~al.(2020){Li}, {Ingersoll}, {Bolton}, {Levin}, {Janssen} et~al.}]{Li+20}
{Li} C, {Ingersoll} A, {Bolton} S, {Levin} S, {Janssen} M, et~al. 2020.
\textit{Nature Astronomy} 4:609--616

\bibitem[{{Lissauer} et~al.(2011){Lissauer}, {Fabrycky}, {Ford}, {Borucki}, {Fressin} et~al.}]{Lissauer+11}
{Lissauer} JJ, {Fabrycky} DC, {Ford} EB, {Borucki} WJ, {Fressin} F, et~al. 2011.
\textit{\nat} 470:53--58

\bibitem[{{Lissauer} et~al.(2009){Lissauer}, {Hubickyj}, {D'Angelo} \& {Bodenheimer}}]{Lissauer+09}
{Lissauer} JJ, {Hubickyj} O, {D'Angelo} G, {Bodenheimer} P. 2009.
\textit{\icarus} 199:338--350

\bibitem[{{Lubow}, {Seibert} \& {Artymowicz}(1999)}]{Lubow+99}
{Lubow} SH, {Seibert} M, {Artymowicz} P. 1999.
\textit{\apj} 526:1001--1012

\bibitem[{{Machida}, {Inutsuka} \& {Matsumoto}(2011)}]{Machida+11}
{Machida} MN, {Inutsuka} Si, {Matsumoto} T. 2011.
\textit{\apj} 729:42

\bibitem[{{Mah} et~al.(2023){Mah}, {Bitsch}, {Pascucci} \& {Henning}}]{Mah+23}
{Mah} J, {Bitsch} B, {Pascucci} I, {Henning} T. 2023.
\textit{\aap} 677:L7

\bibitem[{{Mahaffy} et~al.(2000){Mahaffy}, {Niemann}, {Alpert}, {Atreya}, {Demick} et~al.}]{Mahaffy+00}
{Mahaffy} PR, {Niemann} HB, {Alpert} A, {Atreya} SK, {Demick} J, et~al. 2000.
\textit{\jgr} 105:15061--15072

\bibitem[{{Mankovich} \& {Fuller}(2021)}]{Mankovich+21}
{Mankovich} CR, {Fuller} J. 2021.
\textit{Nature Astronomy} 5:1103--1109

\bibitem[{{Marleau} et~al.(2017){Marleau}, {Klahr}, {Kuiper} \& {Mordasini}}]{Marleau+17}
{Marleau} GD, {Klahr} H, {Kuiper} R, {Mordasini} C. 2017.
\textit{\apj} 836:221

\bibitem[{{Marleau}, {Mordasini} \& {Kuiper}(2019)}]{Marleau+19}
{Marleau} GD, {Mordasini} C, {Kuiper} R. 2019.
\textit{\apj} 881:144

\bibitem[{{Marley}(1991)}]{Marley91}
{Marley} MS. 1991.
\textit{\icarus} 94:420--435

\bibitem[{{Marley} et~al.(2007){Marley}, {Fortney}, {Hubickyj}, {Bodenheimer} \& {Lissauer}}]{Marley+07}
{Marley} MS, {Fortney} JJ, {Hubickyj} O, {Bodenheimer} P, {Lissauer} JJ. 2007.
\textit{\apj} 655:541--549

\bibitem[{{Marois} et~al.(2008){Marois}, {Macintosh}, {Barman}, {Zuckerman}, {Song} et~al.}]{Marios+08}
{Marois} C, {Macintosh} B, {Barman} T, {Zuckerman} B, {Song} I, et~al. 2008.
\textit{Science} 322:1348

\bibitem[{{Marois} et~al.(2010){Marois}, {Zuckerman}, {Konopacky}, {Macintosh} \& {Barman}}]{Marios+10}
{Marois} C, {Zuckerman} B, {Konopacky} QM, {Macintosh} B, {Barman} T. 2010.
\textit{\nat} 468:1080--1083

\bibitem[{{Mayor} \& {Queloz}(1995)}]{Mayor+95}
{Mayor} M, {Queloz} D. 1995.
\textit{\nat} 378:355--359

\bibitem[{{Miguel} et~al.(2022){Miguel}, {Bazot}, {Guillot}, {Howard}, {Galanti} et~al.}]{Miguel+22}
{Miguel} Y, {Bazot} M, {Guillot} T, {Howard} S, {Galanti} E, et~al. 2022.
\textit{\aap} 662:A18

\bibitem[{{Miki}(1982)}]{Miki82}
{Miki} S. 1982.
\textit{Progress of Theoretical Physics} 67:1053--1067

\bibitem[{{Militzer} et~al.(2022){Militzer}, {Hubbard}, {Wahl}, {Lunine}, {Galanti} et~al.}]{Militzer+22}
{Militzer} B, {Hubbard} WB, {Wahl} S, {Lunine} JI, {Galanti} E, et~al. 2022.
\textit{\psj} 3:185

\bibitem[{{Miller} \& {Fortney}(2011)}]{Miller+11}
{Miller} N, {Fortney} JJ. 2011.
\textit{\apjl} 736:L29

\bibitem[{{Mizuno}(1980)}]{Mizuno80}
{Mizuno} H. 1980.
\textit{Progress of Theoretical Physics} 64:544--557

\bibitem[{{Mizuno}, {Nakazawa} \& {Hayashi}(1978)}]{Mizuno+78}
{Mizuno} H, {Nakazawa} K, {Hayashi} C. 1978.
\textit{Progress of Theoretical Physics} 60:699--710

\bibitem[{{Moldenhauer} et~al.(2022){Moldenhauer}, {Kuiper}, {Kley} \& {Ormel}}]{Moldenhauer+22}
{Moldenhauer} TW, {Kuiper} R, {Kley} W, {Ormel} CW. 2022.
\textit{\aap} 661:A142

\bibitem[{{Morbidelli} et~al.(2016){Morbidelli}, {Bitsch}, {Crida}, {Gounelle}, {Guillot} et~al.}]{Morbidelli+16}
{Morbidelli} A, {Bitsch} B, {Crida} A, {Gounelle} M, {Guillot} T, et~al. 2016.
\textit{\icarus} 267:368--376

\bibitem[{{Mordasini}(2014)}]{Mordasini14}
{Mordasini} C. 2014.
\textit{\aap} 572:A118

\bibitem[{{Mordasini} et~al.(2012){Mordasini}, {Alibert}, {Benz}, {Klahr} \& {Henning}}]{Mordasini+12}
{Mordasini} C, {Alibert} Y, {Benz} W, {Klahr} H, {Henning} T. 2012.
\textit{\aap} 541:A97

\bibitem[{{Mousis} et~al.(2012){Mousis}, {Lunine}, {Madhusudhan} \& {Johnson}}]{Mousis+12}
{Mousis} O, {Lunine} JI, {Madhusudhan} N, {Johnson} TV. 2012.
\textit{\apjl} 751:L7

\bibitem[{{Movshovitz} et~al.(2010){Movshovitz}, {Bodenheimer}, {Podolak} \& {Lissauer}}]{Movshovitz+10}
{Movshovitz} N, {Bodenheimer} P, {Podolak} M, {Lissauer} JJ. 2010.
\textit{\icarus} 209:616--624

\bibitem[{{Movshovitz} \& {Podolak}(2008)}]{Movshovitz+08}
{Movshovitz} N, {Podolak} M. 2008.
\textit{\icarus} 194:368--378

\bibitem[{{Mulders} et~al.(2021){Mulders}, {Pascucci}, {Ciesla} \& {Fernandes}}]{Mulders+21}
{Mulders} GD, {Pascucci} I, {Ciesla} FJ, {Fernandes} RB. 2021.
\textit{\apj} 920:66

\bibitem[{{M{\"u}ller}, {Ben-Yami} \& {Helled}(2020)}]{Muller+20}
{M{\"u}ller} S, {Ben-Yami} M, {Helled} R. 2020.
\textit{\apj} 903:147

\bibitem[{{Nakajima} et~al.(1995){Nakajima}, {Oppenheimer}, {Kulkarni}, {Golimowski}, {Matthews} \& {Durrance}}]{Nakajima+95}
{Nakajima} T, {Oppenheimer} BR, {Kulkarni} SR, {Golimowski} DA, {Matthews} K, {Durrance} ST. 1995.
\textit{\nat} 378:463--465

\bibitem[{{Naoz} et~al.(2011){Naoz}, {Farr}, {Lithwick}, {Rasio} \& {Teyssandier}}]{Noaz+11}
{Naoz} S, {Farr} WM, {Lithwick} Y, {Rasio} FA, {Teyssandier} J. 2011.
\textit{\nat} 473:187--189

\bibitem[{{Ndugu}, {Bitsch} \& {Jurua}(2018)}]{Ndugu+18}
{Ndugu} N, {Bitsch} B, {Jurua} E. 2018.
\textit{\mnras} 474:886--897

\bibitem[{{Nettelmann} et~al.(2021){Nettelmann}, {Movshovitz}, {Ni}, {Fortney}, {Galanti} et~al.}]{Nettelmann+21}
{Nettelmann} N, {Movshovitz} N, {Ni} D, {Fortney} JJ, {Galanti} E, et~al. 2021.
\textit{\psj} 2:241

\bibitem[{{Nielsen} et~al.(2019){Nielsen}, {De Rosa}, {Macintosh}, {Wang}, {Ruffio} et~al.}]{Nielsen+19}
{Nielsen} EL, {De Rosa} RJ, {Macintosh} B, {Wang} JJ, {Ruffio} JB, et~al. 2019.
\textit{\aj} 158:13

\bibitem[{{Niemann} et~al.(1998){Niemann}, {Atreya}, {Carignan}, {Donahue}, {Haberman} et~al.}]{Niemann+98}
{Niemann} HB, {Atreya} SK, {Carignan} GR, {Donahue} TM, {Haberman} JA, et~al. 1998.
\textit{\jgr} 103:22831--22846

\bibitem[{{{\"O}berg} \& {Wordsworth}(2019)}]{Oberg+19}
{{\"O}berg} KI, {Wordsworth} R. 2019.
\textit{\aj} 158:194

\bibitem[{{Ohashi} et~al.(2021){Ohashi}, {Kobayashi}, {Nakatani}, {Okuzumi}, {Tanaka} et~al.}]{Ohashi+21}
{Ohashi} S, {Kobayashi} H, {Nakatani} R, {Okuzumi} S, {Tanaka} H, et~al. 2021.
\textit{\apj} 907:80

\bibitem[{{Ohno} \& {Ueda}(2021)}]{Ohno+21}
{Ohno} K, {Ueda} T. 2021.
\textit{\aap} 651:L2

\bibitem[{{Okamura} \& {Kobayashi}(2021)}]{Okamura+21}
{Okamura} T, {Kobayashi} H. 2021.
\textit{\apj} 916:109

\bibitem[{{Ormel}(2014)}]{Ormel14}
{Ormel} CW. 2014.
\textit{\apjl} 789:L18

\bibitem[{{Ormel}, {Kuiper} \& {Shi}(2015)}]{Ormal+15a}
{Ormel} CW, {Kuiper} R, {Shi} JM. 2015.
\textit{\mnras} 446:1026--1040

\bibitem[{{Ormel} \& {Liu}(2018)}]{Ormel+18}
{Ormel} CW, {Liu} B. 2018.
\textit{\aap} 615:A178

\bibitem[{{Ormel}, {Shi} \& {Kuiper}(2015)}]{Ormel+15b}
{Ormel} CW, {Shi} JM, {Kuiper} R. 2015.
\textit{\mnras} 447:3512--3525

\bibitem[{{Owen} et~al.(1999){Owen}, {Mahaffy}, {Niemann}, {Atreya}, {Donahue} et~al.}]{Owen+99}
{Owen} T, {Mahaffy} P, {Niemann} HB, {Atreya} S, {Donahue} T, et~al. 1999.
\textit{\nat} 402:269--270

\bibitem[{{Paardekooper} et~al.(2023){Paardekooper}, {Dong}, {Duffell}, {Fung}, {Masset} et~al.}]{Paardekooper+23}
{Paardekooper} S, {Dong} R, {Duffell} P, {Fung} J, {Masset} FS, et~al. 2023.
\textit{{Planet-Disk Interactions and Orbital Evolution}}. In \textit{Protostars and Planets VII}, eds. S~{Inutsuka}, Y~{Aikawa}, T~{Muto}, K~{Tomida}, M~{Tamura}, vol. 534 of \textit{Astronomical Society of the Pacific Conference Series}

\bibitem[{{Papaloizou} \& {Lin}(1984)}]{Papaloizou+84}
{Papaloizou} J, {Lin} DNC. 1984.
\textit{\apj} 285:818--834

\bibitem[{{Perri} \& {Cameron}(1974)}]{Perri+74}
{Perri} F, {Cameron} AGW. 1974.
\textit{\icarus} 22:416--425

\bibitem[{{Pinhas} et~al.(2019){Pinhas}, {Madhusudhan}, {Gandhi} \& {MacDonald}}]{Pinhas+19}
{Pinhas} A, {Madhusudhan} N, {Gandhi} S, {MacDonald} R. 2019.
\textit{\mnras} 482:1485--1498

\bibitem[{{Podolak}(2003)}]{Podolak03}
{Podolak} M. 2003.
\textit{\icarus} 165:428--437

\bibitem[{{Podolak}, {Pollack} \& {Reynolds}(1988)}]{Podolak+88}
{Podolak} M, {Pollack} JB, {Reynolds} RT. 1988.
\textit{\icarus} 73:163--179

\bibitem[{{Pollack} et~al.(1996){Pollack}, {Hubickyj}, {Bodenheimer}, {Lissauer}, {Podolak} \& {Greenzweig}}]{Pollack+96}
{Pollack} JB, {Hubickyj} O, {Bodenheimer} P, {Lissauer} JJ, {Podolak} M, {Greenzweig} Y. 1996.
\textit{\icarus} 124:62--85

\bibitem[{{Polman} et~al.(2023){Polman}, {Waters}, {Min}, {Miguel} \& {Khorshid}}]{Polman+23}
{Polman} J, {Waters} LBFM, {Min} M, {Miguel} Y, {Khorshid} N. 2023.
\textit{\aap} 670:A161

\bibitem[{{Popovas} et~al.(2018){Popovas}, {Nordlund}, {Ramsey} \& {Ormel}}]{Popovas+18}
{Popovas} A, {Nordlund} {\r{A}}, {Ramsey} JP, {Ormel} CW. 2018.
\textit{\mnras} 479:5136--5156

\bibitem[{{Rasio} \& {Ford}(1996)}]{Rasio+96}
{Rasio} FA, {Ford} EB. 1996.
\textit{Science} 274:954--956

\bibitem[{{Rustamkulov} et~al.(2023){Rustamkulov}, {Sing}, {Mukherjee}, {May}, {Kirk} et~al.}]{Rustamkulov+23}
{Rustamkulov} Z, {Sing} DK, {Mukherjee} S, {May} EM, {Kirk} J, et~al. 2023.
\textit{\nat} 614:659--663

\bibitem[{{Santos}, {Israelian} \& {Mayor}(2004)}]{Santos+04}
{Santos} NC, {Israelian} G, {Mayor} M. 2004.
\textit{\aap} 415:1153--1166

\bibitem[{{Sato} et~al.(2005){Sato}, {Fischer}, {Henry}, {Laughlin}, {Butler} et~al.}]{Sato+05}
{Sato} B, {Fischer} DA, {Henry} GW, {Laughlin} G, {Butler} RP, et~al. 2005.
\textit{\apj} 633:465--473

\bibitem[{{Saumon} \& {Guillot}(2004)}]{Saumon+04}
{Saumon} D, {Guillot} T. 2004.
\textit{\apj} 609:1170--1180

\bibitem[{{Schneider} \& {Bitsch}(2021)}]{Schneider+21}
{Schneider} AD, {Bitsch} B. 2021.
\textit{\aap} 654:A71

\bibitem[{{Semenov} et~al.(2003){Semenov}, {Henning}, {Helling}, {Ilgner} \& {Sedlmayr}}]{Semenov+03}
{Semenov} D, {Henning} T, {Helling} C, {Ilgner} M, {Sedlmayr} E. 2003.
\textit{\aap} 410:611--621

\bibitem[{{Shibata}, {Helled} \& {Ikoma}(2020)}]{Shibata+20}
{Shibata} S, {Helled} R, {Ikoma} M. 2020.
\textit{\aap} 633:A33

\bibitem[{{Shibata}, {Helled} \& {Ikoma}(2022)}]{Shibata+22}
{Shibata} S, {Helled} R, {Ikoma} M. 2022.
\textit{\aap} 659:A28

\bibitem[{{Shibata}, {Helled} \& {Kobayashi}(2023)}]{Shibata+23}
{Shibata} S, {Helled} R, {Kobayashi} H. 2023.
\textit{\mnras} 519:1713--1731

\bibitem[{{Shibata} \& {Ikoma}(2019)}]{Shibata+19}
{Shibata} S, {Ikoma} M. 2019.
\textit{\mnras} 487:4510--4524

\bibitem[{{Shiraishi} \& {Ida}(2008)}]{Shiraishi+08}
{Shiraishi} M, {Ida} S. 2008.
\textit{\apj} 684:1416--1426

\bibitem[{{Sing} et~al.(2016){Sing}, {Fortney}, {Nikolov}, {Wakeford}, {Kataria} et~al.}]{Sing+16}
{Sing} DK, {Fortney} JJ, {Nikolov} N, {Wakeford} HR, {Kataria} T, et~al. 2016.
\textit{\nat} 529:59--62

\bibitem[{{Slattery}(1977)}]{Slattery77}
{Slattery} WL. 1977.
\textit{\icarus} 32:58--72

\bibitem[{{Smith} \& {Terrile}(1984)}]{Smith+84}
{Smith} BA, {Terrile} RJ. 1984.
\textit{Science} 226:1421--1424

\bibitem[{{Smoluchowski}(1916)}]{Smoluchowski16}
{Smoluchowski} MV. 1916.
\textit{Zeitschrift fur Physik} 17:557--585

\bibitem[{{Snellen} \& {Brown}(2018)}]{Snellen+18}
{Snellen} IAG, {Brown} AGA. 2018.
\textit{Nature Astronomy} 2:883--886

\bibitem[{{Stevenson}(1982)}]{Stevenson82}
{Stevenson} DJ. 1982.
\textit{\planss} 30:755--764

\bibitem[{{Suyama} et~al.(2012){Suyama}, {Wada}, {Tanaka} \& {Okuzumi}}]{Suyama+12}
{Suyama} T, {Wada} K, {Tanaka} H, {Okuzumi} S. 2012.
\textit{\apj} 753:115

\bibitem[{{Szul{\'a}gyi} et~al.(2016){Szul{\'a}gyi}, {Masset}, {Lega}, {Crida}, {Morbidelli} \& {Guillot}}]{Szulagyi+16}
{Szul{\'a}gyi} J, {Masset} F, {Lega} E, {Crida} A, {Morbidelli} A, {Guillot} T. 2016.
\textit{\mnras} 460:2853--2861

\bibitem[{{Tajima} \& {Nakagawa}(1997)}]{Tajima+97}
{Tajima} N, {Nakagawa} Y. 1997.
\textit{\icarus} 126:282--292

\bibitem[{{Takasao}, {Aoyama} \& {Ikoma}(2021)}]{Takasao+21}
{Takasao} S, {Aoyama} Y, {Ikoma} M. 2021.
\textit{\apj} 921:10

\bibitem[{{Tanaka}, {Anayama} \& {Tazaki}(2023)}]{Tanaka+23}
{Tanaka} H, {Anayama} R, {Tazaki} R. 2023.
\textit{\apj} 945:68

\bibitem[{{Tanaka}, {Murase} \& {Tanigawa}(2020)}]{Tanaka+20}
{Tanaka} H, {Murase} K, {Tanigawa} T. 2020.
\textit{\apj} 891:143

\bibitem[{{Tanaka}, {Takeuchi} \& {Ward}(2002)}]{Tanaka+02}
{Tanaka} H, {Takeuchi} T, {Ward} WR. 2002.
\textit{\apj} 565:1257--1274

\bibitem[{{Tanigawa} \& {Ikoma}(2007)}]{Tanigawa+07}
{Tanigawa} T, {Ikoma} M. 2007.
\textit{\apj} 667:557--570

\bibitem[{{Tanigawa}, {Ohtsuki} \& {Machida}(2012)}]{Tanigawa+12}
{Tanigawa} T, {Ohtsuki} K, {Machida} MN. 2012.
\textit{\apj} 747:47

\bibitem[{{Tanigawa} \& {Tanaka}(2016)}]{Tanigawa_Tanaka16}
{Tanigawa} T, {Tanaka} H. 2016.
\textit{\apj} 823:48

\bibitem[{{Tanigawa} \& {Watanabe}(2002)}]{Tanigawa+02}
{Tanigawa} T, {Watanabe} Si. 2002.
\textit{\apj} 580:506--518

\bibitem[{{Tazaki} \& {Dominik}(2022)}]{Tazaki+22}
{Tazaki} R, {Dominik} C. 2022.
\textit{\aap} 663:A57

\bibitem[{{Tazaki}, {Ginski} \& {Dominik}(2023)}]{Tazaki+23}
{Tazaki} R, {Ginski} C, {Dominik} C. 2023.
\textit{\apjl} 944:L43

\bibitem[{{Tazaki} et~al.(2019){Tazaki}, {Tanaka}, {Kataoka}, {Okuzumi} \& {Muto}}]{Tazaki+19}
{Tazaki} R, {Tanaka} H, {Kataoka} A, {Okuzumi} S, {Muto} T. 2019.
\textit{\apj} 885:52

\bibitem[{{Thorngren} et~al.(2016){Thorngren}, {Fortney}, {Murray-Clay} \& {Lopez}}]{Thorngren+16}
{Thorngren} DP, {Fortney} JJ, {Murray-Clay} RA, {Lopez} ED. 2016.
\textit{\apj} 831:64

\bibitem[{{Tinetti} et~al.(2007){Tinetti}, {Vidal-Madjar}, {Liang}, {Beaulieu}, {Yung} et~al.}]{Tinetti+07}
{Tinetti} G, {Vidal-Madjar} A, {Liang} MC, {Beaulieu} JP, {Yung} Y, et~al. 2007.
\textit{\nat} 448:169--171

\bibitem[{{Trinquier}, {Birck} \& {All{\`e}gre}(2007)}]{Trinquier+07}
{Trinquier} A, {Birck} JL, {All{\`e}gre} CJ. 2007.
\textit{\apj} 655:1179--1185

\bibitem[{{Tychoniec} et~al.(2020){Tychoniec}, {Manara}, {Rosotti}, {van Dishoeck}, {Cridland} et~al.}]{Tychoniec20}
{Tychoniec} {\L}, {Manara} CF, {Rosotti} GP, {van Dishoeck} EF, {Cridland} AJ, et~al. 2020.
\textit{\aap} 640:A19

\bibitem[{{Udry} et~al.(2002){Udry}, {Mayor}, {Naef}, {Pepe}, {Queloz} et~al.}]{Udry+02}
{Udry} S, {Mayor} M, {Naef} D, {Pepe} F, {Queloz} D, et~al. 2002.
\textit{\aap} 390:267--279

\bibitem[{{Valletta} \& {Helled}(2019)}]{Valletta+19}
{Valletta} C, {Helled} R. 2019.
\textit{\apj} 871:127

\bibitem[{{van der Marel} \& {Mulders}(2021)}]{van_der_Marel+21}
{van der Marel} N, {Mulders} GD. 2021.
\textit{\aj} 162:28

\bibitem[{{van Dishoeck} et~al.(2014){van Dishoeck}, {Bergin}, {Lis} \& {Lunine}}]{van_Dishoeck+14}
{van Dishoeck} EF, {Bergin} EA, {Lis} DC, {Lunine} JI. 2014.
\textit{{Water: From Clouds to Planets}}. In \textit{Protostars and Planets VI}, eds. H~{Beuther}, RS~{Klessen}, CP~{Dullemond}, T~{Henning}

\bibitem[{{van Dishoeck} et~al.(2021){van Dishoeck}, {Kristensen}, {Mottram}, {Benz}, {Bergin} et~al.}]{van_Dishoeck+21}
{van Dishoeck} EF, {Kristensen} LE, {Mottram} JC, {Benz} AO, {Bergin} EA, et~al. 2021.
\textit{\aap} 648:A24

\bibitem[{{Venturini}, {Alibert} \& {Benz}(2016)}]{Venturini+16}
{Venturini} J, {Alibert} Y, {Benz} W. 2016.
\textit{\aap} 596:A90

\bibitem[{{Venturini} et~al.(2015){Venturini}, {Alibert}, {Benz} \& {Ikoma}}]{Venturini+15}
{Venturini} J, {Alibert} Y, {Benz} W, {Ikoma} M. 2015.
\textit{\aap} 576:A114

\bibitem[{{Visscher}, {Lodders} \& {Fegley}(2006)}]{Visscher+06}
{Visscher} C, {Lodders} K, {Fegley} Bruce J. 2006.
\textit{\apj} 648:1181--1195

\bibitem[{{Wada} et~al.(2013){Wada}, {Tanaka}, {Okuzumi}, {Kobayashi}, {Suyama} et~al.}]{Wada+13}
{Wada} K, {Tanaka} H, {Okuzumi} S, {Kobayashi} H, {Suyama} T, et~al. 2013.
\textit{\aap} 559:A62

\bibitem[{{Wagner} et~al.(2018){Wagner}, {Follete}, {Close}, {Apai}, {Gibbs} et~al.}]{Wagner+18}
{Wagner} K, {Follete} KB, {Close} LM, {Apai} D, {Gibbs} A, et~al. 2018.
\textit{\apjl} 863:L8

\bibitem[{{Wahl} et~al.(2017){Wahl}, {Hubbard}, {Militzer}, {Guillot}, {Miguel} et~al.}]{Wahl+17}
{Wahl} SM, {Hubbard} WB, {Militzer} B, {Guillot} T, {Miguel} Y, et~al. 2017.
\textit{\grl} 44:4649--4659

\bibitem[{{Walsh} et~al.(2011){Walsh}, {Morbidelli}, {Raymond}, {O'Brien} \& {Mandell}}]{Walsh+11}
{Walsh} KJ, {Morbidelli} A, {Raymond} SN, {O'Brien} DP, {Mandell} AM. 2011.
\textit{\nat} 475:206--209

\bibitem[{{Warren}(2011)}]{Warren11}
{Warren} PH. 2011.
\textit{Earth and Planetary Science Letters} 311:93--100

\bibitem[{{Wilson} \& {Militzer}(2010)}]{Wilson+10}
{Wilson} HF, {Militzer} B. 2010.
\textit{\prl} 104:121101

\bibitem[{{Wong} et~al.(2004){Wong}, {Mahaffy}, {Atreya}, {Niemann} \& {Owen}}]{Wong+04}
{Wong} MH, {Mahaffy} PR, {Atreya} SK, {Niemann} HB, {Owen} TC. 2004.
\textit{\icarus} 171:153--170

\bibitem[{{Wright} et~al.(2012){Wright}, {Marcy}, {Howard}, {Johnson}, {Morton} \& {Fischer}}]{Wright+12}
{Wright} JT, {Marcy} GW, {Howard} AW, {Johnson} JA, {Morton} TD, {Fischer} DA. 2012.
\textit{\apj} 753:160

\bibitem[{{Wuchterl}(1993)}]{Wuchterl93}
{Wuchterl} G. 1993.
\textit{\icarus} 106:323--334

\bibitem[{{Yasui} et~al.(2014){Yasui}, {Kobayashi}, {Tokunaga} \& {Saito}}]{Yasui+14}
{Yasui} C, {Kobayashi} N, {Tokunaga} AT, {Saito} M. 2014.
\textit{\mnras} 442:2543--2559

\bibitem[{{Zhou} et~al.(2019){Zhou}, {Huang}, {Bakos}, {Hartman}, {Latham} et~al.}]{Zhou+19}
{Zhou} G, {Huang} CX, {Bakos} G{\'A}, {Hartman} JD, {Latham} DW, et~al. 2019.
\textit{\aj} 158:141

\end{thebibliography}
\bibliographystyle{ar-style2}

\end{document}